\documentclass[oneside,11pt]{article}

\usepackage[utf8]{inputenc}
\usepackage[T1]{fontenc}

\usepackage{amsthm}
\usepackage{amsmath}
\usepackage{amsfonts}
\usepackage{amssymb}

\usepackage{graphicx} 
\usepackage{float}
\usepackage{booktabs}
\usepackage{multirow}
\usepackage{rotating}
\usepackage{array}
\usepackage{xcolor}
\usepackage{subfig}
\usepackage[ruled]{algorithm2e}
\newcolumntype{R}[1]{>{\raggedleft\let\newline\\\arraybackslash\hspace{0pt}}m{#1}}
\newcolumntype{L}[1]{>{\raggedright\let\newline\\\arraybackslash\hspace{0pt}}m{#1}}
\newcolumntype{C}[1]{>{\centering\let\newline\\\arraybackslash\hspace{0pt}}m{#1}}

\usepackage{breakcites}

\usepackage{nowidow}

\usepackage{enumitem}
\newlist{inlinelist}{enumerate*}{1}
\setlist[inlinelist]{label=(\textit{\alph*}),ref=(\textit{\alph*})}

\usepackage[top=1.5cm,bottom=2cm,right=2.5cm,left=2.5cm]{geometry}
\usepackage{titling}

\usepackage{natbib}

\usepackage{ifplatform}

\usepackage{textcomp}

\ifwindows
\fi

\allowdisplaybreaks

\newcommand{\Sq}{\mathbb{S}^q}

\newcommand{\rd}{\mathrm{d}}

\newcommand{\bx}{\mathbf{x}}
\newcommand{\be}{\mathbf{e}}

\newcommand{\by}{\mathbf{y}}
\newcommand{\bX}{\mathbf{X}}

\newcommand{\bmu}{\boldsymbol\mu}
\newcommand{\bu}{\mathbf{u}}
\newcommand{\bv}{\mathbf{v}}

\newcommand{\bxi}{\boldsymbol\xi}

\newcommand{\btheta}{\boldsymbol\theta}

\newcommand{\Ical}{\mathcal{I}}

\newcommand{\Hcal}{\mathcal{H}}

\newcommand{\bB}{\mathbf{B}}

\newcommand{\equald}{\stackrel{d}{=}}

\newcommand{\inlawH}{\stackrel{\Hcal_0}{\rightsquigarrow}}

\newcommand{\lrp}[1]{\left(#1\right)}

\newcommand{\om}[1]{\omega_{#1}}

\DeclareFontFamily{OT1}{pzc}{}
\DeclareFontShape{OT1}{pzc}{m}{it}{<-> s * [1.10] pzcmi7t}{}
\DeclareMathAlphabet{\mathpzc}{OT1}{pzc}{m}{it}

\newtheorem{definition}{Definition}
\newtheorem{theorem}{Theorem}
\newtheorem{corollary}{Corollary}
\newtheorem{remark}{Remark}
\newtheorem{proposition}{Proposition}
\newtheorem{lemma}{Lemma}

\usepackage[pdftex,final,bookmarksnumbered,bookmarksopen=false,breaklinks,colorlinks]{hyperref}
\hypersetup{
	final,
	bookmarksnumbered=true,
	bookmarksopen=true,
	bookmarksopenlevel=0,
	unicode=false,
	pdftoolbar=true,
	pdfmenubar=true,
	pdffitwindow=false,
	pdftitle={On new omnibus tests of uniformity on the hypersphere},
	pdfdisplaydoctitle=true,
	pdftoolbar=true,
	pdfmenubar=true,
	pdflang={English},
	pdfauthor={Alberto Fernandez-de-Marcos, Eduardo Garcia-Portugues},
	pdfsubject={arXiv paper},
	pdfcreator={Alberto Fernandez-de-Marcos},
	pdfproducer={Alberto Fernandez-de-Marcos},
	pdfkeywords={Directional statistics}{Poisson kernel}{Smooth maximum}{Sobolev tests},
	pdfnewwindow=true,
	breaklinks=true,
	hidelinks,       
	linkcolor=black,
	citecolor=black,
	filecolor=magenta,
	urlcolor=cyan
}

\newif\ifmain
\maintrue
\newif\ifsupplement
\supplementtrue

\newif\iffigstabs
\figstabstrue

\newcommand{\myref}[2]{
#2} 

\begin{document}

\ifmain

\title{On new omnibus tests of uniformity on the hypersphere}
\setlength{\droptitle}{-1cm}
\predate{}%
\postdate{}%
\date{}

\author{Alberto Fern\'andez-de-Marcos$^{1}$ and Eduardo Garc\'ia-Portugu\'es$^{1,2}$}
\footnotetext[1]{Department of Statistics, Universidad Carlos III de Madrid (Spain).}
\footnotetext[2]{Corresponding author. e-mail: \href{mailto:edgarcia@est-econ.uc3m.es}{edgarcia@est-econ.uc3m.es}.}
\maketitle

\begin{abstract}
Two new omnibus tests of uniformity for data on the hypersphere are proposed. The new test statistics exploit closed-form expressions for orthogonal polynomials, feature tuning parameters, and are related to a ``smooth maximum'' function and the Poisson kernel. We obtain exact moments of the test statistics under uniformity and rotationally symmetric alternatives, and give their null asymptotic distributions. We consider approximate oracle tuning parameters that maximize the power of the tests against known generic alternatives and provide tests that estimate oracle parameters through cross-validated procedures while maintaining the significance level. Numerical experiments explore the effectiveness of null asymptotic distributions and the accuracy of inexpensive approximations of exact null distributions. A simulation study compares the powers of the new tests with other tests of the Sobolev class, showing the benefits of the former. The proposed tests are applied to the study of the (seemingly uniform) nursing times of wild polar bears.
\end{abstract}
\begin{flushleft}
\end{flushleft}

\section{Introduction}
\label{sec:intro}

Directional statistics deals with data for which magnitude is not of interest, but the direction is. This kind of data is supported on the unit hypersphere $\Sq:=\{\bx\in\mathbb{R}^{q+1}:\|\bx\|=\bx'\bx=1\}$, $q\geq1$. Classical instances of ``circular data'' on the unit circle $\mathbb{S}^1$ are given by wind direction and animal direction movements, yet less obvious appearances, such as the direction of cracks in medical hip prostheses \citep{Mann2003}, are possible. Examples of ``spherical data'' commonly appear in astronomy. Higher-dimensional directional data arise in text mining \citep{Banerjee2005} or in genetics \citep{Eisen1998}. Suitable statistical methods have been developed for the analysis of directional data; see the books by \cite{Mardia1999a} and \cite{Ley2017a}, and \cite{Pewsey2021} for a review of recent advances.

Arguably, one of the most fundamental analyses one can perform on an independent and identically distributed (iid) sample $\bX_1,\ldots,\bX_n$ following a common absolutely continuous distribution $\mathrm{P}$ on $\Sq$, with $q\geq1$, is testing
\begin{align*}
    \Hcal_0:\mathrm{P}=\mathrm{Unif}(\Sq)\text{ vs. }\Hcal_1:\mathrm{P}\neq\mathrm{Unif}(\Sq),
\end{align*}
with $\mathrm{Unif}(\Sq)$ denoting the uniform distribution on $\Sq$. Testing uniformity on $\Sq$ is also the basis for further inference, such as testing spherical symmetry of distributions on $\mathbb{R}^d$, $d\geq1$ \citep[e.g., ][]{Cai2013} or testing rotational symmetry about a direction $\bmu\in\Sq$ \citep[e.g., ][]{Garcia-Portugues2020}. It is a widely studied problem and still with relevance today. An incomplete list of recent works includes the geometric mean tests of \cite{Pycke2007, Pycke2010}, projection-based tests \citep{Cuesta-Albertos2009,Garcia-Portugues2020b}, and high-dimensional tests \citep[e.g.,][]{Cai2012, Cai2013}.

Uniformity tests on $\Sq$ are diverse and have different characteristics. The most classical is the \cite{Rayleigh1919} test, which rejects $\mathcal{H}_0$ for large values of the statistic $R_n:=(q+1)/n\sum_{i,j=1}^n \bX_i'\bX_j$. It is designed to show power under unimodal alternatives, but is ``blind'' to centro-symmetric alternatives. 
\cite{Bingham1974} test detects such symmetric alternatives using second-order moments. However, it is not consistent against all the alternatives (i.e., it is ``non-omnibus''). These two tests belong to the general class of \emph{Sobolev tests} pioneered by \cite{Beran1968,Beran1969}, \cite{Gine1975}, and \cite{Prentice1978}, which provides a general framework for more tests that can be designed to be omnibus against any alternative. Some newly introduced tests belong to the class of \emph{projected-ecdf} tests \citep{Garcia-Portugues2020b}, which extends several $\mathbb{S}^1$-only tests and introduces a hyperspherical version of the Anderson--Darling test. The Poisson kernel-based test introduced in \cite{Pycke2010} was built to be powerful against multimodal distributions, but only defined in $\mathbb{S}^1$. It depends on a parameter that must be pre-specified heuristically based on the number of modes of the alternative, typically unknown in practice.

The aim of this paper is to provide two new omnibus tests of uniformity on $\Sq$ induced by tuning parameter-dependent kernels related to two hyperspherical distributions and an automated procedure to approximately maximize its power. We propose the \emph{smooth maximum} test statistic, induced by the von Mises--Fisher distribution, and a test statistic based on the \emph{Poisson kernel} that is related to the spherical Cauchy distribution \citep{Kato2020}. For the smooth maximum statistic, we show its equivalence to the ``LogSumExp'' smooth maximum function, and to Rayleigh and \cite{Cai2013}'s tests for limit cases of its parameter. On the other hand, the Poisson kernel statistic extends \cite{Pycke2010} circular-only tests to the hypersphere, and we show its connection to the Rayleigh test. Next, we derive the exact moments of the test statistics under the uniform and rotationally symmetric alternatives in terms of orthogonal polynomials. The latter are needed to benchmark our tests and study the sensitivity of their parameters under this wide set of alternatives. After obtaining the asymptotic null distribution, we find approximated oracle tuning parameters (being oracle because the underlying distribution is already known) through the maximization of a score intimately related to the power against each particular distribution. As these oracle parameters cannot be determined without knowing the alternative distribution, we provide a cross-validated testing methodology using $p$-value aggregation through the harmonic mean $p$-value \citep{Wilson2019} that estimates the oracle parameter under an unknown distribution, while maintaining the significance level. We investigate with numerical experiments the precision of the null asymptotic distribution and of an approximation of the null distribution based on a gamma match to the first two exact moments. Simulations demonstrate the competitiveness of the automated parameter-tuning cross-validation tests compared to other uniformity tests under rotationally symmetric alternatives and their superiority under multimodal mixture distributions. A data application tests the uniformity of starting times when wild polar bears nurse their cubs.

The rest of this paper is organized as follows. Section \ref{sec:tests} presents the new uniformity test statistics. The exact moments under the null and rotationally symmetric alternatives, along with the asymptotic null distribution, are obtained in Section \ref{sec:theory}. Section \ref{sec:oracle} provides approximate oracle parameters against known alternatives. In Section \ref{sec:estimated-stat}, an automated testing procedure based on $K$-fold cross-validation is introduced. Several simulation studies are given in Section \ref{sec:num}: the precision of asymptotic $p$-values and a fast gamma-match approximation, the effect of $K$ on $K$-fold tests, and a comparison with other uniformity tests. A data application using the new tests is provided  in Section \ref{sec:bears}. Proofs and additional numerical results can be found in the Supplementary Materials~(SM). \nowidow[3]

\section{Test statistics}
\label{sec:tests}

\subsection{Common framework}
\label{sec:frame}

We denote by $\om{q}:=2\pi^{(q+1)/2}/\allowbreak\Gamma((q+1)/2)$ the area of $\Sq$. The notations $\sigma_q$ and $\nu_q$ are used for the Lebesgue and uniform measures on $\Sq$, respectively, such that $\sigma_q(\Sq)=\om{q}$ and $\nu_q(\Sq)=1$.

Our new test statistics belong to the ``Sobolev class'' of tests \citep{Beran1968,Beran1969,Gine1975}. Operationally, the tests belonging to the Sobolev class are based on $U$-statistics with kernels $\psi$ that (\textit{i}) depend on the shortest pairwise angles of the sample $\theta_{ij}:=\cos^{-1}(\mathbf{X}_i'\mathbf{X}_j)$ (hence are rotation-invariant) and (\textit{ii}) admit a basis expansion in terms of specific orthogonal polynomials. Precisely, Sobolev test statistics have the form
\begin{align}
    S_{n,q}(\{w_{k,q}\})=&\;\frac{1}{n}\sum_{i,j=1}^n \psi(\theta_{ij}),\nonumber\\ \psi(\theta)=&\;\begin{cases}\sum_{k=1}^\infty 2w_{k,1}T_k(\cos\theta),&q=1,\\
    \sum_{k=1}^\infty (1+2k/(q-1))w_{k,q}C_k^{(q-1)/2}(\cos\theta), & q\geq 2,
    \end{cases}\label{eq:canoSob}
\end{align}
where $\big\{C_k^{(q-1)/2}\big\}_{k=0}^\infty$ and $\big\{T_k\big\}_{k=0}^\infty$ represent Gegenbauer and Chebyshev polynomials, respectively. For $q\geq2$, Gegenbauer polynomials form an orthogonal basis on $L_q^2([-1,1])$, our notation for the space of square-integrable real functions on $[-1,1]$ with respect to the weight $x\mapsto (1-x^2)^{q/2-1}$, $q\geq1$, that is induced by the projected uniform distribution on $\Sq$. The orthogonality condition reads
\begin{align*}
	\int_{-1}^1 C_k^{(q-1)/2}(x)C_\ell^{(q-1)/2}(x)(1-x^2)^{q/2-1}\,\mathrm{d}x=\delta_{k,\ell}c_{k,q}
\end{align*}
for $k,\ell\geq0$, where $\delta_{k,\ell}$ represents the Kronecker delta, and
\begin{align}
	c_{k,q}:=\frac{\om{q}}{\om{q-1}}\lrp{1+\frac{2k}{q-1}}^{-2}d_{k,q},\quad d_{k,q}:=\frac{2k+q-1}{q-1}C_k^{(q-1)/2}(1). \label{eq:dkq}
\end{align}
Chebyshev polynomials, expressible as $T_k(\cos\theta)=\cos(k \theta)$ for $\theta\in[0,\pi]$, form an orthonormal basis on $L_1^2([-1,1])$ with normalizing constants $c_{k,1}=(1+\delta_{k,0})\pi/2$. Chebyshev polynomials can be regarded as the limit of Gegenbauer polynomials for $q=1$, since 
\begin{align}
	\lim_{\alpha\to0^+}\frac{1}{\alpha}C_k^\alpha(x)=\frac{2}{k}T_k(x)\;\;\text{for}\;\; k\geq1. \label{eq:Ck0}
\end{align}

The following definition sets the structure of the two test statistics introduced in the paper, with $\psi_\ell$ presented in Sections \ref{sec:sm} and \ref{sec:pois}.

\begin{definition}[Test statistics]\label{def:stats}
    Given an iid sample $\mathbf{X}_1,\ldots,\mathbf{X}_n$ on $\Sq$, $q\geq1$, and $\ell=1,2$, the \emph{smooth maximum} ($\ell=1$) and \emph{Poisson kernel} ($\ell=2$) test statistics $T_{n,\ell}$ are
\begin{align*}
    T_{n, \ell}:=\frac{2}{n}\sum_{1\leq i<j\leq n} \tilde{\psi}_{\ell}(\theta_{ij}),
\end{align*}
where $\tilde{\psi}_{\ell}(\theta) := \psi_{\ell}(\theta) - \mathbb{E}_{\Hcal_0}[\psi_{\ell}(\theta_{12})]$.
 Equivalently,
\begin{align}
    T_{n, \ell}:=\frac{2}{n}\sum_{1\leq i<j\leq n} \psi_{\ell}(\theta_{ij}) - (n-1)\mathbb{E}_{\Hcal_0}[\psi_{\ell}(\theta_{12})].\label{eq:stat_def}
\end{align}
\end{definition}

\begin{remark}
    The specification of $\tilde{\psi}_{\ell}$ implies that $\mathbb{E}_{\Hcal_0}[\tilde{\psi}_{\ell}(\theta_{12})]=\mathbb{E}_{\Hcal_0}[\tilde{\psi}_{\ell}(\theta_{12})\mid\bX_1]=0$ (i.e., under $\Hcal_0$, the statistic is centered and the kernel $\tilde{\psi}_{\ell}$ is degenerate) and makes $T_{n, \ell}$ non-diverging under~$\Hcal_0$.
\end{remark}

\subsection{Smooth maximum}
\label{sec:sm}

Our first statistic, $T_{n,1}(\kappa)$, uses the kernel 
\begin{align*}
    \psi_1(\theta; \kappa):=\exp\lrp{\kappa (\cos \theta - 1)},\quad \kappa>0
\end{align*}
that is related to the von Mises--Fisher probability density function (pdf), see Section \myref{sec:oracle-sim}{B.1} in the SM. Large values of $T_{n, 1}(\kappa)$ indicate clustering of the sample, and therefore the test based on $T_{n, 1}$ rejects $\Hcal_0$ for large values. The term $-\kappa$ in the exponent of $\psi_1$ serves to numerically stabilize the evaluation for large $\kappa$'s.

For the sake of notational simplicity, we will omit the dependence of $T_{n,1}(\kappa)$ on its parameter $\kappa$ unless it is strictly necessary. The same consideration holds for other test statistics.

The statistic $T_{n,1}$ generated by this kernel is also related to the ``smooth maximum'' given by a generalization of the LogSumExp (LSE) of the sample $\{\bX_i'\bX_j\}_{1\leq i<j\leq n}$:
\begin{align}
    \mathrm{LSE}_\kappa(\{\bX_i'\bX_j\}_{1\leq i<j\leq n}):=&\;\frac{1}{\kappa}\log\bigg(\sum_{1\leq i<j\leq n} \exp\lrp{\kappa\bX_i'\bX_j}\bigg)\nonumber\\
    =&\;\frac{1}{\kappa}\log\left(\frac{n \exp\lrp{\kappa} (T_{n,1}+C)}{2}\right),\label{eq:lse}
\end{align}
with $C = (n-1)\mathbb{E}_{\mathcal{H}_0}[\psi_1(\theta_{12})]$. If $\kappa=1$, then $\mathrm{LSE}_\kappa$ is the standard LSE commonly used in machine learning. Since $\mathrm{LSE}_\kappa(\{\bX_i'\bX_j\}_{1\leq i<j\leq n})$ is a monotone transformation of $T_{n,1}$ by $t\mapsto \kappa^{-1} \log(n \exp\lrp{\kappa} (t + C) / 2)$, the two statistics yield the same test. Additionally, since the LSE is a smooth approximation of the maximum, it happens that $\lim_{\kappa\to\infty}\mathrm{LSE}_\kappa(\{\bX_i'\bX_j\}_{1\leq i<j\leq n})=\max_{1\leq i<j\leq n}\bX_i'\bX_j$.

From a different perspective, $T_{n,1}$ can be exactly expressed as
\begin{align*}
    \frac{1}{\kappa}\log\bigg(\sum_{1\leq i<j\leq n} \exp\lrp{-\kappa\|\bX_i-\bX_j\|^2}\bigg)=-2+\frac{1}{\kappa}\log\bigg(\sum_{1\leq i<j\leq n} \exp\lrp{2\kappa \bX_i'\bX_j}\bigg),
\end{align*}
which points to $T_{n, 1}$ featuring the somehow ``opposite'' averaging of distances to \cite{Pycke2007}'s geometric mean of the chordal distances:
\begin{align*}
	\Gamma_n^2=e^{2/(n(n-1))}\exp\bigg(\sum_{1\leq i<j\leq n}\log(\|\bX_i-\bX_j\|)\bigg).
\end{align*}

The parameter $\kappa$ determines the behavior of the smooth maximum test and raises some interesting connections to other tests (Table~\ref{tbl:stat_connections}).

\begin{proposition}[Connections of the smooth maximum test]\label{prp:lse_connection}
The \emph{smooth maximum test} that rejects for large values of $T_{n, 1}$ is equivalent to:
\begin{enumerate}[label=(\textit{\roman*}),ref=(\textit{\roman*})]
    \item The Rayleigh test based on the statistic $R_n:=(q+1)/n\sum_{i,j=1}^n \bX_i'\bX_j$, when $\kappa\to0$. \label{prp:lse_rayleigh}
    \item \cite{Cai2013}'s maximum test based on the statistic $C_n:=\max_{1\leq i<j\leq n}\bX_i'\bX_j$, when $\kappa\to\infty$.\label{prp:lse_cai_jiang}
\end{enumerate}
\end{proposition}

\subsection{Poisson kernel}
\label{sec:pois}

\citet[Equation 11]{Pycke2010} proposed a test on $\mathbb{S}^1$ based on the \emph{Poisson kernel}. An extension to $\Sq$ of this kernel is possible through
\begin{align*}
    \psi_2(\theta; \rho):=\frac{1-\rho^2}{(1-2 \rho\cos\theta+\rho^2)^{(q+1)/2}},\quad 0<\rho<1.
\end{align*}
Note that $\psi_2$ is similar to \cite{Kato2020}'s spherical Cauchy distribution on $\Sq$, although it differs in the power of the denominator (see Section \myref{sec:oracle-sim}{B.1} in the SM for an alternative Cauchy distribution). A possible numerical stabilization of the kernel is $(1-\rho)^{(q+1)}/(1-\rho^2)\psi_2(\theta; \rho)$.

The test based on $T_{n,2}$ also rejects $\mathcal{H}_0$ for large values of $T_{n,2}$, which indicate sample clustering, and is controlled by the parameter $\rho$ which leads to the following connection with other tests (Table~\ref{tbl:stat_connections}).

\begin{proposition}[Connections of the Poisson kernel test]\label{prp:poisson_rayleigh}
The \emph{Poisson kernel test} that rejects for large values of $T_{n,2}$ is equivalent to the Rayleigh test when $\rho\to0$.
\end{proposition}

\begin{table}[htpb!]
    \centering
    \begin{tabular}{l | l | l}
    \toprule
        \multirow{3}{12em}{Smooth maximum $T_{n,1}(\kappa)$} & $\kappa\to0$ & \cite{Rayleigh1919} \\
        & $\kappa>0$ & --- \\
        & $\kappa\to\infty$& \cite{Cai2013} \\
        \midrule
        \multirow{3}{12em}{Poisson kernel $T_{n,2}(\rho)$} & $\rho\to0$ & \cite{Rayleigh1919} \\
        & $\rho\in(0,1)$ & \cite{Pycke2010} if $q=1$\\
        & $\rho\to1$ & --- \\
    \bottomrule
    \end{tabular}
    \caption{\small Extensions and connections of the tests based on $T_{n,\ell}$, $\ell=1,2$.}
    \label{tbl:stat_connections}
\end{table}

\section{Theoretical properties}
\label{sec:theory}

We denote by $\{b_{k,q}(\psi_\ell)\}_{k=0}^\infty$ the \emph{Gegenbauer coefficients} of $\psi_\ell$ such that
\begin{align}
\psi_\ell(\theta)=\begin{cases}
\sum_{k=0}^\infty b_{k,1}(\psi_\ell) T_k(\cos\theta),&q=1,\\    
\sum_{k=0}^\infty b_{k,q}(\psi_\ell) C_k^{(q-1)/2}(\cos\theta),&q\geq2
\end{cases}\label{eq:psi_ell}
\end{align}
almost everywhere for $\theta\in[0,\pi]$. %
The Gegenbauer coefficients correspond bijectively to the $\{w_{k,q,\ell}\}_{k=1}^\infty$ coefficients that connect $T_{n,\ell}$ with its Sobolev form~\eqref{eq:canoSob}:
\begin{align}
w_{k,q,\ell}=\begin{cases}
2^{-1}b_{k,q}(\psi_{\ell}),& q=1,\\
(1+2k/(q-1))^{-1}b_{k,q}(\psi_{\ell}),& q\geq2.
\end{cases}\label{eq:ws}
\end{align}

The Gegenbauer coefficients in the following proposition are key for the results in Sections \ref{sec:th_null} and \ref{sec:th_exp}.

\begin{proposition}\label{prp:gegen}
Let $q\geq1$ and $k\geq0$. The Gegenbauer coefficients of the kernels $\psi_\ell$, $\ell=1,2$, are:
\begin{enumerate}
    \item $b_{k,q}(\psi_1)=\begin{cases}
	\displaystyle (2-\delta_{k,0})\, e^{-\kappa}\, \Ical_{k}(\kappa),\!\!\!\!&q=1,\\
	\displaystyle (2/\kappa)^{(q-1)/2}\Gamma((q-1)/2)(k+(q-1)/2)&\\
 \qquad\times\,e^{-\kappa}\,\Ical_{k+(q-1)/2}(\kappa),\!\!\!\!&q\geq 2.
    \end{cases}$
    \item $b_{k,q}(\psi_2)=\begin{cases}
	\displaystyle (2-\delta_{k,0})\rho^k,&q=1,\\
	\displaystyle ((2k+q-1)/(q-1))\rho^k,&q\geq 2,
    \end{cases}$
\end{enumerate}
where $\Ical_{k}$ is the modified Bessel function of the first kind and $k$th order.
\end{proposition}

\subsection{Asymptotic properties}
\label{sec:th_null}

Once the Gegenbauer coefficients are known, the first two moments of $T_{n,\ell}$ under $\mathcal{H}_0$ can be obtained directly.

\begin{proposition}[Expectation and variance under $\mathcal{H}_{0}$]\label{prp:moments_null}
Let $q\geq1$ and $\ell=1,2$. Then $\mathbb{E}_{\Hcal_0}[\psi_{\ell}(\theta_{12})]=b_{0,q}(\psi_{\ell})$ and 
\begin{align}
    \mathbb{E}_{\Hcal_0}[T_{n, \ell}]=0, \quad \mathbb{V}\mathrm{ar}_{\Hcal_0}[T_{n, \ell}]=\frac{2(n-1)}{n}\left(b_{0,q}(\psi_{\ell}^2) - b_{0,q}^2(\psi_{\ell})\right).\label{eq:expvar_null}
\end{align}
\end{proposition}

A useful property of the smooth maximum kernel is that $b_{0,q}(\psi_{1}^2(\kappa)) = b_{0,q}(\psi_1(2\kappa))$, which gives an analytical expression for its variance. Numerical integration must be performed to compute $b_{0,q}(\psi_{2}^2(\rho))$.

Since $\tilde{\psi}_\ell$ is
a degenerate kernel, the asymptotic distribution of $T_{n, \ell}$ is a centered infinite weighted sum of independent chi-squared random variables.

\begin{theorem}[Null asymptotic distribution]\label{thm:asymp_null}
Let $q\geq1$ and $\ell=1,2$. Then, under $\Hcal_0$ and as $n\to\infty$, 
\begin{align}
    T_{n, \ell}\inlawH T_{\infty,\ell}^{(0)}     \equald\sum_{k=1}^\infty w_{k,q,\ell} Y_{d_{k,q}} - \tilde{\psi}_\ell(0),\label{eq:null_asymp_dist}
\end{align}
with $\{Y_{d_{k,q}}\}_{k=1}^\infty$ being a sequence of independent random variables such that $Y_{d_{k,q}}\sim \chi^2_{d_{k,q}}$ and $\equald$ denoting equality in distribution.
\end{theorem}

\begin{corollary}[Omnibusness]\label{cor:omnibus}
The tests that, respectively, reject $\Hcal_0$ for large values of $T_{n, 1}(\kappa)$, $\kappa>0$, and $T_{n,2}(\rho)$, $\rho>0$, are consistent against all alternatives with square-integrable pdf on $\Sq$.
\end{corollary}

\subsection{Moments under rotationally symmetric alternatives}
\label{sec:th_exp}

Rotationally symmetric distributions are expressible in terms of Gegenbauer expansions. As a consequence, they offer a wide set of specifications for $\Hcal_1$ where the performance of our statistics can be precisely benchmarked in Section \ref{sec:oracle}. For that purpose, knowledge of the moments of $T_{n,\ell}$ is required.

Let $f_1$ be a rotationally symmetric pdf under $\Hcal_1$: $f_1(\bx)=g(\bx'\bmu)$, with $\bx, \bmu\in\mathbb{S}^q$, where we assume that $g\in L^2_q([-1,1])$. Then, $g$ can be expressed for almost every $t\in[-1,1]$ as 
\begin{align*}
    g(t) = \begin{cases}
    \sum_{k=0}^{\infty} e_{k,1} \,T_k(t), & q=1,\\
    \sum_{k=0}^{\infty} e_{k,q} \,C_{k}^{(q-1)/2}(t), & q\geq 2.
    \end{cases}
\end{align*}

\begin{proposition}[Expectation under rotationally symmetric alternatives]\label{prp:exp_alt}
Let
\begin{align*}
    \tau_{k, q} := \begin{cases}
    (1+\delta_{k,0})^2/4,& q=1,\\
    \left(1 + 2k/(q-1)\right)^{-2} (q-1)_k/k!,& q\geq2,\\\end{cases}
\end{align*}
where $(a)_k := a \cdot (a+1) \cdots (a+k-1)$ is the Pochhammer symbol. Then, 
\begin{align}
    \mathbb{E}_{\Hcal_1}[T_{n, \ell}] = \omega_q^2 (n-1) \sum_{k=1}^{\infty} \tau_{k, q} \, b_{k, q}(\psi_{\ell}) \,e_{k, q}^2.\label{eq:exp_alt}
\end{align}
\end{proposition}

\begin{proposition}[Variance under rotationally symmetric alternatives]\label{prp:var_alt}
Let
\begin{align*}
    \zeta_{k_1,k_2, q} := \begin{cases}
    (1+\delta_{k_1,0})\,(1+\delta_{k_2,0})/2,& q=1,\\
    \omega_{q-1} \left(1 + 2k_1/(q-1)\right)^{-1} \left(1 + 2k_2/(q-1)\right)^{-1},& q\geq2,\\
    \end{cases}
\end{align*}
and $\tau_{k,q}$ and $t_{k_1,k_2,k_3;q}$ be defined as in Proposition \ref{prp:exp_alt} and Lemma \myref{lem:triplet}{1} in the SM, respectively. Then, 
\begin{align}
    \mathbb{V}\mathrm{ar}_{\mathcal{H}_1}[T_{n, \ell}] = \frac{2(n-1)}{n}\left(2 (n-2) \eta_{1, \ell} + \eta_{2, \ell}\right),\label{eq:var_stat}
\end{align}
where $\eta_{1, \ell} = %
\alpha_{1,\ell} - \beta_{\ell}^2$ and $\eta_{2, \ell} = %
\alpha_{2,\ell} - \beta_{\ell}^2$  with
\begin{align*}
\alpha_{1,\ell} =&\; \omega_{q}^2 \sum_{k_1,k_2,k_3=0}^{\infty} \zeta_{k_1,k_2,q} \,b_{k_1,q}(\psi_\ell)\,b_{k_2,q}(\psi_\ell)\,e_{k_1,q}\,e_{k_2,q}\,e_{k_3,q} \,t_{k_1,k_2,k_3;q}, \\
\alpha_{2,\ell} =&\;  \omega_q^2 \sum_{k=0}^{\infty} \tau_{k,q} \,b_{k,q}(\psi_\ell^2) \,e_{k, q}^2,\qquad
\beta_{\ell} = \omega_q^2 \sum_{k=0}^{\infty} \tau_{k,q} \,b_{k,q}(\psi_\ell) \,e_{k,q}^2.
\end{align*}
\end{proposition}

\begin{remark}
As might be expected, both \eqref{eq:exp_alt} and \eqref{eq:var_stat} diverge as $n\to\infty$ due to the standardization of $T_{n,\ell}$ in \eqref{eq:stat_def}. Since $\tilde{\psi}_\ell$ is a degenerate kernel under $\mathcal{H}_0$, $\eta_{1,\ell}=0$ and both expectation and variance converge.
\end{remark}

\section{Approximate oracle parameters}
\label{sec:oracle}

The statistic $T_{n, \ell}(\lambda)$ depends on the parameter $\lambda$ ($\kappa$ or $\rho$ for $\ell=1,2$), and so does its behavior under different distributions of the sample. Therefore, the selection of $\lambda$ to ensure a powerful test is crucial in practice. Although finding the asymptotic power of the test under a certain alternative $\mathcal{H}_1$ would yield an exact theoretical response, those expressions are not obtainable in general. Hence, as \cite{Gregory1977}, we consider the score
\begin{align*}
    q_{n,\ell, \mathcal{H}_1}(\lambda) := \frac{\mathbb{E}_{\mathcal{H}_1}[T_{n,\ell}(\lambda)]}{\sqrt{\mathbb{V}\mathrm{ar}_{\mathcal{H}_0}[T_{n,\ell}(\lambda)]}}
\end{align*}
as an approximate indicator of the power of $T_{n, \ell}(\lambda)$ against $\mathcal{H}_1$.

This score is based on the idea that the power of a test based on $T_{n, \ell}(\lambda)$, the tail probability $\mathbb{P}_{\Hcal_1}[T_{n, \ell}(\lambda)>c_{\alpha, n, \ell}(\lambda)]$ with $c_{\alpha, n, \ell}(\lambda)$ being the exact-$n$ upper $\alpha$-quantile of $T_{n,\ell}(\lambda)$ under $\Hcal_0$, is greater the further apart its distributions under $\Hcal_0$ and $\Hcal_1$ are. A fairly simple approach to quantifying that distance is through the difference of expectations $\mathbb{E}_{\mathcal{H}_1}[T_{n,\ell}(\lambda)] - \mathbb{E}_{\mathcal{H}_0}[T_{n,\ell}(\lambda)] = \mathbb{E}_{\mathcal{H}_1}[T_{n,\ell}(\lambda)]$, scaled by the standard deviation of $T_{n,\ell}(\lambda)$ under $\mathcal{H}_0$. Then, 
\begin{align}
    \tilde{\lambda}_{\mathcal{H}_1}:=\arg\max_{\lambda\in\Lambda} q_{n,\ell, \mathcal{H}_1}(\lambda)\label{eq:opt-lambda}
\end{align}
gives an approximation to the oracle parameter that maximizes the power of the test based on $T_{n, \ell}(\lambda)$ against $\mathcal{H}_1$ within a (discrete) grid of parameters $\Lambda$. We use $\Lambda=\{0.01, 0.1, 0.2, \ldots, 5, 5.2, \ldots, 10, 11, \allowbreak \ldots, 20, 25, \ldots, 50\}$ for $\ell=1$ and $\Lambda=\{a/50: a = 1, \ldots, 49\}$ for $\ell=2$.

The theoretical value $\tilde{\lambda}_{\mathcal{H}_1}$ can be found only when the distribution under $\mathcal{H}_1$ is known. Obviously, this is unrealistic in practice; in Section \ref{sec:estimated-stat} an estimation approach for $\tilde{\lambda}_{\mathcal{H}_1}$ is proposed.

The analysis of the precision of $\tilde{\lambda}_{\mathcal{H}_1}$ as an approximate oracle parameter in several simulation scenarios can be found in the SM. The results demonstrate that these oracle parameters accurately approximate the actual oracle parameter: the median power differences are just $0.15\%$ and $0.03\%$ for $T_{n,1}$ and $T_{n,2}$, respectively. Among the six different simulation scenarios, there are four well-known distributions on $\Sq$:
\begin{inlinelist}
    \item the von Mises--Fisher (vMF) distribution\label{vMF-dist},
    \item a Cauchy-like (Ca) distribution\label{C-dist},
    \item the Watson (Wa) distribution\label{W-dist}, and
    \item the Small Circle (SC) distribution\label{SC-dist},
\end{inlinelist}
which are rotationally symmetric. Consequently, the results obtained in Section \ref{sec:th_exp} are used to compute $q_{n,\ell,\Hcal_1}(\lambda)$. In addition, two multimodal mixtures, which are centro-symmetric about the origin, are built from 
\begin{inlinelist}[start=5]
    \item the vMF distribution (mixture denoted MvMF) and\label{MvMF-dist}
    \item the Cauchy-like distribution (denoted MCa)\label{MC-dist}.
\end{inlinelist}
All distributions depend on a concentration parameter $\kappa_{\mathrm{dev}}>0$. For a formal definition of distributions \ref{vMF-dist}--\ref{MC-dist} see Section \myref{sec:oracle-sim}{B.1} in the SM.

\section{\texorpdfstring{Tuning parameter selection: $K$-fold tests}{Tuning parameter selection: K-fold tests}}
\label{sec:estimated-stat}

A plug-in estimator of \eqref{eq:opt-lambda} is $\widehat{\lambda} := \arg \max_{\lambda\in\Lambda} T_{n, \ell}(\lambda)\big/\sqrt{\mathbb{V}\mathrm{ar}_{\mathcal{H}_0}[T_{n, \ell}(\lambda)]}$. However, on the one hand, the test statistic $T_{n,\ell}(\widehat{\lambda})$ does not converge to the asymptotic null distribution in Theorem \ref{thm:asymp_null}. Ignoring this fact leads to a liberal test when using the asymptotic distribution \eqref{eq:null_asymp_dist}, as using $\widehat{\lambda}$ in $T_{n,\ell}(\widehat{\lambda})$ implies maximizing the discrepancy with respect to $\mathcal{H}_0$. On the other hand, sound Monte Carlo approaches for $T_{n,\ell}(\widehat{\lambda})$ are expensive, as they require refitting $\widehat{\lambda}$.

An effective solution to the above problem is to split the sample $\mathcal{S}=\{\bX_1, \ldots, \bX_n\}$ into two disjoint subsamples $\mathcal{S}_1$ and $\mathcal{S}_2$, then estimate $\tilde{\lambda}_{\mathcal{H}_1}$ using $\mathcal{S}_1$, obtaining $\widehat{\lambda}(\mathcal{S}_1)$, and then apply the test on the remaining part $\mathcal{S}_2$. We denote the statistic computed through this two-step procedure by
\begin{align}
    T_{\vert\mathcal{S}_2\vert,\ell}(\widehat{\lambda}(\mathcal{S}_1),\mathcal{S}_2).\label{eq:TnS1S2}
\end{align}
Clearly, conditionally on $\mathcal{S}_1$ and due to the iid assumption, \eqref{eq:TnS1S2} has the asymptotic null distribution of Theorem \ref{thm:asymp_null} featuring $\widehat{\lambda}(\mathcal{S}_1)$ instead of $\lambda$ inside the weights $\{w_{k,q,l}\}_{k=1}^\infty$. As shown below, the $\lambda$-dependent asymptotic $\alpha$-critical levels of $T_{n,\ell}(\lambda)$ given by Theorem \ref{thm:asymp_null} can be readily used in the testing procedure based on \eqref{eq:TnS1S2}, and the resulting Type I error asymptotically respects the significance level~$\alpha$.

\begin{proposition}\label{prp:rejection_level}
    Let $\mathcal{S}=\mathcal{S}_1\cup \mathcal{S}_2$ be a disjoint partition of the sample such that $\vert\mathcal{S}_2\vert \to\infty$ as $n\to\infty$, $\widehat{\lambda}(\mathcal{S}_1) := \arg \max_{\lambda\in\Lambda} T_{\vert \mathcal{S}_1 \vert, \ell}(\lambda)\big/\sqrt{\mathbb{V}\mathrm{ar}_{\mathcal{H}_0}[T_{\vert \mathcal{S}_1 \vert, \ell}(\lambda)]}$, and $\Lambda$ is a discrete set of tuning parameters. Consider $\lambda\mapsto c_{\alpha}(\lambda)$ such that $\lim_{n\to\infty}\mathbb{P}_{\mathcal{H}_0}[T_{n,\ell}(\lambda)\leq c_\alpha(\lambda)]=\alpha$. Then,
    \begin{align*}
        \lim_{n\to\infty}&\mathbb{P}_{\mathcal{H}_0}[T_{\vert\mathcal{S}_2\vert,\ell}(\widehat{\lambda}(\mathcal{S}_1),\mathcal{S}_2) \leq c_\alpha(\widehat{\lambda}(\mathcal{S}_1))] = \alpha.
    \end{align*}
\end{proposition}

The assumption of finiteness in $\Lambda$ honors the empirical application of the test procedure, where finite grids are considered for $\Lambda$ (see below \eqref{eq:opt-lambda}). Initial investigations showed that the power of the tests derived from the two-subsample split approach was far from the optimal power attained with the oracle parameters in Section \ref{sec:oracle} (see SM for further details). 
Thus, a $K$-fold cross-validated approach is advocated: perform $K$ tests that exclude one out of $K$ partitions, each test performed with an estimate of $\tilde{\lambda}_{\mathcal{H}_1}$ obtained from the excluded partition, and combine their outcomes afterward. The rationale is to devote most of the data strength to testing and reduce the dependence on the specific partition compared to a single-partition test. Due to Proposition \ref{prp:rejection_level}, each of the $K$ tests respects the nominal level asymptotically.

\cite{Fisher1925}'s method is arguably the simplest for combining multiple test $p$-values. However, it relies on the assumption that the tests are independent, which is not fulfilled in our setting, and in our experiments yielded a Type I error above the significance level. Recently, the Harmonic Mean $P$-value (HMP), a method that tests whether no $p$-value is significant in a set of $p$-values, has been proposed in \cite{Wilson2019}. This method is more robust to dependence between $p$-values, while controlling the family-wise error rate, and builds an agglomerating test through the HMP. It has been empirically shown \citep{Wilson2019} to be more powerful than other methods, such as \cite{Benjamini1995}'s false discovery rate correction. Therefore, the proposed test uses HMP to combine the $K$ tests into a single $p$-value.

The following definition describes the proposed testing procedure.
\begin{definition}[$K$-fold test based on $T_{n, \ell}$]\label{def:K}
Given an iid sample $\mathcal{S}=\{\mathbf{X}_1,\ldots,\mathbf{X}_n\}$, the \emph{$K$-fold test} for $\Hcal_0$ based on $T_{n, \ell}$, $T_{n, \ell}^{(K)}$, proceeds as follows:

    \begin{enumerate}
    \item Split $\mathcal{S}$ into $K$ (disjoint) subsamples $\mathcal{S}_{1}, \ldots, \mathcal{S}_{K}$ of (roughly) equal sizes.
    \item ($K$-fold testing) For $k = 1,\ldots, K$:\label{step:K}
    \begin{enumerate}[label=\textit{(\alph{*})},ref=2\textit{(\alph{*})}]
        \item Compute $\widehat{\lambda}(\mathcal{S}_{k}) = \arg \max_{\lambda\in\Lambda}\allowbreak T_{\lvert \mathcal{S}_{k}\rvert, \ell}(\lambda)\big/\sqrt{\mathbb{V}\mathrm{ar}_{\mathcal{H}_0}[T_{\lvert \mathcal{S}_{k}\rvert, \ell}(\lambda)]}$, the estimate of $\tilde{\lambda}_{\Hcal_1}$ using $\mathcal{S}_{k}$.
    \item Use the remaining $K-1$ subsamples to perform the test based on $T_{n-\lvert \mathcal{S}_{k}\rvert,\ell}(\widehat{\lambda}(\mathcal{S}_{k}),\mathcal{S}\backslash \mathcal{S}_{k})$ and obtain its (Monte Carlo or asymptotic) $p$-value $p_k$. \label{step:2}
    \end{enumerate}
    \item ($p$-value aggregation)
    \begin{enumerate}[label=\textit{(\alph{*})},ref=3\textit{(\alph{*})}]
        \item Compute $
            \overset{\circ}{p}:=\big[\sum_{k=1}^{K}(K p_{k})^{-1}\big]^{-1}$, the harmonic mean of $\{K p_k\}_{k=1}^{K}$. \label{step:3}
        \item Compute the asymptotically-exact HMP \citep{Wilson2019}, 
        \begin{align*}
        p_{\mathring{p}}=\int_{1 / \mathring{p}}^{\infty} f_{\mathrm {Landau}}(x; \log K+0.874, \pi/2) \,\rd x,
        \end{align*}
        where $f_{\mathrm{Landau}}(x; \mu, \sigma)=(\pi \sigma)^{-1} \int_{0}^{\infty} \exp\{-t (x-\mu)/\sigma-(2/\pi) t \log t\}\allowbreak \sin (2 t) \,\rd t$ is the Landau pdf, and set it as the $p$-value of the test.
    \end{enumerate}
\end{enumerate} 
\end{definition}

If Monte Carlo $p$-values are computed in Step \ref{step:2}, the potential null $p$-values must be replaced by $1/M$ to allow for the computation of the harmonic mean in Step \ref{step:3}.

We highlight that our $K$-fold test is applicable to other goodness-of-fit tests depending on a tuning parameter $\lambda$, as long as the null asymptotic distribution is known to apply Step \ref{step:K} in Definition \ref{def:K}. Other testing procedures with automatic selection of $\lambda$ are possible; see the review in \cite{Tenreiro2019} and his bootstrap-based approach in his Equations (10) and (12). Our $K$-fold approach has the appeal of avoiding costly bootstrap resampling and a recalculation of the critical value. Using the $O(n^2)$-statistic $T_{n,\ell}(\lambda)$, the $K$-fold test involves $O(K\{\vert\Lambda\vert(n/K)^2+(n(K-1)/K)^2\})$ computations.

\section{Simulation studies}
\label{sec:num}

\subsection{\texorpdfstring{Computation of asymptotic $p$-values}{Computation of asymptotic p-values}}
\label{sec:asymp_dist}

The asymptotic distribution \eqref{eq:null_asymp_dist} is usable in practice. In particular, the exact method developed by \cite{Imhof1961} allows computing asymptotic $p$-values by evaluating the truncated-series tail probability function:
\begin{align}
    x \mapsto \mathbb{P}\left[\sum_{k=1}^{K_\mathrm{tr}} w_{k,q,\ell} \, Y_{d_{k,q}} > x + \tilde{\psi}_\ell(0)\right].\label{eq:imhof}
\end{align}

We performed numerical experiments to measure the convergence speed of the $K_\mathrm{tr}$-truncated series in the probability \eqref{eq:imhof} by comparing it to the probabilities for $K_{\max}=10^4$ terms. Table \ref{tbl:imhof_convergence} demonstrates that $K_\mathrm{tr}=50$ ensures a negligible uniform error for $T_{n,\ell}$, $\ell=1,2$. This uniform error is the maximum error that is committed on a grid for $x$ formed by the asymptotic quantiles $0.01,\ldots,0.99$  and for $q=1,\ldots,5, 10$. When $\rho\approx 1$, we noticed that the asymptotic distribution of $T_{n,2}$ is hard to approximate with a limited $K_\mathrm{tr}$ since $k\mapsto b_{k,q}(\psi_2)$ in Proposition \ref{prp:gegen} slowly decreases in this limiting case. Therefore, throughout our empirical studies, we use $K_\mathrm{tr}=50$ for the truncation in \eqref{eq:imhof} that is fed to Imhof's method.

\begin{table}[htpb!]
    \centering
    \scalebox{1}{
    \begin{tabular}{ l | l l l l l | l l l}
    \toprule
        \multirow{2}{1em}{$K_\mathrm{tr}$} & \multicolumn{5}{c|}{$\kappa$} & \multicolumn{3}{c}{$\rho$} \\
        & $0.1$ & $1$ & $5$ & $30$ & $60$ & $0.25$ & $0.5$ & $0.75$ \\
        \midrule
        $10$ & $0$ & $2\cdot{10^{-16}}$ & $4\cdot{10^{-10}}$ & $2\cdot{10^{-2}}$ & $2\cdot{10^{-1}}$ & $3\cdot{10^{-9}}$ & $7\cdot{10^{-4}}$ & $2\cdot{10^{-1}}$ \\
        $50$ & $0$ & $0$ & $0$ & $5\cdot{10^{-14}}$ & $3\cdot{10^{-13}}$ & $0$ & $2\cdot{10^{-14}}$ & $1\cdot{10^{-7}}$ \\
        $100$ & $0$ & $0$ & $0$ & $0$ & $0$ & $0$ & $0$ & $4\cdot{10^{-12}}$ \\
        $1000$ & $0$ & $0$ & $0$ & $0$ & $0$ & $0$ & $0$ & $0$ \\
    \bottomrule
    \end{tabular}
    }
    \caption{\small Uniform errors of the tail probability \eqref{eq:imhof} with $K_\mathrm{tr}$ terms, relative to considering $K_{\max}=10^4$.}
    \label{tbl:imhof_convergence}
\end{table}

Another numerical experiment evaluates the convergence speed of \eqref{eq:null_asymp_dist}. The rows labeled ``Asymp'' in Table \ref{tbl:null_asymp_dist_05} show the rejection proportions for the significance level $\alpha=0.05$ when the asymptotic critical values are used in the test decision, approximated by $10^5$ Monte Carlo replicates for $q\in\{1,2,3,5\}$, $n\in\{10, 50, 200\}$, $\kappa\in\{0.1, 1, 5\}$, and $\rho\in\{0.25, 0.5, 0.75\}$. In addition, the rejected frequencies that lie within the $95\%$ confidence interval around $\alpha$ are highlighted. Although dependent on parameters $\kappa$ and $\rho$, the convergence starts to manifest itself consistently when $n \geq 50$. For $n=50,200$, the proportions not included in the $95\%$ confidence interval are quite close to it. Empirical rejection proportions for $\alpha = 0.01, 0.1$ can be found in the SM.

\subsection{\texorpdfstring{Fast approximation of $p$-values}{Fast approximation of p-values}}
\label{sec:exact_approx}

\cite{Imhof1961}'s method, as well as Monte Carlo simulations, can be time consuming. Several methods have been proposed to approximate the asymptotic distribution \eqref{eq:null_asymp_dist}, see \cite{Bodenham2016} for a review. %

\begin{table}[htpb!]
\iffigstabs
\centering
\small
\scalebox{1}{
\begin{tabular}{ m{0.35cm} m{0.35cm} m{0.9cm} R{1.2cm} R{1.2cm} R{1.2cm} R{1.2cm}  R{1.3cm} R{1.3cm} R{1.3cm} R{1.2cm} } 
    \toprule
    $q$ & $n$ & Type & $\kappa=0.1$ & $\kappa=1$ & $\kappa=5$ & Time & $\rho=0.25$ & $\rho=0.5$ & $\rho=0.75$ & Time \\
      \midrule
    \multirow{9}{0.6cm}{$1$} & \multirow{3}{1cm}{$10$} & Asymp. & $0.0460$ & $0.0459$ & $0.0428$ & $2\times 10^5$ & $0.0457$ & $0.0445$ & $0.0430$ & $4\times 10^4$ \\ 
    & & Gamma & $\mathbf{0.0511}$ & $0.0519$ & $\mathbf{0.0504}$ & $4\times 10^1$ & $0.0532$ & $0.0532$ & $0.0517$ & $3\times 10^2$ \\ 
    & & MC & $\mathbf{0.0500}$ & $\mathbf{0.0504}$ & $0.0514$ & $3\times 10^6$ & $\mathbf{0.0507}$ & $\mathbf{0.0508}$ & $\mathbf{0.0511}$ & $2\times 10^6$ \\ 
    \cline{2-11}
    & \multirow{3}{1cm}{$50$} & Asymp. & $\mathbf{0.0500}$  & $0.0482$ & $0.0483$ & $2\times 10^5$ & $\mathbf{0.0494}$ & $0.0478$ & $\mathbf{0.0489}$ & $4\times 10^4$ \\ 
    & & Gamma & $\mathbf{0.0500}$ & $\mathbf{0.0499}$ & $0.0514$ & $4\times 10^1$ & $\mathbf{0.0510}$ & $\mathbf{0.0506}$ & $0.0514$ & $2\times 10^2$ \\ 
    & & MC & $\mathbf{0.0510}$ & $\mathbf{0.0506}$ & $\mathbf{0.0502}$ & $2\times 10^7$ & $\mathbf{0.0506}$ & $\mathbf{0.0497}$ & $\mathbf{0.0502}$ & $2\times 10^7$ \\ 
    \cline{2-11}
    & \multirow{3}{1cm}{$200$} & Asymp. & $\mathbf{0.0493}$ & $\mathbf{0.0488}$ & $\mathbf{0.0499}$ & $2\times 10^5$ & $\mathbf{0.0492}$ & $\mathbf{0.0496}$ & $\mathbf{0.0494}$ & $4\times 10^4$ \\ 
    & & Gamma & $\mathbf{0.0496}$ & $\mathbf{0.0502}$ & $\mathbf{0.0500}$ & $5\times 10^1$ & 0.0523 &$\mathbf{0.0512}$ & $0.0523$ & $2\times 10^2$ \\ 
    & & MC & $\mathbf{0.0504}$ & $\mathbf{0.0504}$ & $\mathbf{0.0505}$ & $1\times 10^8$ & $\mathbf{0.0505}$ & $\mathbf{0.0504}$ & $\mathbf{0.0507}$ & $2\times 10^8$ \\ 
      \midrule
    \multirow{9}{0.6cm}{$2$} & \multirow{3}{1cm}{$10$} & Asymp. & $0.0446$ & $0.0438$ & $0.0451$ & $4\times 10^4$ & $0.0445$ & $0.0443$ & $0.0515$ & $2\times 10^4$ \\ 
    & & Gamma & $0.0516$ & $0.0528$ & $0.0538$ & $6\times 10^1$ & $0.0540$ & $0.0551$ & $0.0607$ & $6\times 10^2$ \\ 
    & & MC & $\mathbf{0.0509}$ & $\mathbf{0.0510}$ & $\mathbf{0.0509}$ & $2\times 10^6$ & $\mathbf{0.0513}$ & $\mathbf{0.0512}$ & $\mathbf{0.0495}$ & $2\times 10^6$ \\ 
    \cline{2-11}
    & \multirow{3}{1cm}{$50$} & Asymp. & $\mathbf{0.0490}$ & $\mathbf{0.0491}$ & $0.0478$ & $4\times 10^4$ & $\mathbf{0.0490}$ & $\mathbf{0.0494}$ & $\mathbf{0.0494}$ & $2\times 10^4$ \\ 
    & & Gamma & $\mathbf{0.0507}$ & $0.0515$ & $0.0522$ & $6\times 10^1$ & $0.0529$ & $0.0542$ & $0.0547$ & $6\times 10^2$ \\ 
    & & MC & ${0.0524}$ & ${0.0517}$ & ${0.0527}$ & $1\times 10^7$ & ${0.0517}$ & ${0.0522}$ & ${0.0530}$ & $2\times 10^7$ \\ 
    \cline{2-11}
    & \multirow{3}{1cm}{$200$} & Asymp. & $\mathbf{0.0493}$ & $\mathbf{0.0498}$ & $\mathbf{0.0513}$ & $4\times 10^4$ & $\mathbf{0.0500}$ & $\mathbf{0.0496}$ & $\mathbf{0.0509}$ & $2\times 10^4$ \\ 
    & & Gamma & $\mathbf{0.0502}$ & $0.0529$ & $0.0524$ & $5\times 10^1$ & $0.0540$ & $0.0536$ & $0.0535$ & $6\times 10^2$ \\ 
    & & MC & $\mathbf{0.0502}$ & $\mathbf{0.0510}$ & ${0.0521}$ & $1\times 10^8$ & $\mathbf{0.0509}$ & ${0.0516}$ & $\mathbf{0.0512}$ & $2\times 10^8$ \\ 
      \midrule
    \multirow{9}{0.6cm}{$3$} & \multirow{3}{1cm}{$10$} & Asymp. & $0.0436$ & $0.0439$ & $0.0467$ & $3\times 10^4$ & $0.0443$ & $0.0468$ & $0.0635$ & $2\times 10^4$ \\ 
    & & Gamma & $\mathbf{0.0509}$ & $0.0528$ & $0.0561$ & $5\times 10^1$ & $0.0554$ & $0.0576$ & $0.0738$ & $7\times 10^2$ \\ 
    & & MC & $\mathbf{0.0505}$ & $\mathbf{0.0498}$ & $\mathbf{0.0503}$ & $2\times 10^6$ & $\mathbf{0.0498}$ & $\mathbf{0.0504}$ & $\mathbf{0.0511}$ & $2\times 10^6$ \\ 
    \cline{2-11}
    & \multirow{3}{1cm}{$50$} & Asymp. & $0.0479$ & $0.0485$ & $0.0483$ & $3\times 10^4$ & $0.0479$ & $\mathbf{0.0494}$ & $0.0549$ & $2\times 10^4$ \\ 
    & & Gamma & $\mathbf{0.0506}$ & $0.0527$ & $0.0539$ & $6\times 10^1$ & $0.0538$ & $0.0549$ & $0.0586$ & $7\times 10^2$ \\ 
    & & MC & ${0.0520}$ & ${0.0524}$ & ${0.0517}$ & $1\times 10^7$ & ${0.0528}$ & ${0.0519}$ & $\mathbf{0.0511}$ & $2\times 10^7$ \\ 
    \cline{2-11}
    & \multirow{3}{1cm}{$200$} & Asymp. & $\mathbf{0.0499}$ & $\mathbf{0.0496}$ & $\mathbf{0.0504}$ & $3\times 10^4$ & $\mathbf{0.0500}$ & $\mathbf{0.0511}$ & $0.0514$ & $2\times 10^4$ \\ 
    & & Gamma & $\mathbf{0.0510}$ & $0.0527$ & $0.0531$ & $5\times 10^1$ & $0.0553$ & $0.0548$ & $0.0545$ & $7\times 10^2$ \\ 
    & & MC & $\mathbf{0.0503}$ & $\mathbf{0.0503}$ & $\mathbf{0.0502}$ & $1\times 10^8$ & $\mathbf{0.0504}$ & $\mathbf{0.0504}$ & $\mathbf{0.0508}$ & $2\times 10^8$ \\ 
      \midrule
    \multirow{9}{0.6cm}{$5$} & \multirow{3}{1cm}{$10$} & Asymp. & $0.0426$ & $0.0421$ & $\mathbf{0.0501}$ & $2\times 10^4$ & $0.0452$ & $0.0549$ & $0.0375$ & $2\times 10^4$ \\ 
    & & Gamma & $\mathbf{0.0503}$ & $0.0546$ & $0.0603$ & $5\times 10^1$ & $0.0577$ & $0.0664$ & $0.0383$ & $7\times 10^2$ \\ 
    & & MC & $\mathbf{0.0507}$ & $\mathbf{0.0507}$ & $\mathbf{0.0508}$ & $2\times 10^6$ & $\mathbf{0.0504}$ & $\mathbf{0.0510}$ & $\mathbf{0.0494}$ & $2\times 10^6$ \\ 
    \cline{2-11}
    & \multirow{3}{1cm}{$50$} & Asymp. & $\mathbf{0.0488}$ & $\mathbf{0.0491}$ & $\mathbf{0.0499}$ & $2\times 10^4$ & $\mathbf{0.0491}$ & $0.0522$ & $0.0667$ & $2\times 10^4$ \\ 
    & & Gamma & $\mathbf{0.0505}$ & $0.0540$ & $0.0569$ & $6\times 10^1$ & $0.0572$ & $0.0575$ & $0.0692$ & $7\times 10^2$ \\ 
    & & MC & $\mathbf{0.0495}$ & $\mathbf{0.0501}$ & $\mathbf{0.0506}$ & $1\times 10^7$ & $\mathbf{0.0500}$ & $\mathbf{0.0511}$ & $\mathbf{0.0513}$ & $2\times 10^7$ \\ 
    \cline{2-11}
    & \multirow{3}{1cm}{200} & Asymp. & $0.0483$ & $\mathbf{0.0512}$ & $\mathbf{0.0501}$ & $2\times 10^4$ & $0.0481$ & $0.0522$ & $0.0568$ & $2\times 10^4$ \\ 
    & & Gamma & $\mathbf{0.0509}$ & $0.0547$ & $0.0562$ & $4\times 10^1$ & $0.0559$ & $0.0562$ & $0.0584$ & $7\times 10^2$ \\ 
    & & MC & $\mathbf{0.0512}$ & $\mathbf{0.0508}$ & $\mathbf{0.0505}$ & $1\times 10^8$ & $\mathbf{0.0505}$ & $\mathbf{0.0505}$ & $\mathbf{0.0495}$ & $2\times 10^8$ \\ 
      \bottomrule
      \end{tabular}
}
\fi
    \caption{\small Empirical rejection proportion for significance level $\alpha = 0.05$ of $T_{n,1}(\kappa)$ and $T_{n,2}(\rho)$ computed with $M = 10^5$ Monte Carlo samples using each approximation method: asymptotic distribution computed by Imhof's method, our gamma-match approximation, and Monte Carlo ($M=10^5$). Boldface denotes that the empirical rejection proportion lies within the $95\%$ confidence interval $(0.0486, 0.0513)$. In addition, execution times (in microseconds) for the three approximation methods are shown. The median of $10^2$ evaluations is averaged for the three $\kappa$ and $\rho$ values.}\label{tbl:null_asymp_dist_05}
    \vspace{-0.5cm}
\end{table}

Due to the rapid convergence of the null distribution to the asymptotic one, we propose a method to approximate exact-$n$ tail probabilities, which is shown to be faster than Imhof's exact method while maintaining accuracy, or even yielding a better approximation for small sample sizes $n < 50$. Our approximation consists in matching the first two exact moments of $T_{n, \ell}$ under $\mathcal{H}_0$ to a gamma distribution. To that end, we use the $V$-statistic form of $T_{n, \ell}$, $V_{n, \ell} = T_{n, \ell} + \tilde{\psi}_{\ell}(0)$, since $\mathbb{E}_{\mathcal{H}_0}[T_{n,\ell}]=0$. Matching the first exact two moments of $V_{n, \ell}$ to those of $\Gamma(k_0, \theta_0)$, we get the gamma parameters
\begin{align*}
    \hat{k}_0 %
    = \frac{\psi_\ell(0)-b_{0,q}(\psi_\ell)}{\hat{\theta}_0}\text{ and }
    \hat{\theta}_0 %
    = \frac{2(n-1)}{n}\frac{b_{0,q}(\psi_\ell^2) - b_{0,q}^2(\psi_\ell)}{\psi_\ell(0)-b_{0,q}(\psi_\ell)}
\end{align*}
that give the gamma match to the null distribution of $T_{n,\ell}$.

The performance of the gamma-match approximation is studied empirically in the same setting described in Section \ref{sec:asymp_dist}. The rows labeled ``Gamma'' in Table \ref{tbl:null_asymp_dist_05} present the rejection proportions for the significance level $\alpha=0.05$ when the critical values are approximated by the gamma-match critical values in the test decision, for $q\in\{1,2,3,5\}$, $n\in\{10, 50, 200\}$, $\kappa\in\{0.1, 1, 5\}$, and $\rho\in\{0.25, 0.5, 0.75\}$. Table \ref{tbl:null_asymp_dist_05} allows us to compare our approximation with the asymptotic distribution and the Monte Carlo ($M = 10^5$) approximation (rows labeled ``MC''). The accuracy differs between $T_{n,1}$ and $T_{n,2}$ and for different dimensions $q$. In the circular case ($q=1$), the gamma match presents lower errors than the asymptotic distribution for $n\in\{10, 50\}$, and similar results in $n=200$. In the spherical case ($q=2$), the gamma match improves the asymptotic distribution when $n=10$. For higher dimensions, the asymptotic approximation seems to work better for moderate sample sizes. As expected, the Monte Carlo ($M=10^5$) approximation achieves higher accuracy than our gamma-match approximation in general, and than the asymptotic distribution for small sample sizes.

However, the main advantage of the gamma match lies in the speed of computation, in particular for applications or simulations that do not need remarkable precision, such as estimating the statistics' power under different distributions. Table \ref{tbl:null_asymp_dist_05} shows the median execution time of the asymptotic distribution via Imhof's method with $K_\mathrm{tr}=50$, our gamma-match approximation, and the Monte Carlo ($M=10^5$) approximation. On average, compared to Imhof's method, the gamma match is 3--4 orders of magnitude faster for $T_{n,1}$, and 2 orders of magnitude faster for $T_{n,2}$. With respect to Monte Carlo, it is 5--7 orders of magnitude faster for $T_{n,1}$, and 4--6 orders of magnitude faster for $T_{n,2}$, depending on the sample size.

\subsection{\texorpdfstring{Effect of $K$ in $K$-fold tests}{Effect of K in K-fold tests}}
\label{sec:num:estimated-stat}

To quantify the dependence of the $K$-fold test according to the number of partitions $K$, an empirical study on the power of these tests is presented in Figure \ref{fig:k-fold-test} for $K \in \{2,4,10,20\}$. Each $K$-test is performed (using gamma-match $\alpha$-critical values) on $10^3$ Monte Carlo replicates of size $n=100$ for distributions \ref{vMF-dist}--\ref{MC-dist} depending on $\kappa_{\mathrm{dev}}$. The results show that the larger the number of partitions $K$, the higher the power achieved, although for large $K\geq10$ the power curve increases at a lower rate. Table \ref{tbl:null_rejection}, which contains the rejection proportion under uniformity (using asymptotic and gamma-match $\alpha$-critical values), shows that $K$ does not appear to significantly influence the empirical Type I error, since using asymptotic $\alpha$-critical values, the rejection frequency is maintained at the significance level $\alpha$. Notice that using the gamma-match approximation slightly changes the rejection proportion in certain dimensions, but, in general, it remains within the $95\%$ confidence intervals. For these reasons, in practice, we regard $K=10$ as a sensible choice to run the test, bearing also in mind that a larger $K$ risks exhausting the sample.

\begin{figure}[htpb!]
\iffigstabs
    \centering
    \scalebox{0.9}{
        \subfloat[][]{
            \includegraphics[width=0.5\textwidth]{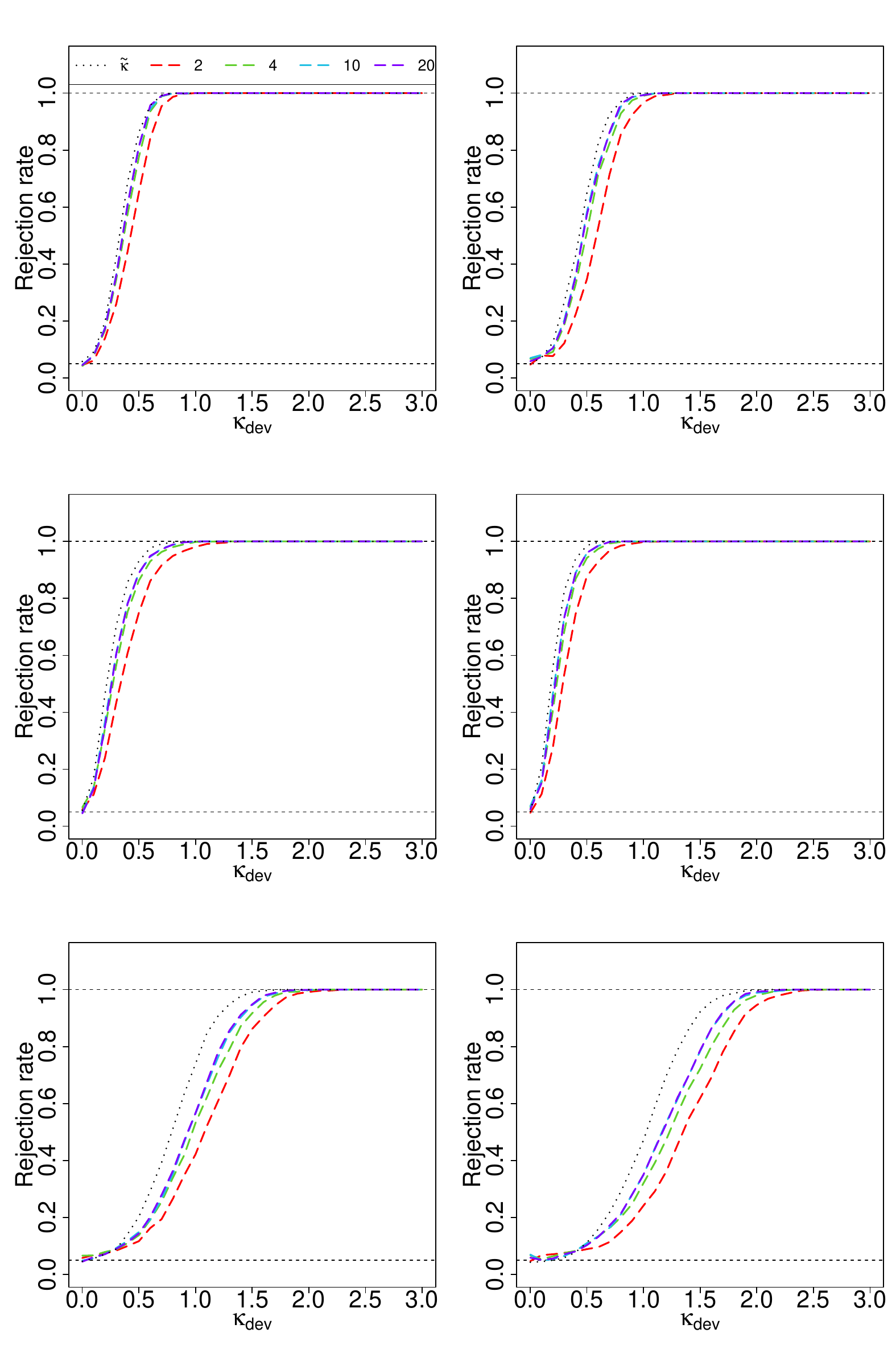}
                \vrule
            \includegraphics[width=0.5\textwidth]{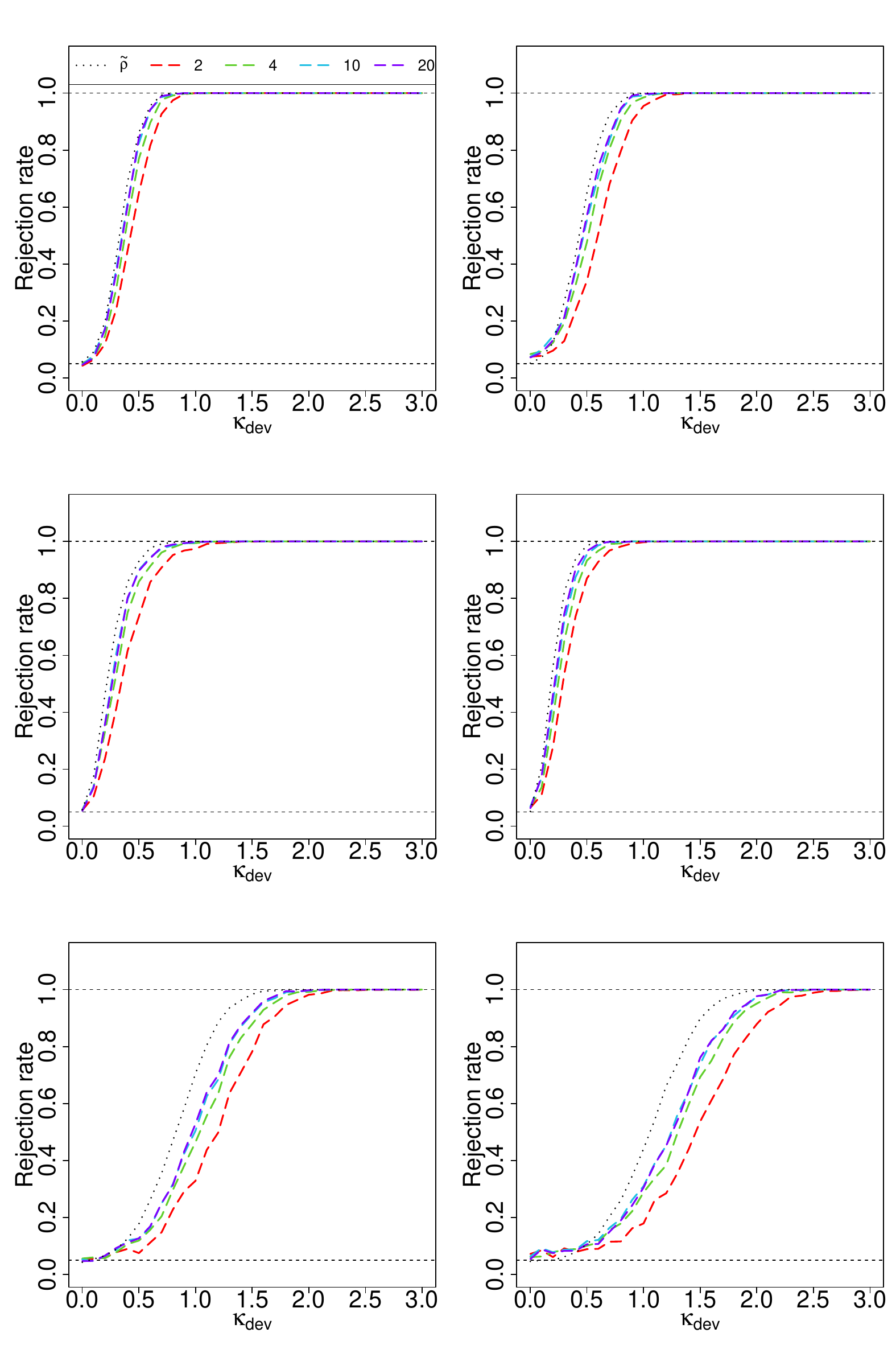}
        }
    }
	
    \scalebox{0.9}{
        \subfloat[][]{
    	\includegraphics[width=0.5\textwidth]{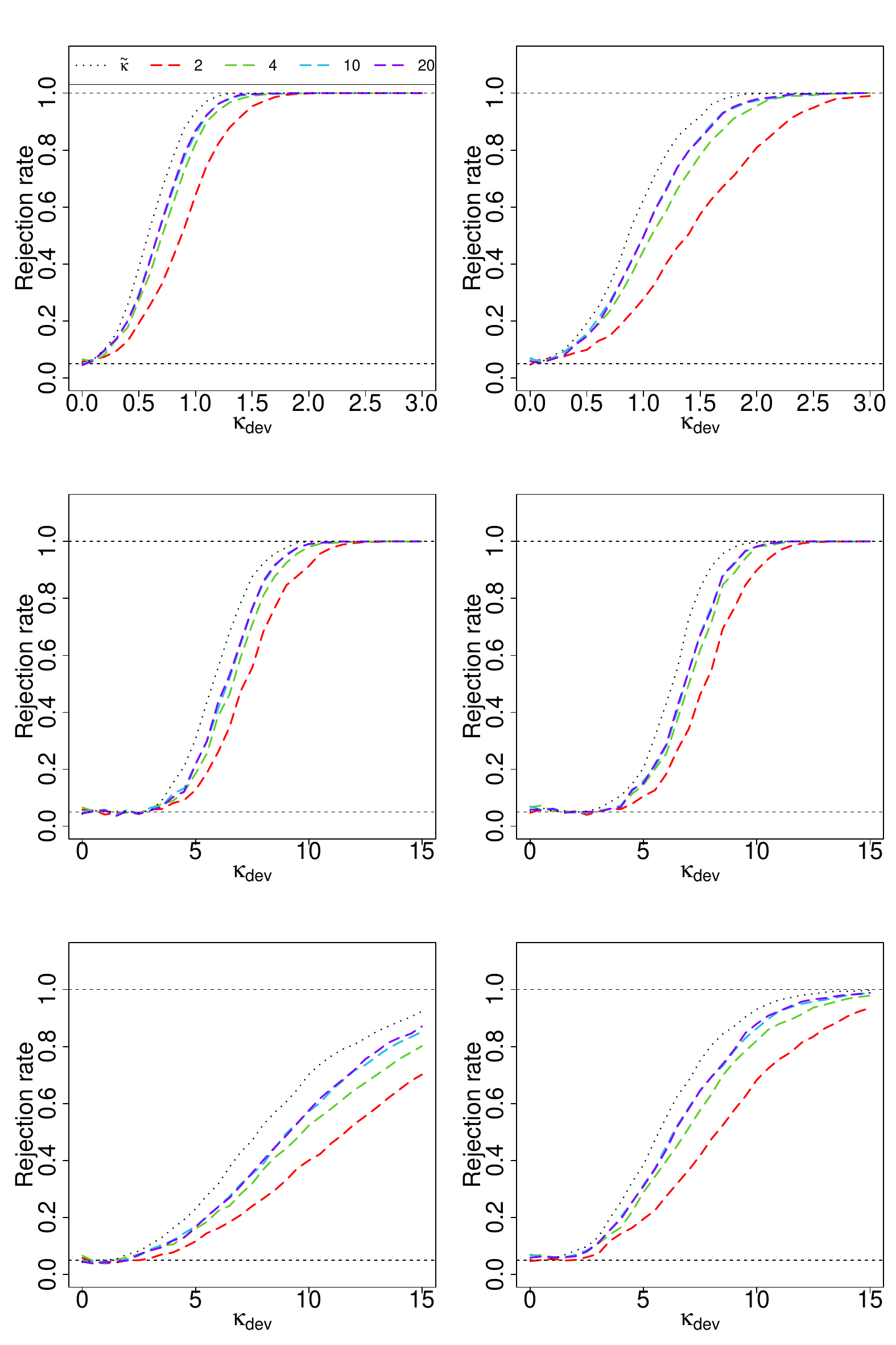}
            \vrule
    	\includegraphics[width=0.5\textwidth]{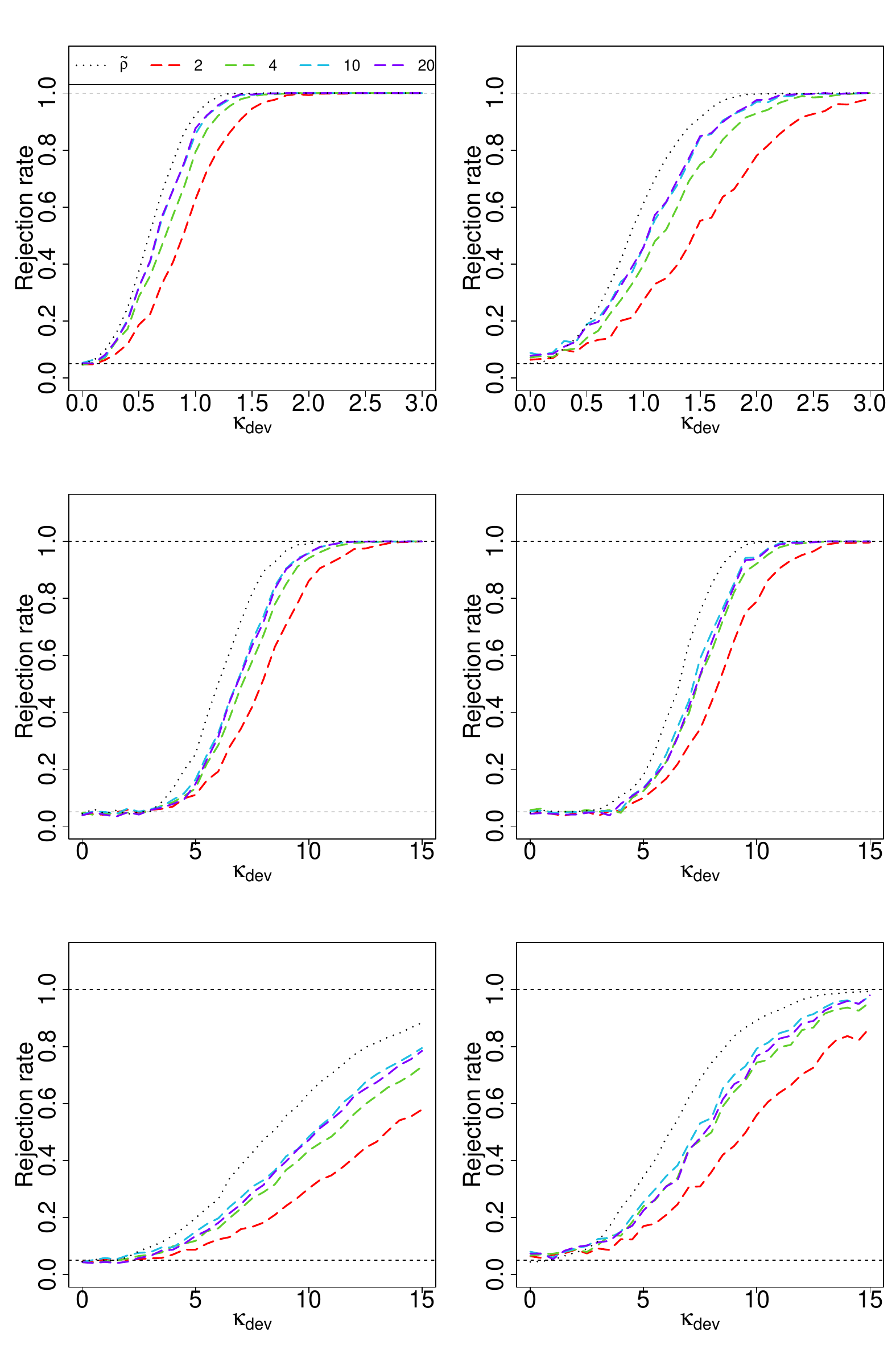}
        }
    }
\fi
	\caption{\small Empirical rejection frequency of the $K$-fold tests (Definition \ref{def:K}) $\smash{T_{n, 1}^{(K)}}$ (left panel) and $\smash{T_{n, 2}^{(K)}}$ (right panel) at significance level $\alpha=0.05$ for alternative distributions \ref{vMF-dist}--\ref{MC-dist} with concentration $\kappa_\mathrm{dev}$. The number of partitions, $K$, is indicated in the legend. $M=10^3$ samples of size $n=100$ were drawn from the alternative distribution. Dotted curves indicate the power of the oracle test based on $T_{n, \ell}(\tilde{\lambda}_{\mathcal{H}_1})$.} \label{fig:k-fold-test}
\end{figure}

\begin{table}[htpb!]
    \centering
    \scalebox{1}{
    \begin{tabular}{L{1.2cm} R{1.2cm} R{1.2cm} R{1.2cm} R{1.2cm} | R{1.2cm} R{1.2cm} R{1.2cm} R{1.2cm}}
    \toprule
        \multirow{2}{2em}{$K$} & \multicolumn{4}{c |}{$\smash{T_{n, 1}^{(K)}}$} & \multicolumn{4}{c}{$\smash{T_{n, 2}^{(K)}}$} \\
        \cline{2-9}
        & $q=1$ & $q=2$ & $q=3$ & $q=5$ & $q=1$ & $q=2$ & $q=3$ & $q=5$ \\
        \midrule
        & \multicolumn{8}{c}{Asymptotic $\alpha$-critical values} \\
        $2$ & $\mathbf{0.047}$ & $\mathbf{0.052}$ & $\mathbf{0.057}$ & $\mathbf{0.061}$ & $\mathbf{0.047}$ & $\mathbf{0.052}$ & $\mathbf{0.045}$ & $\mathbf{0.039}$ \\
        $4$ & $\mathbf{0.042}$ & $\mathbf{0.049}$ & $\mathbf{0.055}$ & $\mathbf{0.061}$ & $\mathbf{0.051}$ & $\mathbf{0.059}$ & $\mathbf{0.067}$ & $0.028$ \\
        $10$ & $\mathbf{0.043}$ & $\mathbf{0.040}$ & $\mathbf{0.048}$ & $0.082$ & $\mathbf{0.046}$ & $\mathbf{0.064}$ & $\mathbf{0.063}$ & $0.020$ \\
        $20$ & $\mathbf{0.041}$ & $\mathbf{0.038}$ & $\mathbf{0.044}$ & $\mathbf{0.060}$ & $\mathbf{0.043}$ & $\mathbf{0.064}$ & $\mathbf{0.058}$ & $0.018$ \\
        \midrule
        & \multicolumn{8}{c}{Gamma-match $\alpha$-critical values} \\
        $2$ & $\mathbf{0.056}$ & $\mathbf{0.047}$ & $\mathbf{0.054}$ & $\mathbf{0.063}$ & $\mathbf{0.048}$ & $0.066$ & $\mathbf{0.057}$ & $\mathbf{0.047}$ \\
        $4$ & $\mathbf{0.062}$ & $0.067$ & $\mathbf{0.058}$ & $0.073$ & $\mathbf{0.050}$ & $0.068$ & $\mathbf{0.062}$ & $\mathbf{0.044}$ \\
        $10$ & $\mathbf{0.049}$ & $0.070$ & $\mathbf{0.049}$ & $0.082$ & $\mathbf{0.050}$ & $0.071$ & $\mathbf{0.063}$ & $\mathbf{0.038}$ \\
        $20$ & $\mathbf{0.045}$ & $\mathbf{0.059}$ & $\mathbf{0.055}$ & $0.069$ & $\mathbf{0.047}$ & $\mathbf{0.064}$ & $\mathbf{0.054}$ & $0.031$ \\
    \bottomrule
    \end{tabular}
    }
    \caption{\small Empirical rejection frequency of the $K$-fold tests $\smash{T_{n, 1}^{(K)}}$ and $\smash{T_{n, 2}^{(K)}}$ at significance level $\alpha = 0.05$ under $\Hcal_0$, for different values of $K$. $M=10^3$ samples of size $n=100$ were drawn from the uniform distribution. Critical values were computed using the asymptotic distribution and the gamma-match approximation. Boldface denotes that the empirical rejection proportion lies within the $95\%$ confidence interval $(0.036, 0.064)$.}
    \label{tbl:null_rejection}
\end{table}

\begin{figure}[htpb!]
\iffigstabs
    \centering
    \subfloat[][]{
    	\includegraphics[width=0.45\textwidth]{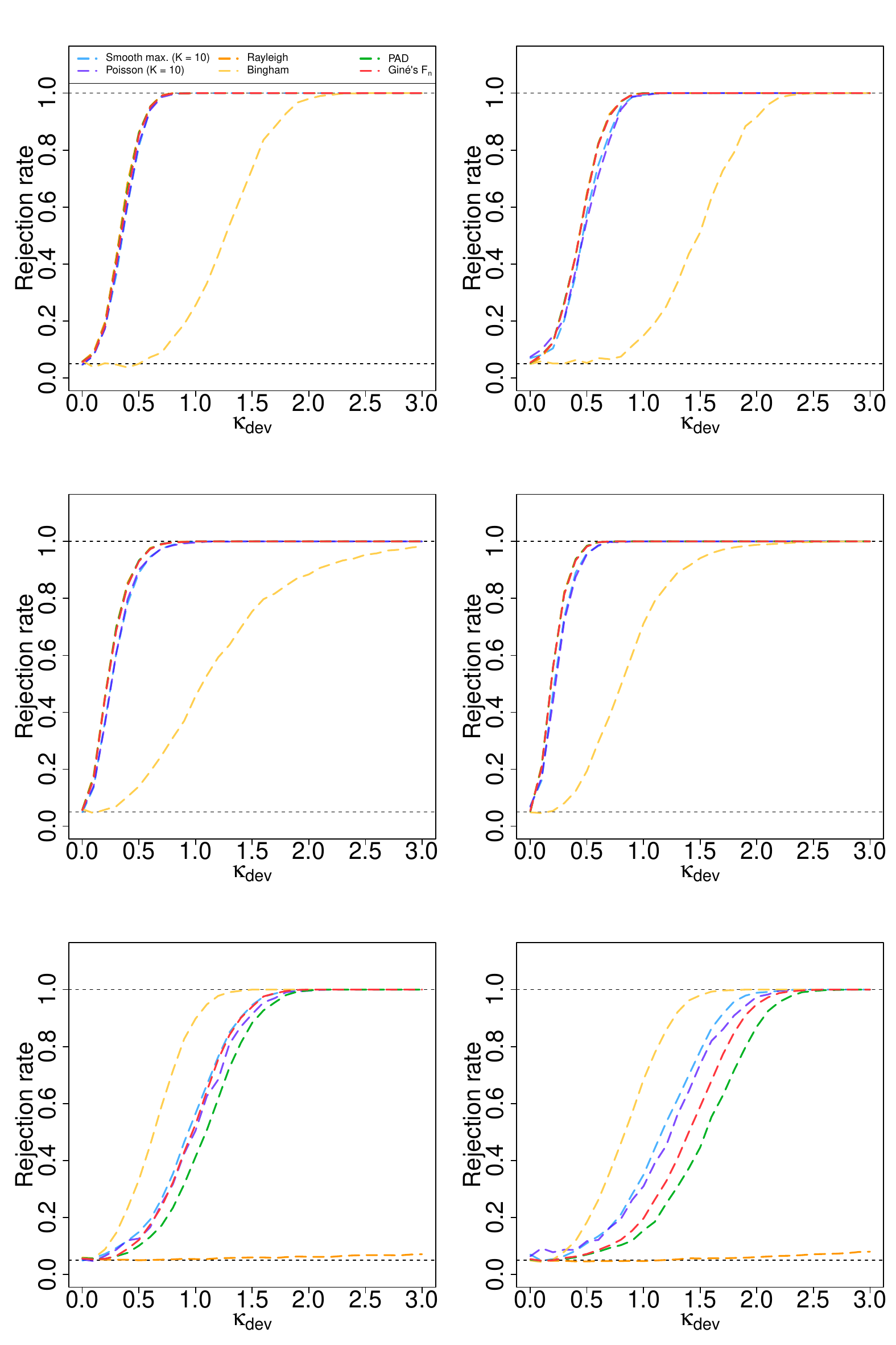}
            \includegraphics[width=0.45\textwidth]{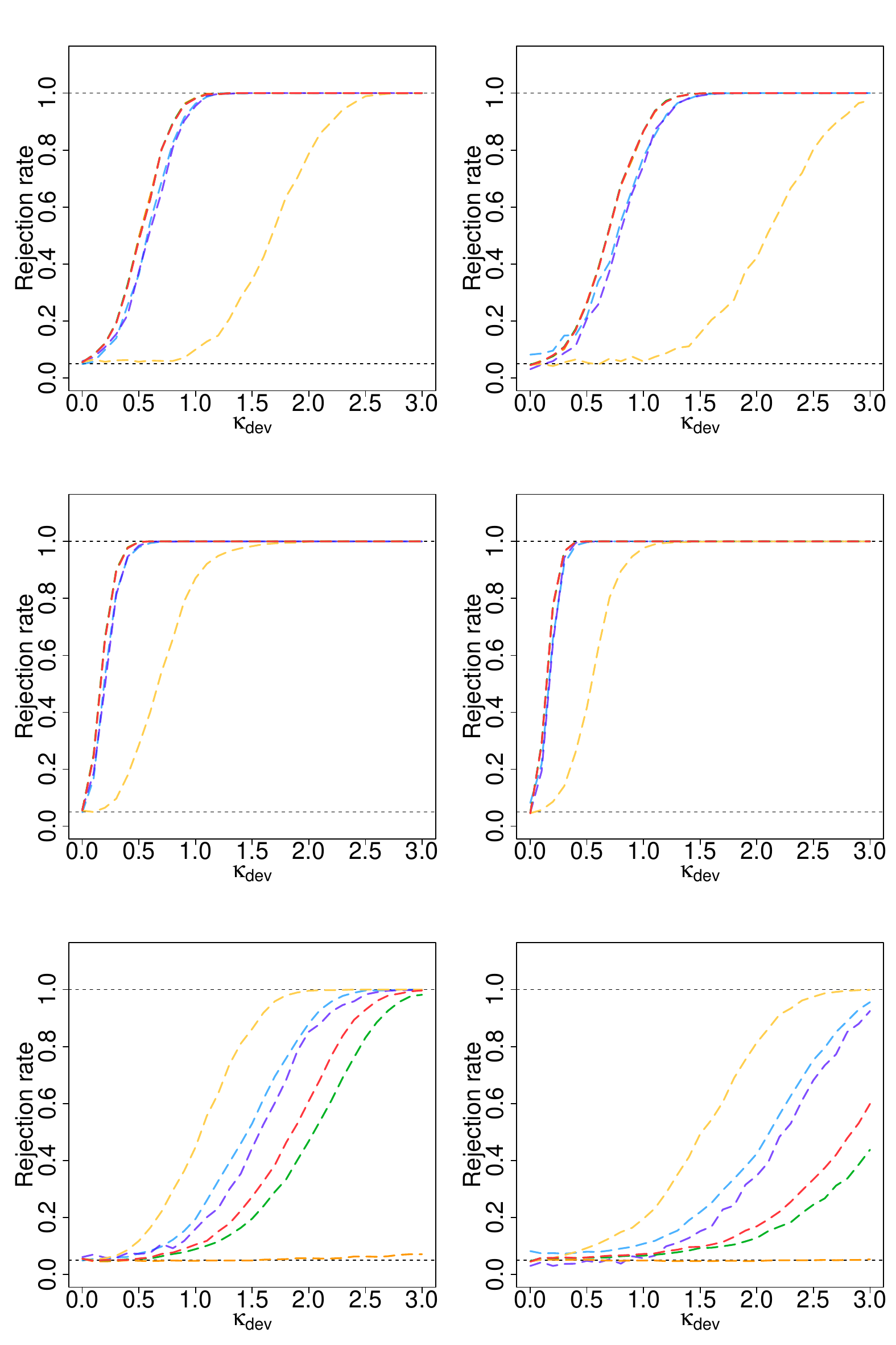}
    }
    
    \subfloat[][]{
    	\includegraphics[width=0.45\textwidth]{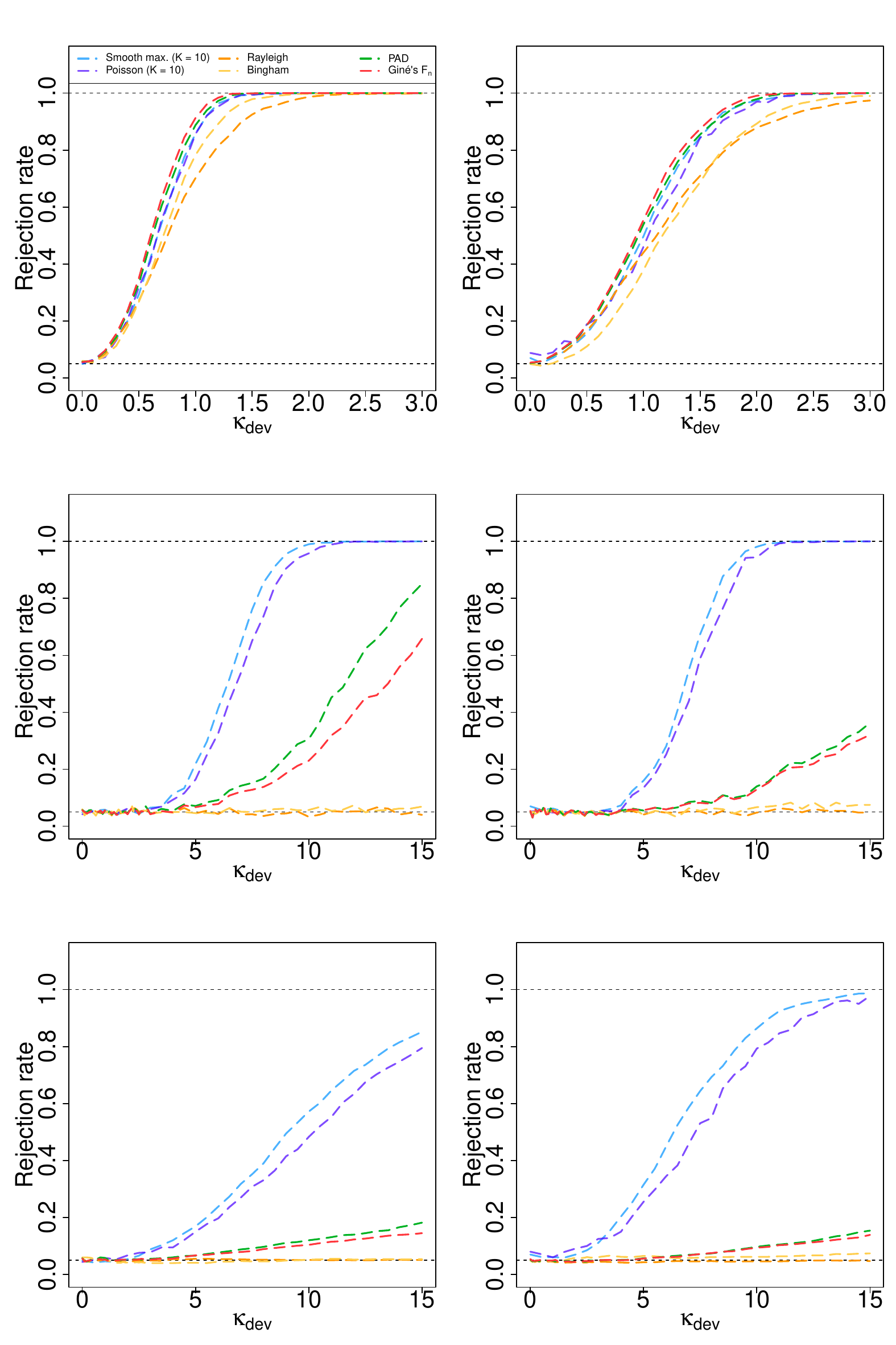}
            \includegraphics[width=0.45\textwidth]{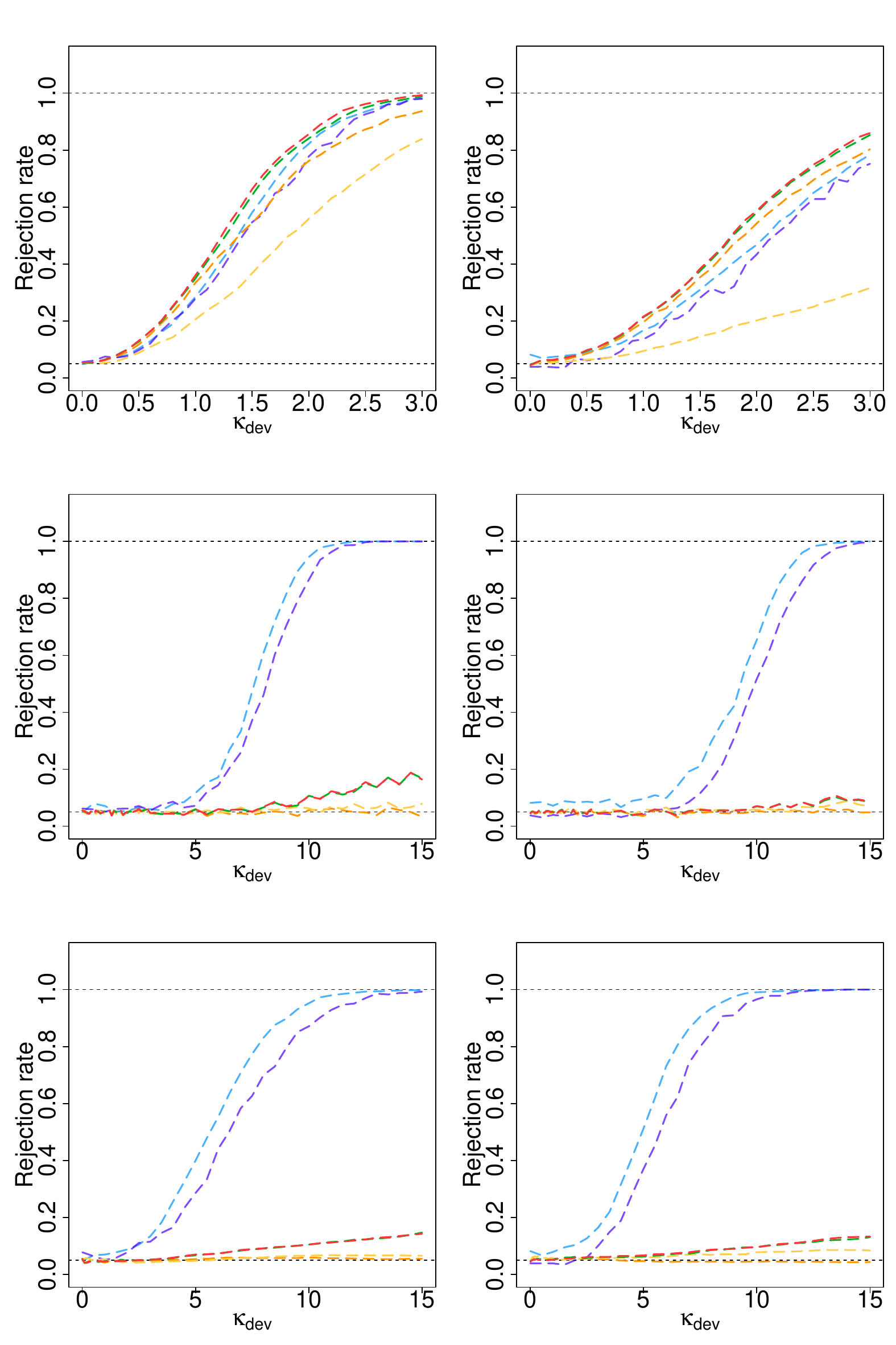}
    }
	
\fi
	\caption{\small Empirical rejection frequency of different tests of uniformity at significance level $\alpha=0.05$ for varying concentrations $\kappa_\mathrm{dev}$. From top to bottom, rows represent the alternative distributions \ref{vMF-dist}--\ref{MC-dist}. From left to right, columns stand for $q=1,2,3,5$. $10$-fold tests $\smash{T_{n, 1}^{(10)}}$ and $\smash{T_{n, 2}^{(10)}}$, Rayleigh, Bingham, PAD, and Giné's $F_n$ tests are compared.} \label{fig:test-comparison}
\end{figure}

\subsection{Comparison with other tests}
\label{sec:comp}

We compare through simulations the empirical performance of the tests proposed by comparing their empirical powers with Sobolev tests of uniformity on $\Sq$. The $10$-fold tests $\smash{T_{n, 1}^{(10)}}$ and $\smash{T_{n, 2}^{(10)}}$ are compared to the following tests: Rayleigh ($R_n$), Bingham ($B_n$), Projected Anderson--Darling (PAD), and Giné's $F_n$ \citep{Gine1975, Prentice1978}.

For each $q\in\{1,2,3,5\}$, $M = 10^3$ independent random samples of size $n = 100$ were generated from each of the \ref{vMF-dist}--\ref{SC-dist} distributions with concentration parameters $\kappa_\mathrm{dev} \in \{k/10: k=1,\ldots,30\}$ and from \ref{MvMF-dist}--\ref{MC-dist} with $\kappa_\mathrm{dev} \in \{k/2: k=1,\ldots,30\}$. The compared tests reject $\mathcal{H}_0$ based on exact-$n$ critical values approximated by $M=10^5$ Monte Carlo replicates at the significance level $\alpha = 0.05$. The $\alpha$-critical values of our tests based on $T_{n, \ell}$ were computed using the gamma-match (Section \ref{sec:exact_approx}) to improve its speed.

The empirical rejection frequencies at $\alpha = 0.05$ are presented in Figure \ref{fig:test-comparison}. The following conclusions are drawn from the study:
\begin{enumerate}[label=(\textit{\roman*}),ref=(\textit{\roman*})]
    \item The $10$-tests $T_{n,1}^{(10)}$ and $T_{n,2}^{(10)}$ perform very similarly in all alternatives and dimensions $q$. However, there is a consistent advantage in terms of power for $\smash{T_{n, 1}^{(10)}}$.
    \item Unimodal alternatives \ref{vMF-dist} and \ref{C-dist} are well detected by $T_{n,1}^{(10)}$ and $T_{n,2}^{(10)}$. In addition, the power difference relative to the Rayleigh test is minimal, despite being the optimal one in \ref{vMF-dist}.
    \item The axial alternative \ref{W-dist}, where Bingham is optimal, is harder to detect by our tests, although they are more powerful than PAD and Giné's $F_n$, especially in $q>1$.
    \item In the non-unimodal and non-axial distribution \ref{SC-dist}, $T_{n,1}^{(10)}$ and $T_{n,2}^{(10)}$ perform similarly to PAD and Giné's $F_n$.
    \item In the mixtures \ref{MvMF-dist} and \ref{MC-dist} distributions, $T_{n,1}^{(10)}$ and $T_{n,2}^{(10)}$ outperform all the competing tests by a large difference.
\end{enumerate}

According to the previous conclusions, we regard the $10$-fold tests $\smash{T_{n,1}^{(10)}}$ and $\smash{T_{n,2}^{(10)}}$ as competing tests to other tests of uniformity due to their omnibusness, good performance against unimodal alternatives, relative robustness against non-unimodal alternatives, and outstanding power against multimodal mixtures.

\section{Nursing times of wild polar bears}
\label{sec:bears}

A long-term study of the behavior of wild polar bears was conducted between 1973 and 1999, when wild polar bears were observed during spring and summer in the Canadian Arctic \citep{Stirling2022}. The data collected have been used to analyze different aspects of the life of adult and young polar bears in their habitat, such as hunting and feeding \citep{Stirling1974, Stirling1978} or breeding behavior \citep{Stirling2016}. 

The data records also collect the nursing behavior of female adults with their cubs. The relevant variable consists of 225 individual observations of the times at which the adult was nursing her cubs from the years 1973 to 1999: 69 observed during spring and 156 in summer (Figure \ref{fig:polar-bears}). 

Some interesting behavioral questions are whether females nurse continuously throughout the day and night, without any distinguishable peak period of nursing activity, and whether the nursing behavior is consistent between seasons. These questions can be addressed by testing the uniformity of the start nursing times around the 24-hour clock, translated to the unit circle $\mathbb{S}^1$, both for the whole dataset, and for spring and summer separately.

\begin{figure}[htpb!]
    \centering
    \includegraphics[trim={0cm 2cm 0cm 2cm},clip,width=\textwidth]{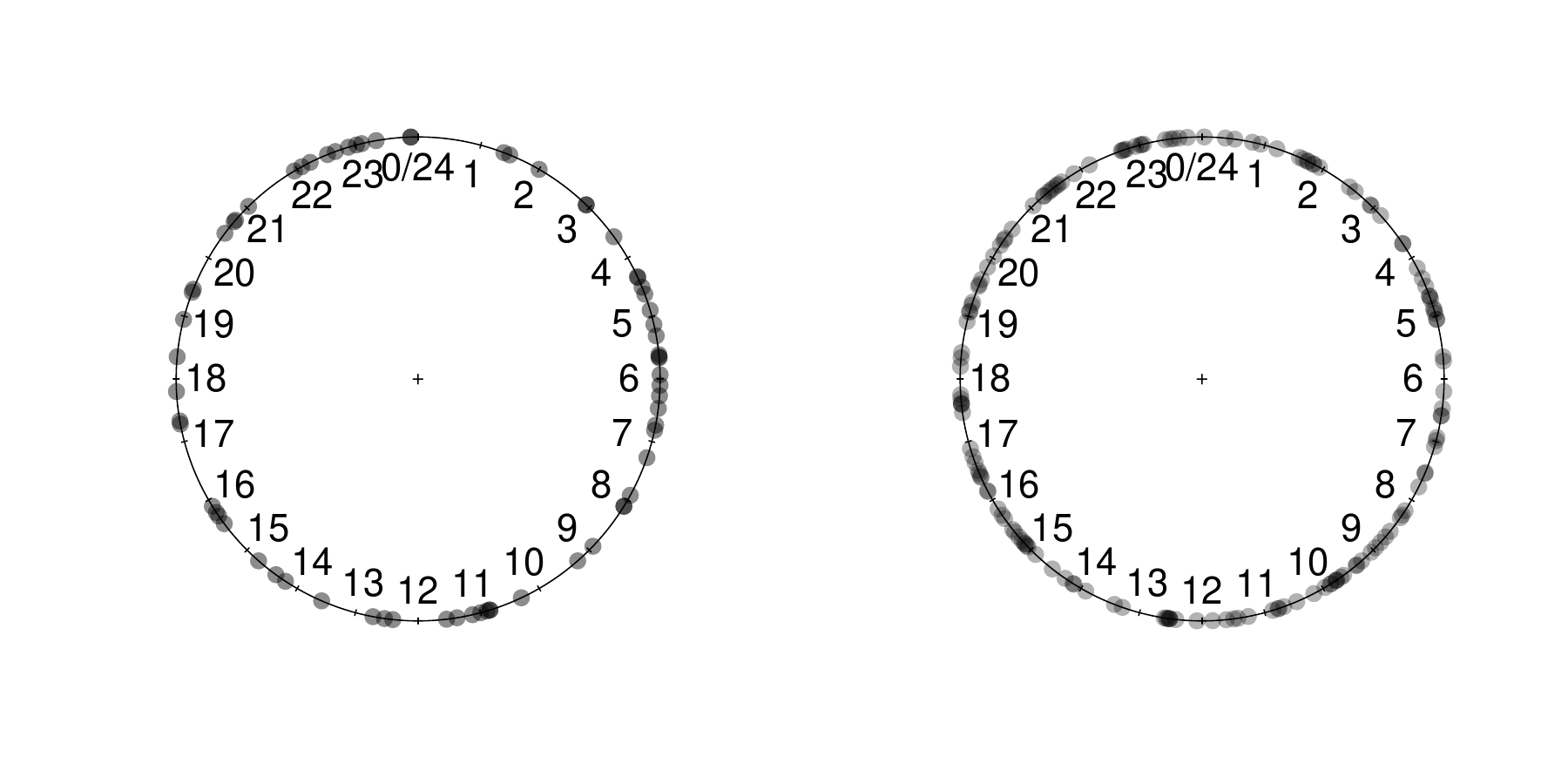}
    \caption{\small Starting times of nursing periods observed on wild polar bears of the Canadian Arctic during spring (left) and summer (right) of the time span 1973--1999.}
    \label{fig:polar-bears}
\end{figure}

Table \ref{tbl:polar-bears} shows that uniformity is not rejected at the significance level $\alpha=0.05$ for the tests based on $T_{n,1}$ and $T_{n,2}$, neither for specific values of $\kappa$ and $\rho$, nor for the $10$-fold version, across all periods considered. A similar behavior is observed for the classical uniformity tests, which do not reject uniformity for either season or for the full field season. Overall, the results reinforce the conclusion that the observed nursing times are compatible with uniformity on the circle.

\begin{table}[htpb!]
    \centering
    \scalebox{0.95}{
    \begin{tabular}{L{1.5cm} | L{0.6cm} | C{0.7cm} C{0.7cm} C{0.7cm} | C{0.7cm} | C{0.7cm} C{0.7cm} C{0.7cm} | C{1.1cm} | L{0.7cm} L{0.7cm} L{0.75cm} L{0.7cm}}
    \toprule
        \multirow{2}{1cm}{Season} & \multirow{2}{0.45cm}{$n$} & \multicolumn{3}{c|}{$T_{n,1}(\kappa)$} & \multirow{2}{0.7cm}{$T_{n,1}^{(10)}$} & \multicolumn{3}{c|}{$T_{n,2}(\rho)$} & \multirow{2}{1.1cm}{$T_{n,2}^{(10)}$} & \multirow{2}{0.6cm}{$R_n$} & \multirow{2}{0.6cm}{$B_n$} & \multirow{2}{0.75cm}{PAD} & \multirow{2}{0.6cm}{$F_n$} \\
        & & $0.1$ & $1$ & $10$ & & $0.25$ & $0.5$ & $0.75$ & & & & & \\
        \midrule
        Spring & 69 & 0.49 & 0.56 & 0.42 & 0.83 & 0.54 & 0.52 & 0.48 & 0.78 & 0.48 & 0.76 & 0.52 & 0.56 \\
        Summer & 156 & 0.68 & 0.79 & 0.90 & 0.93 & 0.81 & 0.88 & 0.90 & 0.73 & 0.67 & 0.76 & 0.84 & 0.85 \\
        \midrule
        Field & 225 & 0.95 & 0.96 & 0.84 & 0.96 & 0.96 & 0.93 & 0.85 & 0.86 & 0.94 & 0.67 & 0.93 & 0.93 \\
    \bottomrule
    \end{tabular}}
    \caption{\small $p$-values of several uniformity tests applied to the starting times of nursing periods observed in wild polar bears. The $p$-values are obtained from the $T_{n,\ell}$ tests for specific parameters, the $10$-fold tests $\smash{T_{n,\ell}^{(10)}}$, and other available tests of uniformity. The $p$-values were obtained through Monte Carlo ($M = 10^4$).}
    \label{tbl:polar-bears}
\end{table}

\section*{Supplementary materials}

Supplementary Materials (SM) contain the proofs of the paper and further numerical results.

\section*{Acknowledgments}

The authors acknowledge support from grant PID2021-124051NB-I00, funded by MCIN/\-AEI/\-10.13039/\-501100011033 and by ``ERDF A way of making Europe''. The authors greatly acknowledge the data availability from Dr. Ian Stirling (University of Alberta) and Dr. Lynne Burns (Environment and Climate Change Canada), and the stimulating conversations with them. Comments by two anonymous referees and an Associate Editor are greatly acknowledged.



\fi

\ifsupplement

\newpage
\title{Supplementary materials for ``On new omnibus tests of uniformity on the hypersphere''}
\setlength{\droptitle}{-1cm}
\predate{}%
\postdate{}%
\date{}

\author{Alberto Fern\'andez-de-Marcos$^{1}$ and Eduardo Garc\'ia-Portugu\'es$^{1,2}$}
\footnotetext[1]{Department of Statistics, Carlos III University of Madrid (Spain).}
\footnotetext[2]{Corresponding author. e-mail: \href{mailto:edgarcia@est-econ.uc3m.es}{edgarcia@est-econ.uc3m.es}.}
\maketitle

\begin{abstract}
    These supplementary materials contain two sections. Section \ref{app:proofs} provides the proofs of the results of the paper, which are based on two lemmas. Section \ref{app:morenum} contains further numerical results omitted from the main text.
\end{abstract}
\begin{flushleft}
	\small\textbf{Keywords:} Directional statistics; Poisson kernel; Sobolev tests; Smooth maximum.
\end{flushleft}

\appendix

\section{Proofs}
\label{app:proofs}

\begin{proof}[Proof of Proposition \ref{prp:lse_connection}.]
\textit{Proof of \ref{prp:lse_rayleigh}}. We apply the transformation $t\mapsto (q+1)(\kappa^{-1}t + 1)$ to $T_{n,1}$. Since it is monotonic, the tests based on $T_{n,1}$ and based on any statistic resulting from this transformation are equivalent.

First, we address the case $q\geq2$. Using l'H\^{o}pital's rule and $\partial (z^{-\nu}\Ical_{\nu}(z))/\partial z = z^{-\nu}\Ical_{\nu+1}(z)$,
\begin{align}
    \lim_{\kappa\to0}&(q+1)\bigg(\frac{e^{\kappa}}{\kappa} T_{n,1} + 1\bigg)\nonumber\\
    =&\;(q+1)\bigg(1 + \frac{2}{n}\sum_{1\leq i<j\leq n} \bX_i'\bX_j - A(n,q)\lim_{\kappa\to0} \kappa^{-(q-1)/2}\Ical_{(q+1)/2}(\kappa) \bigg),\label{eq:rayleigh-q-gt-2}
\end{align}
where $A(n, q) := 2^{(q-3)/2}(n-1)\Gamma((q-1)/2)(q-1)$. Using $\Ical_{\nu}(z)\asymp (z/2)^\nu/\Gamma(\nu+1)$ when $z\to0$ \cite[Equation 10.25.2]{NIST:DLMF-SM},
\begin{align*}
    \lim_{\kappa\to0} \kappa^{-(q-1)/2}\Ical_{(q+1)/2}(\kappa)
    =&\;\lim_{\kappa\to0} \kappa^{-(q-1)/2} (\kappa/2)^{(q+1)/2}/\Gamma((q+3)/2)=0,
\end{align*}
which makes \eqref{eq:rayleigh-q-gt-2} equal to $R_n$.

Second, we prove the circular case ($q=1$) analogously, where we use that $\mathcal{I}_{0}(0) = 1$ and $\mathcal{I}_{\nu}(0) = 0$, $\nu>0$. Then,
\begin{align*}
    \lim_{\kappa\to0}2\left(\frac{e^{\kappa}}{\kappa} T_{n,1} + 1\right)
    =&\;2\bigg(1 + \frac{2}{n}\sum_{1\leq i<j\leq n} \bX_i'\bX_j - (n-1)\lim_{\kappa\to0} \mathcal{I}_{1}(\kappa) \bigg) = R_n.
\end{align*}

\textit{Proof of \ref{prp:lse_cai_jiang}}. The proposition holds due to the monotonic relation \eqref{eq:lse} between $T_{n,1}$ and the LSE function, and the fact that
    \begin{align*}
        \lim_{\kappa\to\infty}\mathrm{LSE}_\kappa\lrp{\{\bX_i'\bX_j\}_{1\leq i<j\leq n}}=\max_{1\leq i<j\leq n} \bX_i'\bX_j.
    \end{align*}
\end{proof}

\begin{proof}[Proof of Proposition \ref{prp:poisson_rayleigh}.]
We apply the transformation $t\mapsto \rho^{-1}\left(t - 1 + \psi_2(0)\right)$ to $T_{n,2}$. Then,
\begin{align*}
    \lim_{\rho\to0}\frac{T_{n,2} - 1 + \psi_2(0)}{\rho}
    =&\;\lim_{\rho\to0}\frac{2}{n}\sum_{1\leq i<j\leq n} \frac{\partial \psi_2(\theta_{ij})}{\partial \rho} + \frac{\partial \psi_2(0)}{\partial \rho}\\
    =&\;(q+1)\bigg(\frac{2}{n}\sum_{1\leq i<j\leq n} \bX_i'\bX_j + 1\bigg)=R_n.
\end{align*}
Since the transformation is monotonic, the two tests are equivalent.
\end{proof}

\begin{proof}[Proof of Proposition \ref{prp:gegen}.]
The coefficients of $\psi_1$ for $q\geq2$ follow from the second equation in \citet[page 227]{Magnus1966-SM}:
\begin{align}
	e^{\kappa x}=&\;\lrp{\frac{2}{\kappa}}^{(q-1)/2}\frac{\Gamma((q-1)/2)}{2} \sum_{k=0}^{\infty}(2k+q-1) \mathcal{I}_{(2k+q-1)/2}(\kappa) C_k^{(q-1)/2}(x). \label{eq:ekx}
\end{align}
Now, taking the limit $q\to1$ in \eqref{eq:ekx} produces
\begin{align*}
	e^{\kappa x}&=\lim_{q\to1}\bigg\{\frac{(q-1)\Gamma((q-1)/2)}{2} \sum_{k=0}^{\infty}(2k+q-1) \mathcal{I}_{(2k+q-1)/2}(\kappa) (q-1)^{-1}C_k^{(q-1)/2}(x)\bigg\}\\
	&=\sum_{k=0}^{\infty}(2-\delta_{k,0})\mathcal{I}_{k}(\kappa) T_k(x)
\end{align*}
after invoking \eqref{eq:Ck0} and using $\Gamma(s)=\big(s+O(s^2)\big)^{-1}$ as $s\to0^+$.

The coefficients for $\psi_2$ follow by combining several generating equations for the Chebyshev and Gegenbauer polynomials. Equation 18.12.7 in \cite{NIST:DLMF-SM}, together with twice Equation 18.12.5 minus Equation 18.12.4, gives
\begin{align*}
	\frac{1-\rho^2}{\lrp{1+\rho^2-2\rho x}^{(q+1)/2}}=\begin{cases}
		\sum_{k=0}^{\infty} (2-\delta_{k,0})\rho^k T_k(x),&q=1,\\
		\sum_{k=0}^{\infty} \frac{2k+q-1}{q-1}\rho^k C_k^{(q-1)/2}(x),&q\geq 2.
	\end{cases}
\end{align*}
\end{proof}

\begin{proof}[Proof of Proposition \ref{prp:moments_null}.]
To derive the expectation of $T_{n,\ell}$, consider the tangent-normal change of variables $\bx=x\by+(1-x^2)^{1/2}\bB_{\by}\bxi$, with $x:=\bx'\by\in[-1,1]$, $\bxi\in\mathbb{S}^{q-1}$, and $\bB_{\by}$ a $(q+1)\times q$ matrix with orthonormal columns to $\by$ (see Lemma 2 in \cite{Garcia-Portugues2013b-SM}). It then follows that
\begin{align}
    \mathbb{E}_{\Hcal_0}[\psi_{\ell}(\theta_{12})]=&\;\int_{\Sq}\int_{\Sq} \psi_{\ell}(\cos^{-1}(\bx'\by)) \,\rd\nu_q(\bx) \,\rd\nu_q(\by)\nonumber\\
    =&\;\frac{1}{\om{q}}\int_{\Sq}\int_{\mathbb{S}^{q-1}}\int_{-1}^{1} \psi_{\ell}(\cos^{-1}(x))(1-x^2)^{q/2-1} \,\rd x \, \rd\sigma_{q-1}(\bxi) \,\rd\nu_q(\by)\nonumber\\
    =&\;\frac{\om{q-1}}{\om{q}}\int_{-1}^{1} \psi_{\ell}(\cos^{-1}(x))(1-x^2)^{q/2-1} \,\rd x\nonumber\\
    =&\;b_{0,q}(\psi_{\ell}). \label{eq:b0H0}
\end{align}
Plugging this expression into \eqref{eq:stat_def} gives $\mathbb{E}_{\mathcal{H}_0}[T_{n, \ell}]$. Recall that the kernel of the $U$-statistic is centered under $\Hcal_0$: $\tilde{\psi}_{\ell}=\psi_{\ell} - b_{0,q}(\psi_\ell)$.

The variance of a $U$-statistic is \citep[see, e.g.,][Lemma A, page 183]{Serfling1980-SM}:
\begin{align}
    \mathbb{V}\mathrm{ar}[T_{n, \ell}]=&\;\frac{2(n-1)}{n}\left[2(n-2)\eta_{1, \ell} + \eta_{2, \ell}\right],\label{eq:var-u-stat}
\end{align}
where $\eta_{1, \ell} := \mathbb{V}\mathrm{ar}[\mathbb{E}[\tilde{\psi}_\ell(\theta_{12})\mid\bX_1]]$ and $\eta_{2, \ell} := \mathbb{V}\mathrm{ar}[\tilde{\psi}_\ell(\theta_{12})]$. Under $\mathcal{H}_0$, $\eta_{1, \ell} =0$ due to the degeneracy of the kernel, and thus
\begin{align*}
    \mathbb{V}\mathrm{ar}_{\Hcal_0}[T_{n, \ell}]
    =\frac{2(n-1)}{n}(b_{0,q}\big(\psi_{\ell}^2) - b_{0,q}^2(\psi_{\ell})\big).
\end{align*}
\end{proof}

\begin{proof}[Proof of Theorem \ref{thm:asymp_null}.]
The proof follows from \cite{Gine1975-SM} and \cite{Prentice1978-SM}. We use the combined key result from these two references, stated in Theorem 3.1 in \cite{Garcia-Portugues2020b-SM}, which establishes that
\begin{align}
    S_{n,q}(\{w_{k,q}\}) \inlawH \sum_{k=1}^\infty w_{k,q} Y_{d_{k,q}}\label{eq:Soboas}
\end{align}
for a Sobolev test statistic \eqref{eq:canoSob}. Due to the connection \eqref{eq:ws} and \eqref{eq:stat_def},
\begin{align*}
    S_{n,q}(\{w_{k,q,\ell}\})=\frac{1}{n}\sum_{i,j=1}^n [\psi_\ell(\theta_{ij})-b_{0,q}]=\frac{2}{n}\sum_{1\leq i<j\leq n} \psi_\ell(\theta_{ij}) -nb_{0,q} + \psi_\ell(0).
\end{align*}
Then, because of Proposition \ref{prp:moments_null} and $\tilde{\psi}_{\ell}(0) = \psi_{\ell}(0) - \mathbb{E}_{\Hcal_0}[\psi_{\ell}(\theta_{12})]$,
\begin{align*}
    S_{n,q}(\{w_{k,q,\ell}\})=T_{n,\ell}+(n-1)\mathbb{E}_{\Hcal_0}[\psi_\ell(\theta_{12})]-nb_{0,q}+ \psi_\ell(0)=T_{n,\ell}+\tilde{\psi}_{\ell}(0)
\end{align*}
and \eqref{eq:null_asymp_dist} follows from \eqref{eq:Soboas}. Note that \eqref{eq:null_asymp_dist} is also expressible as
\begin{align*}
    T_{n, \ell}\inlawH \sum_{k=1}^\infty w_{k,q,\ell}\big(Y_{d_{k,q}} - d_{k,q}\big),
\end{align*}
since from \eqref{eq:psi_ell} evaluated at $\theta=0$ and expressed in terms of \eqref{eq:dkq}, we obtain
\begin{align*}
\sum_{k=1}^{\infty}w_{k,q,\ell}d_{k,q}=\psi_{\ell}(0) - b_{0,q}(\psi_\ell)=\tilde{\psi}_\ell(0).
\end{align*}
\end{proof}

\begin{proof}[Proof of Corollary \ref{cor:omnibus}.]
Since $b_{k,q}(\psi_{\ell})>0$ for $k \geq 1$, $q\geq1$, $\kappa>0$, $0<\rho<1$, and $\ell=1,2$, the proof follows from \cite{Gine1975-SM} and \cite{Prentice1978-SM}; see, e.g., Theorem 3.1 in \cite{Garcia-Portugues2020b-SM}.
\end{proof}

\begin{proof}[Proof of Proposition \ref{prp:exp_alt}.]
For the sake of simplicity, we denote $b_{k,q}\equiv b_{k,q}(\psi_\ell)$. Following the Gegenbauer expansion of $g$, we can write, for almost every $\bx\in\Sq$,
\begin{align}
    f_1(\bx) = 
    \begin{cases}
        \sum_{k=0}^{\infty} e_{k,1}T_{k}(\bx'\bmu), & q=1,\\
        \sum_{k=0}^{\infty} e_{k,q}C_{k}^{(q-1)/2}(\bx'\bmu), & q\geq2.
    \end{cases}\label{eq:alt-pdf}
\end{align}

First, we derive the expectation of $\psi_\ell$ under $\mathcal{H}_1$ for $q\geq2$:
\begin{align}
    \mathbb{E}_{\Hcal_1}[\psi_{\ell}(\theta_{12})]=&\;\int_{\Sq}\int_{\Sq} \psi_{\ell}(\cos^{-1}(\bx'\by))\,f_1(\bx) \,f_1(\by)\,\rd\sigma_q(\bx) \,\rd\sigma_q(\by)\nonumber\\   =&\;\int_{\Sq}\left[\int_{\Sq}\sum_{k=0}^{\infty} b_{k,q} \,C_k^{(q-1)/2}(\bx'\by) \sum_{m=0}^{\infty} e_{m,q} \,C_m^{(q-1)/2}(\bx'\bmu)\,\rd\sigma_q(\bx)\right]\nonumber\\
    &\times f_1(\by) \,\rd\sigma_q(\by)\nonumber\\
    =&\;\sum_{k,m=0}^{\infty} b_{k,q} \,e_{m,q} \int_{\Sq} \omega_q \left[ \int_{\Sq} C_k^{(q-1)/2}(\bx'\by) \,C_m^{(q-1)/2}(\bx'\bmu)\,\rd\nu_q(\bx)\right]\nonumber\\
    &\times f_1(\by) \,\rd\sigma_q(\by).\label{eq:prp-exp-psi-alt-qgeq2}
\end{align}
Applying Lemma \ref{lem:LB7new} results in
\begin{align}
    \eqref{eq:prp-exp-psi-alt-qgeq2}=&\;\sum_{k=0}^{\infty} b_{k,q} \,e_{k,q} \,\omega_q \left(1+\frac{2k}{q-1}\right)^{-1} \int_{\Sq} C_k^{(q-1)/2}(\by'\bmu) \,f_1(\by) \,\rd\sigma_q(\by)\nonumber\\
    =&\;\sum_{k=0}^{\infty} b_{k,q} \,e_{k,q}^2 \,\omega_q^2 \left(1+\frac{2k}{q-1}\right)^{-1} \int_{\Sq} C_k^{(q-1)/2}(\by'\bmu) \,C_k^{(q-1)/2}(\by'\bmu) \,\rd\nu_q(\by)\nonumber\\
    =&\;\sum_{k=0}^{\infty} b_{k,q} \,e_{k,q}^2 \,\omega_q^2 \left(1+\frac{2k}{q-1}\right)^{-2} \frac{(q-1)_k}{k!},\label{eq:exp-psi-alt-qgeq2}
\end{align}
where we used that $C_k^{(q-1)/2}(1)=(q-1)_k/k!$ (Table 18.6.1 in \cite{NIST:DLMF-SM}).

The case $q=1$ is proved following the same arguments, but using Chebyshev polynomials and the circular version of Lemma \ref{lem:LB7new} twice:
\begin{align}
    \mathbb{E}_{\Hcal_1}[\psi_{\ell}(\theta_{12})]=&\; \sum_{k=0}^{\infty} \frac{\omega_1^2}{4} \,(1+\delta_{k,0})^2 \,b_{k,1} \,e_{k,1}^2 \,T_{k}(1) = \sum_{k=0}^{\infty} \frac{\omega_1^2}{4} \,(1+\delta_{k,0})^2 \,b_{k,1} \,e_{k,1}^2.\label{eq:exp-psi-alt-qeq1}
\end{align}

Since $f_1$ is a pdf, considering a tangent-normal change of variables gives
\begin{align*}
    1=\int_{\mathbb{S}^q} f_1(\bx) \,\rd\sigma_q(\bx) = \omega_{q-1}\int_{-1}^{1} g(x) (1-x^2)^{q/2-1} \,\rd x=\omega_{q-1}c_{0,1}e_{0,q}
\end{align*}
and thus $e_{0,q}=1/\omega_{q}$ due to \eqref{eq:dkq}. Then, to derive the expectation of the statistic, we substitute \eqref{eq:b0H0}, \eqref{eq:exp-psi-alt-qgeq2}, and \eqref{eq:exp-psi-alt-qeq1} into \eqref{eq:stat_def}, resulting in
\begin{align*}
    \mathbb{E}_{\Hcal_1}[T_{n, \ell}] =&\; (n-1)\left(\mathbb{E}_{\Hcal_1}\left[\psi_{\ell}(\theta_{12})\right] - \mathbb{E}_{\Hcal_0}[\psi_{\ell}(\theta_{12})]\right)\\
    =&\; \begin{cases}
        \omega_1^2\,(n-1)/4 \sum_{k=1}^{\infty} \,b_{k,1} \,e_{k,1}^2, & q=1,\\
        \omega_q^2(n-1) \sum_{k=1}^{\infty} b_{k, q} \,e_{k, q}^2 \left(1 + \frac{2k}{q-1}\right)^{-2} C_k^{(q-1)/2}(1), & q\geq2.\\
    \end{cases}
\end{align*}
Using the coefficients $\tau_{k,q}$ defined in Proposition \ref{prp:exp_alt} concludes the proof.
\end{proof}

\begin{proof}[Proof of Proposition \ref{prp:var_alt}.]
The variance of a $U$-statistic is given by \eqref{eq:var-u-stat}, where 
\begin{align*}
    \eta_{1, \ell} &= \mathbb{V}\mathrm{ar}_{\mathcal{H}_1}[\mathbb{E}_{\mathcal{H}_1}[\tilde{\psi}_\ell(\theta_{12})\mid\bX_1]] \\
    &=\mathbb{E}_{\mathcal{H}_1}[\mathbb{E}_{\mathcal{H}_1}[\psi_\ell(\theta_{12})\mid\bX_1]^2]-\mathbb{E}_{\mathcal{H}_1}[\psi_\ell(\theta_{12})]^2\\
    &=:\alpha_{1,\ell} - \beta_{1,\ell}^2
\end{align*}
and similarly $\eta_{2, \ell} = \mathbb{V}\mathrm{ar}_{\mathcal{H}_1}[{\psi}_\ell(\theta_{12})] =: \alpha_{2,\ell} - \beta_{2,\ell}^2$. The terms $\alpha_{i,\ell}$ and $\beta_{i,\ell}$ can be derived through the Gegenbauer coefficients of $\psi_\ell$, denoted by $b_{k,q}$, and the expansion of $f_1$ in \eqref{eq:alt-pdf}.

Applying Lemma \ref{lem:LB7new}, we first determine the following conditional expectation:
\begin{align*}
    \mathbb{E}_{\mathcal{H}_1}[\psi_\ell(\theta_{12})\mid\bX_1] =&\; \int_{\mathbb{S}^q} \psi_{\ell}(\bx'\bX_1)f_1(\bx)\,\rd\sigma_q(\bx)\nonumber \\
    =&\;\begin{cases}
        (\omega_1/2)\sum_{k=0}^{\infty} \,(1+\delta_{k,0}) \,b_{k,1} \,e_{k,1} \,T_k(\bX_1'\bmu), & q=1,\\
        \omega_q \sum_{k=0}^{\infty} b_{k,q} \,e_{k,q} \left(1+\frac{2k}{q-1}\right)^{-1} \,C_{k}^{(q-1)/2}(\bX_1'\bmu), & q\geq2.
    \end{cases}
\end{align*}

Consider the tangent-normal change of variables $\bx=x\bmu+(1-x^2)^{1/2}\bB_{\bmu}\bxi$ and the expansion \eqref{eq:alt-pdf} for $f_1$. Then, for $q\geq2$,
\begin{align*}
    \alpha_{1,\ell}=&\; \mathbb{E}_{\mathcal{H}_1}[(\mathbb{E}_{\mathcal{H}_1}[\psi_\ell(\theta_{12})\mid\bX_1])^2] \\
    =&\; \omega_q^2 \int_{\mathbb{S}^q} \sum_{k_1, k_2=0}^{\infty} \,b_{k_1,q} \,b_{k_2,q} \,e_{k_1, q} \,e_{k_2, q} \left(1+\frac{2k_1}{q-1}\right)^{-1} \left(1+\frac{2k_2}{q-1}\right)^{-1} \\
    &\; \times \,C_{k_1}^{(q-1)/2}(\bx'\bmu) \,C_{k_2}^{(q-1)/2}(\bx'\bmu) \,f_1(\bx)\,\rd\sigma_q(\bx) \\
    =&\; \omega_q^2 \sum_{k_1, k_2, k_3=0}^{\infty} \,b_{k_1,q} \,b_{k_2,q} \,e_{k_1, q} \,e_{k_2, q} \,e_{k_3, q} \left(1+\frac{2k_1}{q-1}\right)^{-1} \left(1+\frac{2k_2}{q-1}\right)^{-1} \\
    &\; \times \int_{\mathbb{S}^{q-1}}\int_{-1}^{1} \,C_{k_1}^{(q-1)/2}(x) \,C_{k_2}^{(q-1)/2}(x) \,C_{k_3}^{(q-1)/2}(x) (1-x^2)^{q/2-1} \,\rd\sigma_{q-1}(\bxi) \,\rd x \\
    =&\; \omega_q^2 \, \omega_{q-1} \sum_{k_1,k_2,k_3=0}^{\infty} \,b_{k_1,q} \,b_{k_2,q} \,e_{k_1, q} \,e_{k_2, q} \,e_{k_3, q} \left(1 + \frac{2k_1}{q-1}\right)^{-1} \left(1 + \frac{2k_2}{q-1}\right)^{-1} \\
    &\; \times\,t_{k_1,k_2,k_3;q},
\end{align*}
where $t_{k_1,k_2,k_3;q}$ refers to the triple product of Gegenbauer polynomials defined in Lemma \ref{lem:triplet} below. Now, 
\begin{align*}
    \beta_{1,\ell}=&\;\mathbb{E}_{\mathcal{H}_1}[\mathbb{E}_{\mathcal{H}_1}[\psi_\ell(\theta_{12})\mid\bX_1]] \\
    =&\; \omega_q \sum_{k,m=0}^{\infty} b_{k,q} \,e_{k,q} \,e_{m,q} \left(1+\frac{2k}{q-1}\right)^{-1} \int_{\mathbb{S}^q} C_{k}^{(q-1)/2}(\bx'\bmu)  \,C_{m}^{(q-1)/2}(\bx'\bmu)\,\rd\sigma_q(\bx) \\
    =&\; \omega_q^2 \sum_{k=0}^{\infty} b_{k,q} \,e_{k,q}^2 \left(1+\frac{2k}{q-1}\right)^{-2} \,C_{k}^{(q-1)/2}(1).
\end{align*}

For $q=1$, $\bxi\in\mathbb{S}^0=\{-1,1\}$ in the tangent-normal change of variables, hence
\begin{align*}
    \alpha_{1,\ell}=&\; \mathbb{E}_{\mathcal{H}_1}[(\mathbb{E}_{\mathcal{H}_1}[\psi_\ell(\theta_{12})\mid\bX_1])^2] \\
    =&\; \frac{\omega_1^2}{2} \sum_{k_1, k_2, k_3=0}^{\infty} \,(1+\delta_{k_1,0}) \,(1+\delta_{k_2,0}) \,b_{k_1,1} \,b_{k_2,1} \,e_{k_1, 1} \,e_{k_2, 1} \,e_{k_3,1} \,t_{k_1,k_2,k_3;1},
\end{align*}
where $t_{k_1,k_2,k_3;1}$ is defined in Lemma \ref{lem:triplet}, and
\begin{align*}
    \beta_{1,\ell}=&\; \frac{\omega_1^2}{4}\sum_{k=0}^{\infty} \,(1+\delta_{k,0})^2 \,b_{k,1} \,e_{k,1}^2.
\end{align*}

Finally, $\alpha_{2, \ell}$ follows from \eqref{eq:exp-psi-alt-qgeq2} ($q\geq2$) and \eqref{eq:exp-psi-alt-qeq1} ($q=1$),
\begin{align*}
    \alpha_{2, \ell}=&\;\mathbb{E}_{\mathcal{H}_1}[\psi^2_\ell(\theta_{12})] 
    =\begin{cases}
        (\omega_1^2/4) \sum_{k=0}^{\infty}  \,(1+\delta_{k,0})^2 \,b_{k,1}(\psi_\ell^2) \,e_{k,1}^2,\!\!\!\! & q=1,\\
        \omega_q^2 \sum_{k=0}^{\infty} b_{k,q}(\psi_\ell^2) \,e_{k,q}^2 \left(1+\frac{2k}{q-1}\right)^{-2}C_k^{(q-1)/2}(1),\!\!\!\! & q\geq2,
    \end{cases}
\end{align*}
while $\beta_{2, \ell}=\mathbb{E}_{\mathcal{H}_1}[{\psi}_\ell(\theta_{12})]$ is analogous.
\end{proof}

\begin{proof}[Proof of Proposition \ref{prp:rejection_level}.]
We have that
\begin{align*}
    \lim_{n\to\infty}&\mathbb{P}_{\mathcal{H}_0}[T_{\vert \mathcal{S}_2 \vert,\ell}(\widehat{\lambda}(\mathcal{S}_1), \mathcal{S}_2)\leq c_\alpha(\widehat{\lambda}(\mathcal{S}_1))]\\
    &= \lim_{n\to\infty}\iint_{\{(\mathbf{x},\lambda)\in(\mathbb{S}^q)^{\vert\mathcal{S}_2\vert}\times\Lambda:T_{\vert \mathcal{S}_2 \vert,\ell}(\mathbf{x},\lambda)\leq c_\alpha(\lambda)\}}\,f_{\mathcal{S}_2}(\mathbf{x})\,\mathrm{d}\mathbf{x}\,\mathrm{d}F_{\widehat{\lambda}(\mathcal{S}_1)}(\lambda)\\
    &= \lim_{n\to\infty}\int_\Lambda \bigg[\int_{\{\mathbf{x}\in(\mathbb{S}^q)^{\vert\mathcal{S}_2\vert}:T_{\vert \mathcal{S}_2 \vert,\ell}(\mathbf{x},\lambda)\leq c_\alpha(\lambda)\}}\,f_{\mathcal{S}_2}(\mathbf{x})\,\mathrm{d}\mathbf{x}\bigg]\,\mathrm{d}F_{\widehat{\lambda}(\mathcal{S}_1)}(\lambda)\\
    &= \lim_{n\to\infty}\int_\Lambda \mathbb{P}_{\mathcal{H}_0}[T_{\vert\mathcal{S}_2\vert,\ell}(\lambda)\leq c_\alpha(\lambda)]\,\mathrm{d}F_{\widehat{\lambda}(\mathcal{S}_1)}(\lambda),
\end{align*}
where $F_{\widehat{\lambda}(\mathcal{S}_1)}$ is the cdf of the random variable $\widehat{\lambda}(\mathcal{S}_1)$.

Since $\Lambda$ is a finite set $\{\lambda_1,\ldots,\lambda_m\}$ by assumption, interchangeability of the limit and integral is trivial:
\begin{align}
    \lim_{n\to\infty}&\mathbb{P}_{\mathcal{H}_0}[T_{\vert \mathcal{S}_2 \vert,\ell}(\widehat{\lambda}(\mathcal{S}_1), \mathcal{S}_2)\leq c_\alpha(\widehat{\lambda}(\mathcal{S}_1))]\nonumber\\
    &=\sum_{j=1}^m \lim_{n\to\infty}\mathbb{P}_{\mathcal{H}_0}[T_{\vert\mathcal{S}_2\vert,\ell}(\lambda_j)\leq c_\alpha(\lambda_j)] p_{\widehat{\lambda}(\mathcal{S}_1)}(\lambda_j)\nonumber\\
    &=\alpha+\sum_{j=1}^m \lim_{n\to\infty}\{\mathbb{P}_{\mathcal{H}_0}[T_{\vert\mathcal{S}_2\vert,\ell}(\lambda_j)\leq c_\alpha(\lambda_j)]-\alpha\} p_{\widehat{\lambda}(\mathcal{S}_1)}(\lambda_j), \label{eq:limnalpha}
\end{align}
where $p_{\widehat{\lambda}(\mathcal{S}_1)}$ is the probability mass function of $\widehat{\lambda}(\mathcal{S}_1)$ and $\lim_{n\to\infty}\mathbb{P}_{\mathcal{H}_0}[T_{\vert\mathcal{S}_2\vert,\ell}(\lambda)\leq c_\alpha(\lambda)]=\alpha$ for all $\lambda\in\Lambda$. From these arguments, $\lim_{n\to\infty}\vert\mathbb{P}_{\mathcal{H}_0}[T_{\vert\mathcal{S}_2\vert,\ell}(\lambda_j)\leq c_\alpha(\lambda_j)]-\alpha\vert p_{\widehat{\lambda}(\mathcal{S}_1)}(\lambda_j)\leq \lim_{n\to\infty}\vert\mathbb{P}_{\mathcal{H}_0}[T_{\vert\mathcal{S}_2\vert,\ell}(\lambda_j)\leq c_\alpha(\lambda_j)]-\alpha\vert=0$, hence the limit in \eqref{eq:limnalpha} is zero.
\end{proof}

\begin{remark}
If $\Lambda$ is a compact or continuous set, it seems necessary from the proof to control the null asymptotic distribution of $\widehat{\lambda}(\mathcal{S}_1)$. This can be done by an application of the argmax theorem \cite[Section 3.2.1]{Vaart1996}, which would require establishing the weak convergence of the degenerate $U$-process $\lambda\mapsto T_{n,\ell}(\lambda)$ \cite[see][Section 5]{Arcones1993}.
\end{remark}

Lemma \ref{lem:LB7new} below is a minor extension of Lemma B.7 in \citet[page 23 of the supplement]{Garcia-Portugues2020b-SM} that benefits the paper completeness. It covers the cases $k=0$ and $m=0$, as needed in the proof of Propositions \ref{prp:exp_alt} and \ref{prp:var_alt}.

\begin{lemma}[Double products of Gegenbauer and Chebyshev polynomials] \label{lem:LB7new}
Let $\bu,\bv\in\Sq$, $q\geq1$. Then, for $k,m\geq 0$ and $q\geq2$,
\begin{align*}
    \int_{\mathbb{S}^1} T_k(\boldsymbol{\gamma}'\bu)\,T_m(\boldsymbol{\gamma}'\bv)\,\rd\nu_1(\boldsymbol{\gamma}) & = \frac{1 + \delta_{k,0}}{2} T_{k}(\bu'\bv)\delta_{k,m},\\
    \int_{\Sq} C_{k}^{(q-1) / 2}\left(\boldsymbol{\gamma}' \mathbf{u}\right) C_{m}^{(q-1) / 2}\left(\boldsymbol{\gamma}' \mathbf{v}\right) \,\rd\nu_q(\boldsymbol{\gamma}) & =\left(1+\frac{2 k}{q-1}\right)^{-1} C_{k}^{(q-1) / 2}\left(\mathbf{u}' \mathbf{v}\right) \delta_{k, m}.
\end{align*}
\end{lemma}

\begin{proof}[Proof of Lemma \ref{lem:LB7new}.]
The case $k,m\geq 1$ corresponds to Lemma B.7 in \citet[page 23 of the supplement]{Garcia-Portugues2020b-SM}. For $k=0$, since $T_0\equiv 1$ and $C_0^{(q-1)/2}\equiv 1$, the orthogonality of the polynomials (when $m\neq0$) or the integration with respect to the uniform measure $\nu_q$ (when $m=0$) proves the result.
\end{proof}

The final lemma allows the computation of triple products of Gegenbauer and Chebyshev polynomials involved in the proof of Proposition~\ref{prp:var_alt}.

\begin{lemma}[Triple products of Gegenbauer and Chebyshev polynomials] \label{lem:triplet}
Let $k_3\geq k_2\geq k_1\geq 0$. Then, for $q\geq2$,
\begin{align*}
	t_{k_1,k_2,k_3;q}:=&\;\int_{-1}^{1}C_{k_1}^{(q-1)/2}(x)C_{k_2}^{(q-1)/2}(x)C_{k_3}^{(q-1)/2}(x)(1-x^2)^{q/2-1}\,\mathrm{d}x\\
	=&\;\begin{cases}
		0,&\!\!\!\!\text{if }s_k\text{ is odd or }k_3>k_1+k_2,\\
		a_{(k_1+k_2-k_3)/2,k_1,k_2,(q-1)/2}\,c_{k_3,q},&\!\!\!\!\text{if }s_k\text{ is even and }k_3\leq k_1+k_2,
	\end{cases}
\end{align*}
where $s_k:=\sum_{i=1}^3k_i$ and
\begin{align*}
	a_{\ell,k_1,k_2,\lambda}:=\frac{k_1+k_2+\lambda-2\ell}{\ell!(k_1+k_2+\lambda-\ell)}\binom{k_1+k_2-2\ell}{k_2-\ell} \frac{\Gamma(\lambda+\ell)\mathrm{B}(\lambda+k_1-\ell,\lambda+k_2-\ell)}{\Gamma(\lambda)\mathrm{B}(\lambda+k_1+k_2-\ell,\lambda)}.
\end{align*}
In addition,
\begin{align*}
	t_{k_1,k_2,k_3;1}:=&\;\int_{-1}^{1}T_{k_1}(x)T_{k_2}(x)T_{k_3}(x)(1-x^2)^{-1/2}\,\mathrm{d}x\\
    =&\;\begin{cases}
		0,&\text{if }k_3\neq k_2\pm k_1,\\
		\frac{1+\delta_{k_1,0}}{2}c_{k_3,1},&\text{if }k_3=k_2\pm k_1.
	\end{cases}
\end{align*}
\end{lemma}

\begin{proof}[Proof of Lemma \ref{lem:triplet}.]
If $k_1+k_2<k_3$, the $(k_1+k_2)$th order polynomial $x\mapsto C_{k_1}^{(q-1)/2}(x)C_{k_2}^{(q-1)/2}(x)$ can be expressed in terms of $\{C_k^{(q-1)/2}\}_{k=0}^{k_1+k_2}$, and therefore is orthogonal to $C_{k_3}^{(q-1)/2}$. Precisely, the linearization formula for the product of Gegenbauer polynomials \citep[Equation 18.18.22]{NIST:DLMF-SM} gives, for $k_2\geq k_1\geq 0$ and $\lambda>0$,
\begin{align*}
	C^{(\lambda)}_{k_1}(x)C^{(\lambda)}_{k_2}(x)=\sum_{\ell=0}^{\min(k_1,k_2)}a_{\ell,k_1,k_2,\lambda}C^{(\lambda)}_{k_1+k_2-2\ell}(x),
\end{align*}
with the positive coefficients
\begin{align*}
	a_{\ell,k_1,k_2,\lambda}
	=&\;\frac{(k_1+k_2+\lambda-2\ell)(k_1+k_2-2\ell)!}{(k_1+k_2+\lambda-\ell)\ell!(k_1- \ell)!(k_2-\ell)!}\frac{(\lambda)_{\ell}(\lambda)_{k_1-\ell}(\lambda)_{k_2-\ell}(2\lambda)_{k_1+k_2-\ell}}{(\lambda)_{k_1+k_2-\ell}(2\lambda)_{k_1+k_2-2\ell}}\\
	=&\;\frac{k_1+k_2+\lambda-2\ell}{\ell!(k_1+k_2+\lambda-\ell)}\binom{k_1+k_2-2\ell}{k_2-\ell} \frac{\Gamma(\lambda+\ell)\mathrm{B}(\lambda+k_1-\ell,\lambda+k_2-\ell)}{\Gamma(\lambda)\mathrm{B}(\lambda+k_1+k_2-\ell,\lambda)}.
\end{align*}

Taking $\lambda=(q-1)/2$ and $k_3\geq k_2\geq k_1\geq 0$ gives
\begin{align*}
	C^{(q-1)/2}_{k_1}(x)C^{(q-1)/2}_{k_2}(x)=\sum_{\ell=0}^{k_1}a_{\ell,k_1,k_2,(q-1)/2}C^{(q-1)/2}_{k_1+k_2-2\ell}(x).
\end{align*}

Hence, it is clear that the integral is null unless $\ell$ equals $(k_1+k_2-k_3)/2$:
\begin{align*}
	\int_{-1}^{1}C_{k_1}^{(q-1)/2}&(x)C_{k_2}^{(q-1)/2}(x)C_{k_3}^{(q-1)/2}(x)(1-x^2)^{q/2-1}\,\mathrm{d}x\\
	=&\;a_{(k_1+k_2-k_3)/2,k_1,k_2,(q-1)/2}\int_{-1}^{1}\big[C_{k_3}^{(q-1)/2}(x)\big]^2(1-x^2)^{q/2-1}\,\mathrm{d}x\\
	=&\;a_{(k_1+k_2-k_3)/2,k_1,k_2,(q-1)/2}\,c_{k_3,q}.
\end{align*}

Analogous derivations are possible for Chebyshev polynomials. From Equation 18.18.21 in \cite{NIST:DLMF-SM}, for $k_2\geq k_1\geq 0$,
\begin{align*}
	T_{k_1}(x)T_{k_2}(x)=\frac{1}{2}(T_{k_2+k_1}(x)+T_{k_2-k_1}(x))
\end{align*}
and, consequently,
\begin{align*}
	\int_{-1}^{1}T_{k_1}(x)T_{k_2}(x)T_{k_3}(x)(1-x^2)^{-1/2}\,\mathrm{d}x
	=&\;\delta_{k_3,k_2\pm k_1}\frac{1+\delta_{k_1,0}}{2}c_{k_3,1}.
\end{align*}
\end{proof}

\section{Additional numerical experiments}
\label{app:morenum}

\subsection{Simulation scenarios}
\label{sec:oracle-sim}

To perform the search for the optimal value $\tilde{\lambda}_{\mathcal{H}_1}$ motivated in Section \ref{sec:oracle}, $q_{n,\ell, \mathcal{H}_1}(\lambda)$ must be determined for each alternative distribution. $\mathbb{V}\mathrm{ar}_{\mathcal{H}_0}[T_{n,\ell}(\lambda)]$ does not depend on the alternative and is given in \eqref{eq:expvar_null}. However, $\mathbb{E}_{\mathcal{H}_1}[T_{n,\ell}(\lambda)]$ does not have a general analytic expression for any $\mathcal{H}_1$. Rotationally symmetric alternatives allow us to directly compute $q_{n,\ell, \mathcal{H}_1}(\lambda)$, since $\mathbb{E}_{\mathcal{H}_1}[T_{n,\ell}(\lambda)]$ is readily available in \eqref{eq:exp_alt}. 

We consider rotationally symmetric distributions on $\mathbb{S}^q$ with pdfs
\begin{align}
    \bx \mapsto \frac{c_{q, \kappa_{\mathrm{dev}}, f}}{\om{q-1}} f(\bx'\btheta; \kappa_{\mathrm{dev}}),\label{eq:kdev1}
\end{align}
where $\btheta\in\mathbb{S}^q$ is a location parameter, which is set without loss of generality to $\btheta = (0, \ldots, 0, 1)'$, $\kappa_{\mathrm{dev}}>0$ is a concentration parameter, $f:[-1,1]\times\mathbb{R}_0^+\rightarrow\mathbb{R}_0^{+}$ satisfies $f(t; 0) = 1$, and $c_{q, \kappa, f}^{-1}:= \int_{-1}^{1} (1-t^2)^{q/2-1} f(t; \kappa)\,\rd t$ is a normalizing constant. Note $\om{0}=2$ when $q=1$. We consider:
\begin{enumerate}[label=(\textit{\alph*})]
    \item The von Mises--Fisher (vMF) distribution, with $f(t; \kappa)=\exp(\kappa t)$.\label{vMF-dist-SM}
    \item A Cauchy-like (Ca) distribution, with $f(t; \kappa)=(1-\rho(\kappa)^2)/(1-2\rho(\kappa) t + \rho(\kappa)^2)^{(q+1)/2}$, where $\rho(\kappa):= ((2\kappa + 1)-\sqrt{4\kappa + 1})/(2\kappa)$.\label{C-dist-SM}
    \item The Watson (Wa) distribution, with $f(t; \kappa)=\exp(\kappa t^2)$.\label{W-dist-SM}
    \item The Small Circle (SC) distribution with $f(t; \kappa, \nu)=\exp(-\kappa(t-\nu)^2)$ and $\nu=0.25$ being the projection along $\btheta$ that controls the modal strip.\label{SC-dist-SM}
\end{enumerate}

In addition, we also consider mixtures consisting of $2(q+1)$ equally-weighted specific distributions with common $\kappa_{\mathrm{dev}}$ and pdfs of the form
\begin{align}
    \bx \mapsto \frac{c_{q, \kappa_{\mathrm{dev}}, f}}{2(q+1)\om{q-1}} \sum_{k=1}^{2(q+1)} f((-1)^r\bx'\be_r; \kappa_{\mathrm{dev}}),\quad r = \left\lceil k/2\right\rceil,\label{eq:kdev2}
\end{align}
with location parameters determined by the $r$th canonical vector $\be_r$. These mixtures are centro-symmetric about the origin. The distributions used to build the mixtures are:
\begin{enumerate}[label=(\textit{\alph*})]
\setcounter{enumi}{4}
    \item The vMF distribution (mixture denoted MvMF).\label{MvMF-dist-SM}
    \item The Cauchy-like distribution (denoted MCa).\label{MC-dist-SM}
\end{enumerate}
For these mixtures $\mathbb{E}_{\mathcal{H}_1}[T_{n,\ell}(\lambda)]$ has no analytic expression available. It is approximated by a Monte Carlo simulation with $10^4$ replicates.

Details on how to simulate the above distributions are given in \citet[Section 4]{Garcia-Portugues2020b}. The alternative \ref{C-dist-SM} was simulated using the inversion method.

\begin{figure}[htpb!]
\iffigstabs
    \centering
    \scalebox{0.9}{
        \subfloat[][]{
            \includegraphics[width=0.5\linewidth]{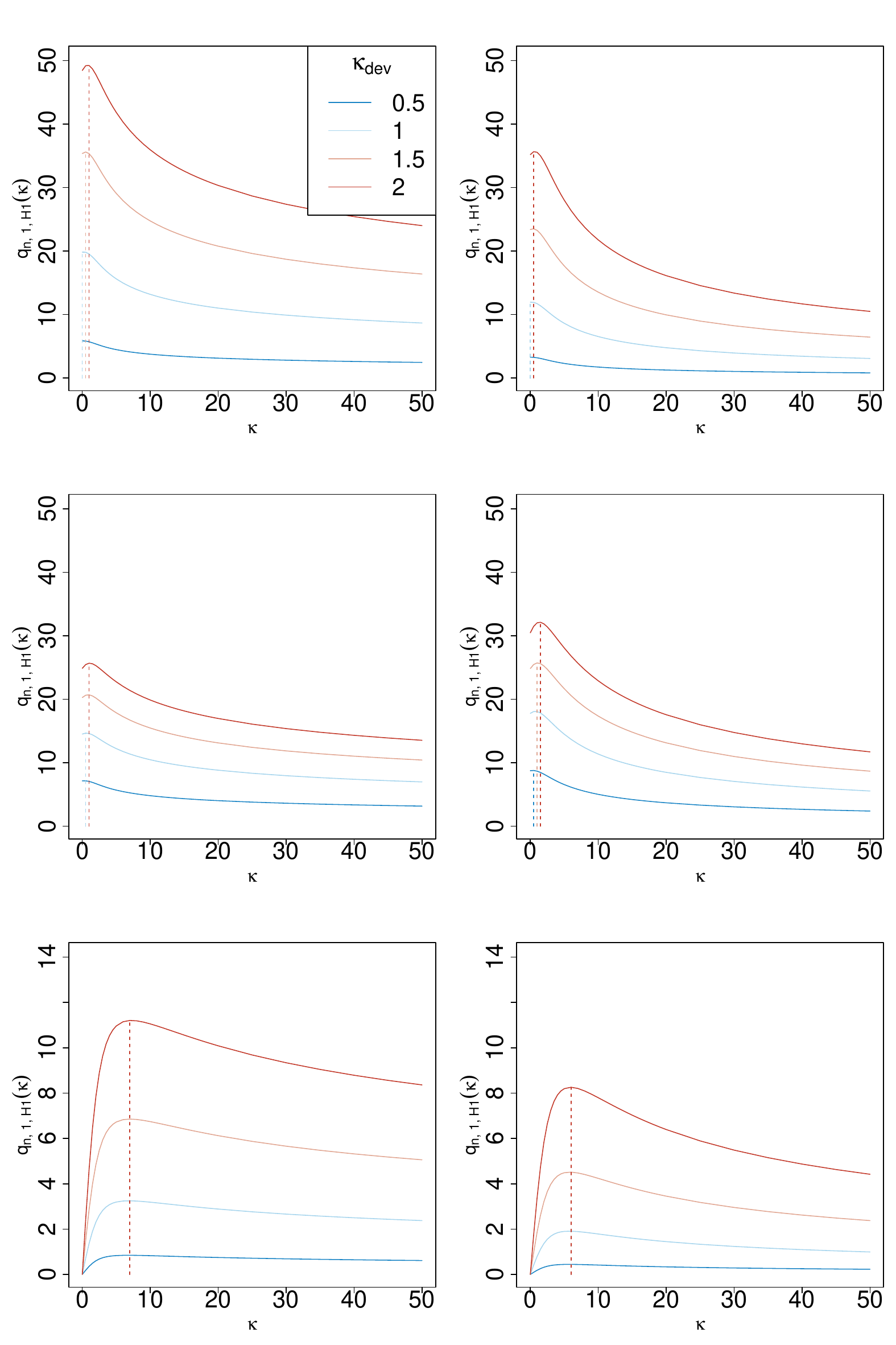}
            \vrule
            \includegraphics[width=0.5\linewidth]{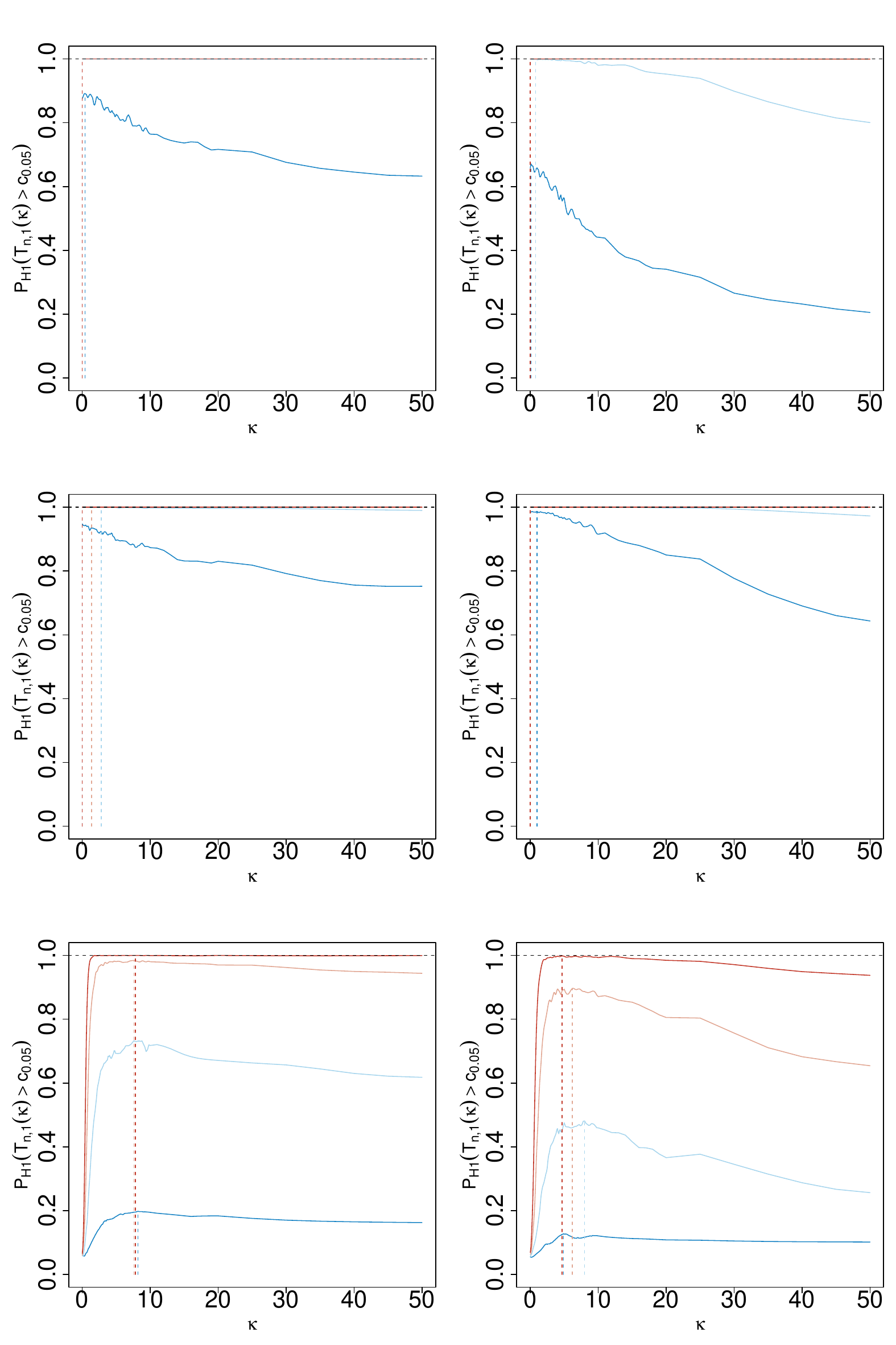}
        }
    }
    
    \scalebox{0.9}{
        \subfloat[][]{
    	\includegraphics[width=0.5\linewidth]{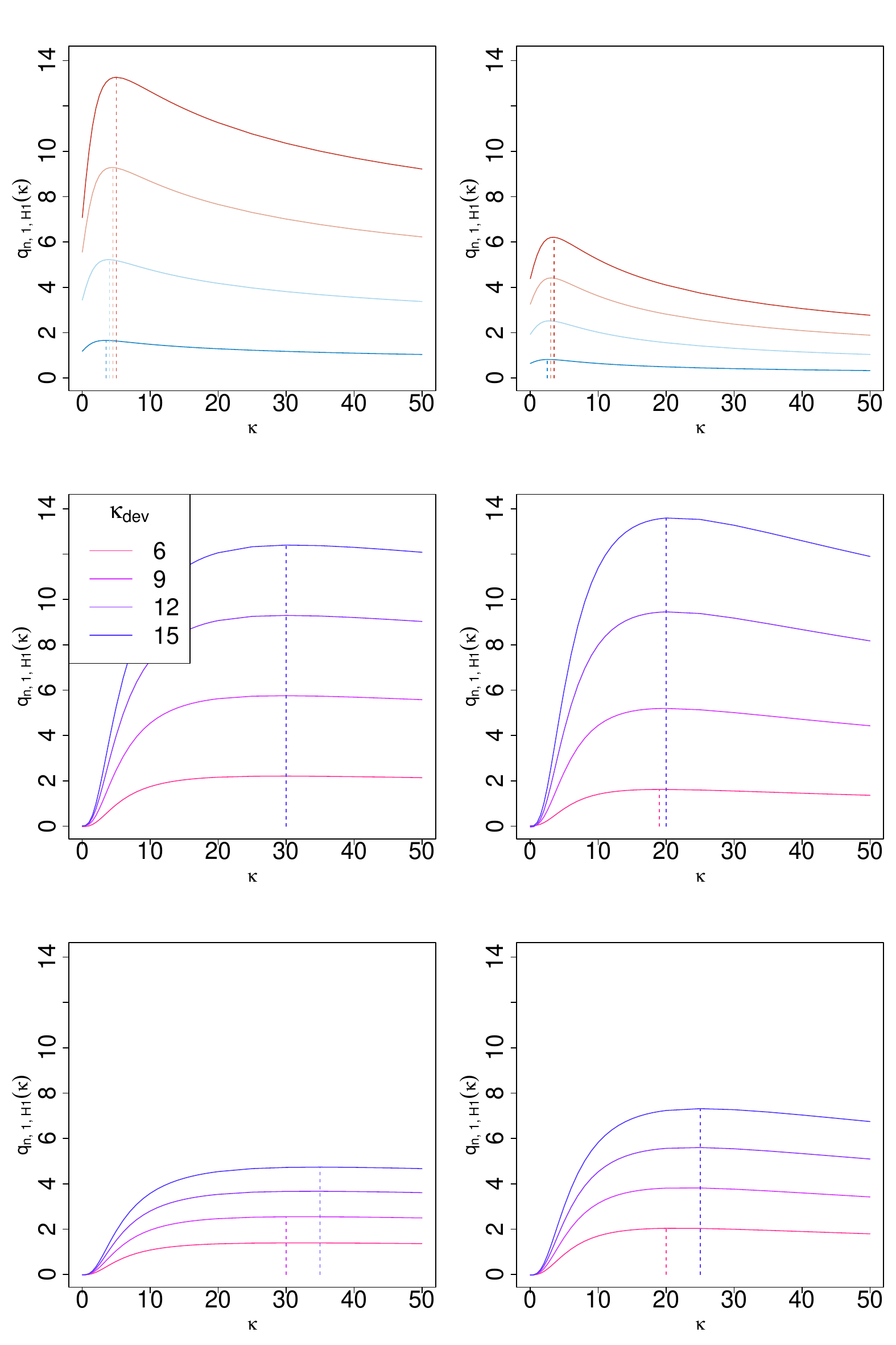}
            \vrule
    	\includegraphics[width=0.5\linewidth]{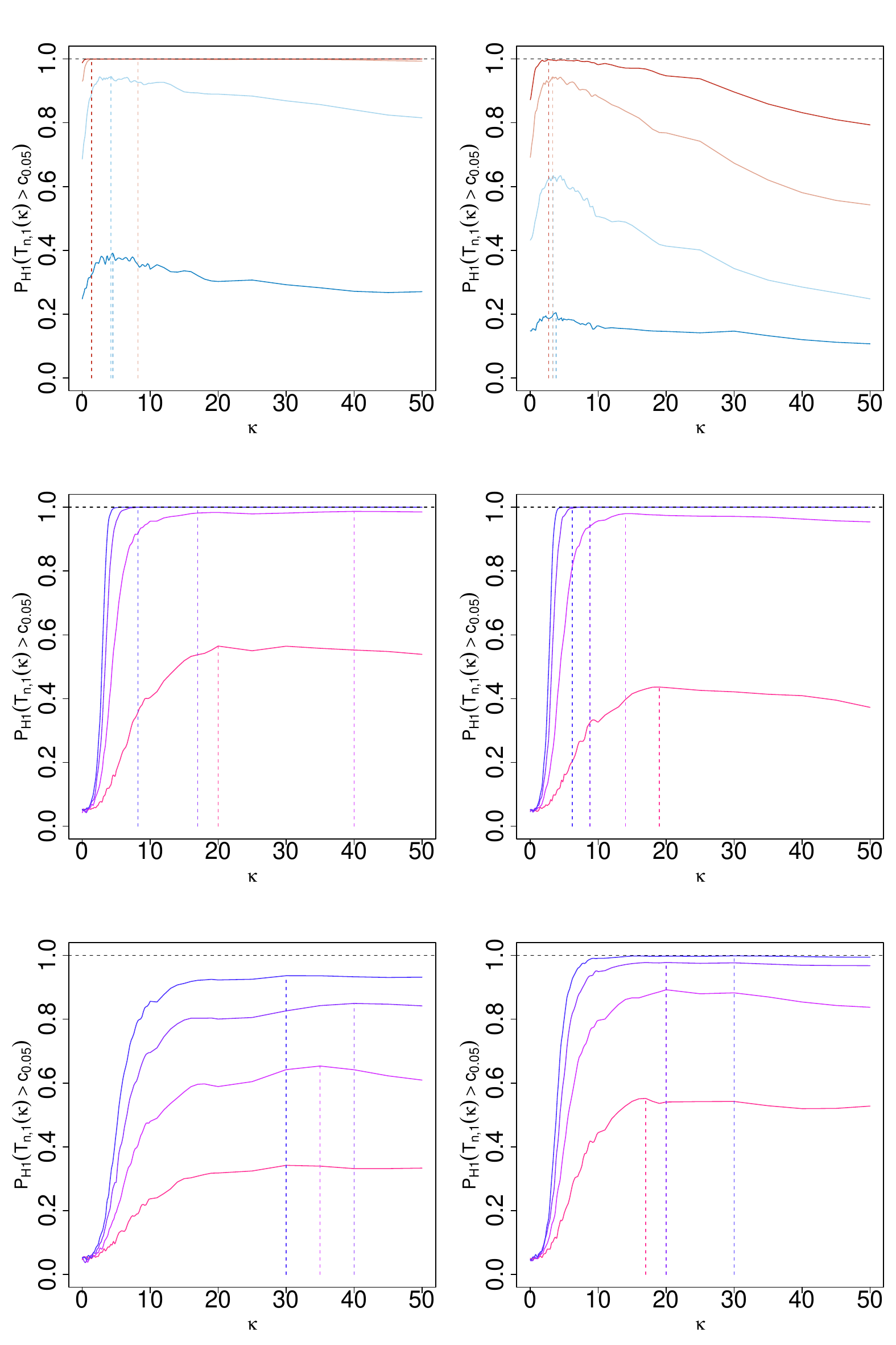}
        }
    }
\fi
	\caption{\small From top to bottom, rows correspond to rotationally symmetric alternatives \ref{vMF-dist-SM}--\ref{SC-dist-SM} and mixture alternatives \ref{MvMF-dist-SM}--\ref{MC-dist-SM}. Left/right columns in each block stand for $q=1,2$. Left block: curves $\kappa\mapsto q_{n,1, \mathcal{H}_1}(\kappa)$ and $\tilde{\kappa}_{\mathcal{H}_1}$ (vertical lines), under alternatives with varying $\kappa_{\mathrm{dev}}$ (acts in \eqref{eq:kdev1}--\eqref{eq:kdev2}). Right block: empirical smoothed power curves $\kappa\mapsto \mathbb{P}_{\mathcal{H}_1}[T_{n,1}(\kappa)>c_{\alpha, n, 1}(\kappa)]$ of the $T_{n, 1}(\kappa)$-tests at significance level $\alpha=0.05$, obtained from $M=10^3$ samples from the alternative with $\kappa_{\mathrm{dev}}$, and $\kappa^{\ast}_{\mathcal{H}_1}$ (vertical lines). The color palettes between the equal-dimension plots in the left and right blocks are shared. The sample sizes are $n=100$.} \label{fig:kappa-sens}
\end{figure}

\begin{figure}[htpb!]
\iffigstabs
    \centering
    \scalebox{0.9}{
        \subfloat[][]{
            \includegraphics[width=0.5\linewidth]{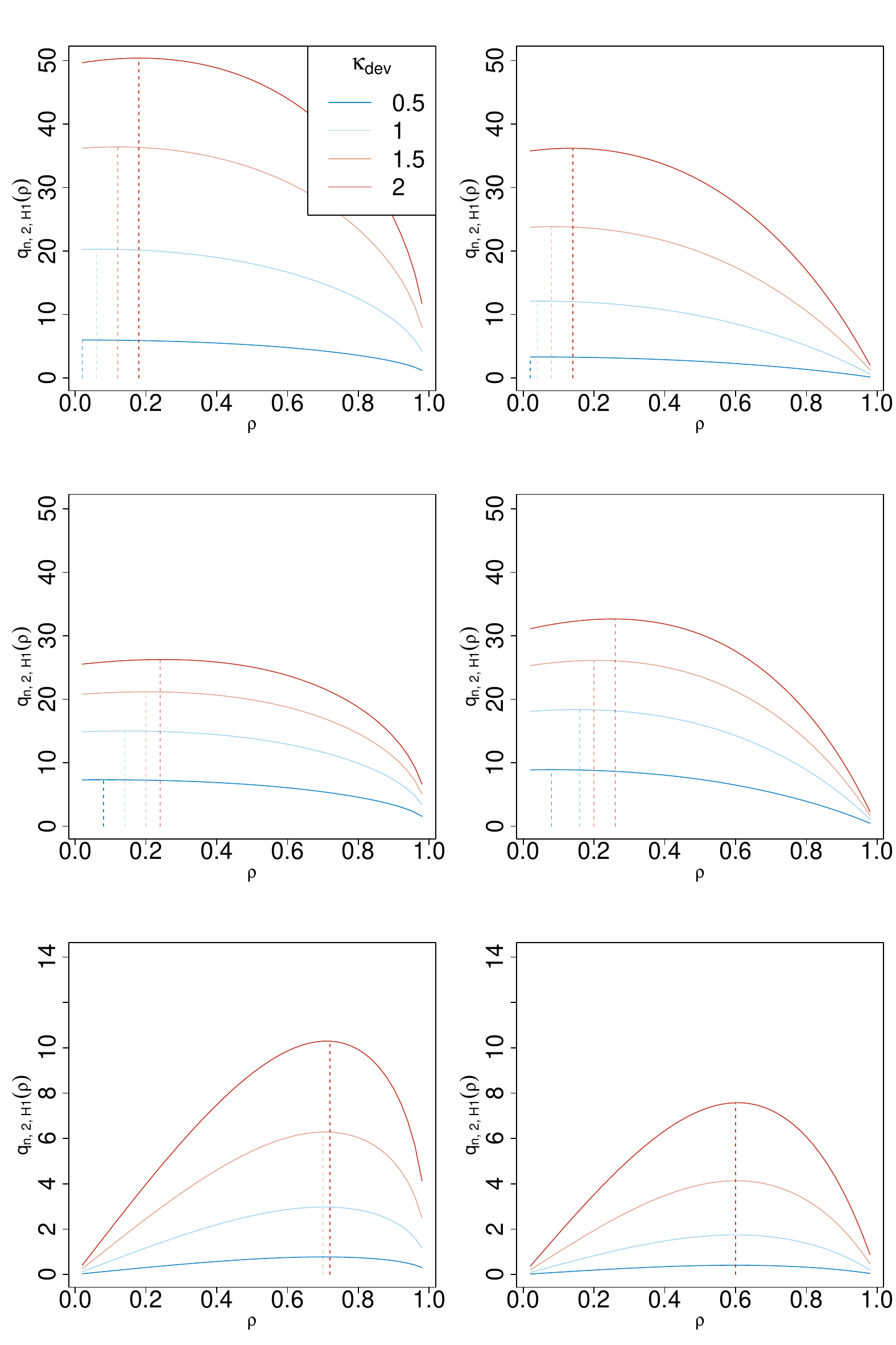}
            \vrule
            \includegraphics[width=0.5\linewidth]{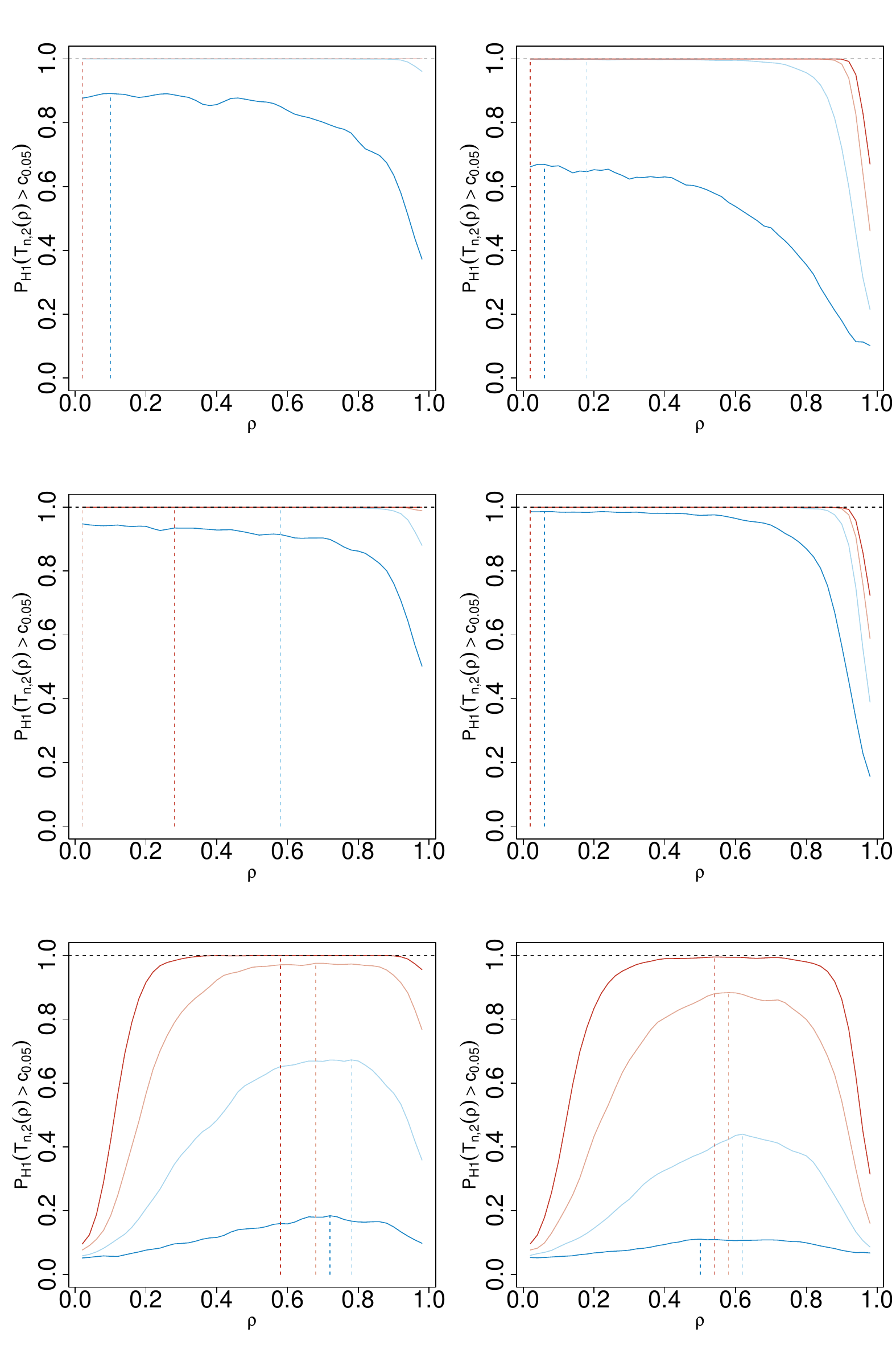}
        }
    }
    
    \scalebox{0.9}{
        \subfloat[][]{
    	\includegraphics[width=0.5\linewidth]{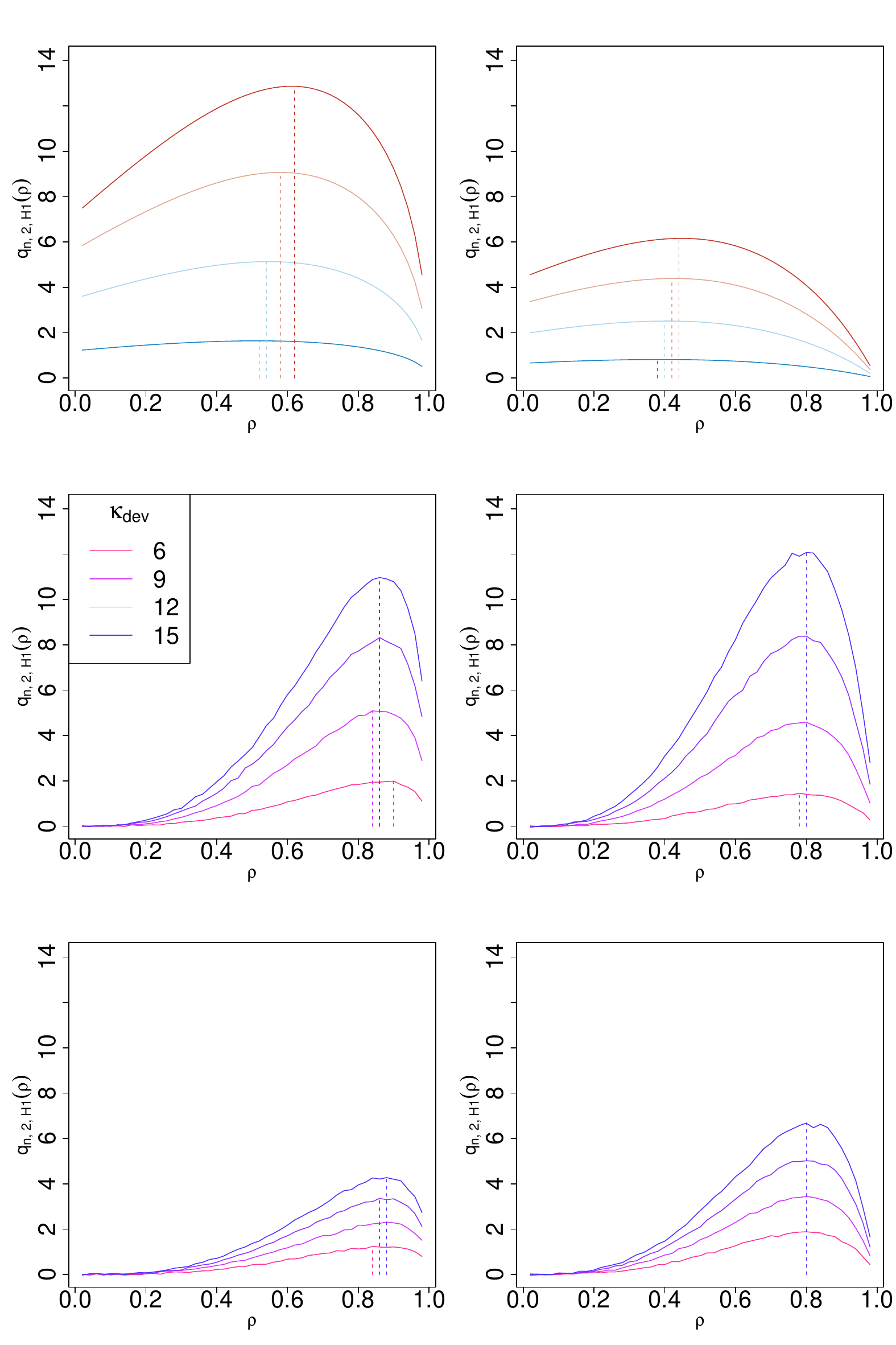}
            \vrule
    	\includegraphics[width=0.5\linewidth]{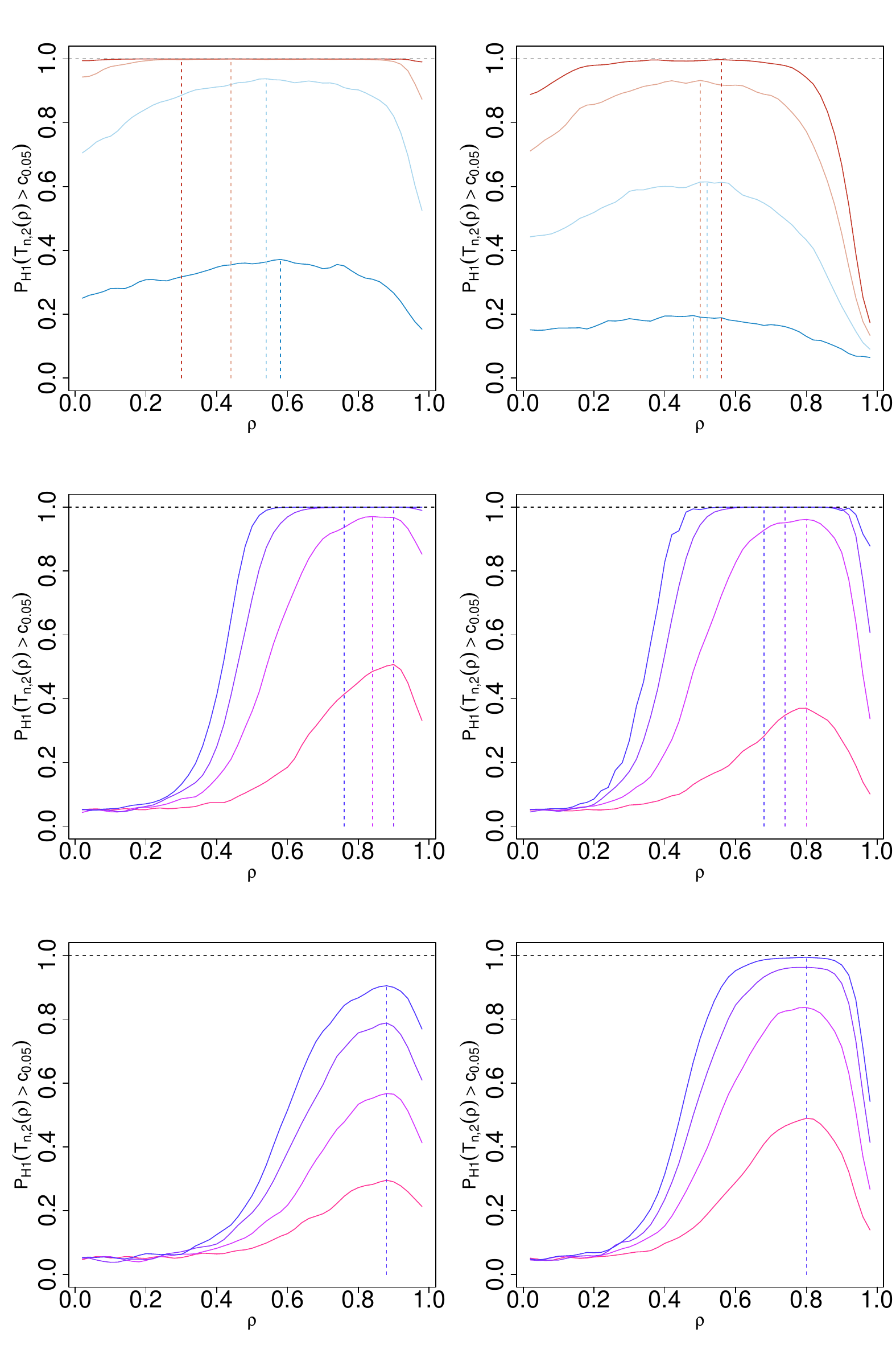}
        }
    }
\fi
	\caption{\small Analogous description as Figure \ref{fig:kappa-sens}, with $\rho\mapsto q_{n,2, \mathcal{H}_1}(\rho)$ and $\rho\mapsto \mathbb{P}_{\mathcal{H}_1}[T_{n,2}(\rho)>c_{\alpha, n, 2}(\rho)]$.} \label{fig:rho-sens}
\end{figure}

\subsection{Precision of approximate oracle parameters}
\label{sec:oracle-pow}

To measure the precision of the oracle parameters in terms of power, we compute the difference between the empirical power attained with $\tilde{\lambda}_{\mathcal{H}_1}$ and the power obtained with the oracle parameter estimated using Monte Carlo, $\lambda^{\ast}_{\mathcal{H}_1}$, for the alternatives described in Section \ref{sec:oracle-sim}. In particular, alternative distributions \ref{vMF-dist-SM}--\ref{SC-dist-SM} with $\kappa_\mathrm{dev} \in\{0.5, 1, 1.5, 2\}$ and mixtures \ref{MvMF-dist-SM}--\ref{MC-dist-SM} with $\kappa_\mathrm{dev} \in\{6, 9, 12, 15\}$ were investigated.

In order to find the oracle parameter $\tilde{\lambda}_{\mathcal{H}_1}$, we consider the grids $\Lambda$ introduced in Section \ref{sec:oracle}. The only sample size studied was $n=100$, since $n$ only affects $q_{n,\ell, \mathcal{H}_1}(\lambda)$ through a constant factor, thus keeping $\tilde{\lambda}_{\mathcal{H}_1}$ unchanged for any $n$. Figures \ref{fig:kappa-sens} and \ref{fig:rho-sens} (left block) show the curves $\lambda\mapsto q_{n,\ell, \mathcal{H}_1}(\lambda)$ suggesting that $\tilde{\lambda}_{\mathcal{H}_1}$ is unique in the alternatives explored.

To estimate the actual oracle parameters, $M=10^3$ Monte Carlo replicates were drawn from the scenarios described. For each replicate, a test based on $T_{n, \ell}(\lambda)$ at the significance level $\alpha = 0.05$ was performed using the $M=10^4$ Monte Carlo $\alpha$-critical values. Then, the empirical power points were smoothed using local constant kernel regression. The right block of Figures \ref{fig:kappa-sens} and \ref{fig:rho-sens} presents the empirical power curve $\lambda\mapsto \mathbb{P}_{\mathcal{H}_1}[T_{n,\ell}(\lambda)>c_{\alpha, n, \ell}(\lambda)]$, which demonstrates that $q_{n,\ell, \mathcal{H}_1}(\lambda)$ captures the behavior of the empirical power of $T_{n, \ell}$-based tests.

Table \ref{tbl:opt-gap} shows the difference between the empirical power obtained with $\smash{\lambda^{\ast}_{\mathcal{H}_1}}$ and $\smash{\tilde{\lambda}_{\mathcal{H}_1}}$. The median power difference for $T_{n,1}$ and $T_{n,2}$ is $0.15\%$ and $0.03\%$, respectively, which supports the adequacy of the oracle parameter $\tilde{\lambda}_{\mathcal{H}_1}$ obtained with $q_{n,\ell, \mathcal{H}_1}(\lambda)$. Additionally, Table \ref{tbl:opt-params} shows the oracle parameters $\tilde{\kappa}_{\mathcal{H}_1}$ and $\tilde{\rho}_{\mathcal{H}_1}$, and the Monte Carlo estimated oracle parameters $\kappa^{\ast}_{\mathcal{H}_1}$ and $\rho^{\ast}_{\mathcal{H}_1}$ found for alternative distributions \ref{vMF-dist-SM}--\ref{MC-dist-SM} for different concentration values $\kappa_\mathrm{dev}$. The results indicate that $\tilde{\lambda}_{\mathcal{H}_1}$ is not significantly influenced by $\kappa_{\mathrm{dev}}$.

\begin{table}[htpb!]
\iffigstabs
    \centering
    \small
    \scalebox{1}{
    \begin{tabular}{ m{4em} L{0.75cm} L{1cm} L{1cm} L{1cm} L{1cm} | L{1cm} L{1cm} L{1cm} L{1cm} } 
      \toprule
      \multirow{2}{4em}{Alternative} & \multirow{2}{0.6cm}{$\kappa_{\mathrm{dev}}$} & \multicolumn{4}{c}{$\tilde{\kappa}_{\mathcal{H}_1}$} & \multicolumn{4}{c}{$\tilde{\rho}_{\mathcal{H}_1}$} \\ 
      \cline{3-10}
       &  & $q=1$ & $q=2$ & $q=3$ & $q=5$ & $q=1$ & $q=2$ & $q=3$ & $q=5$ \\
      \midrule
      \multirow{4}{6em}{\ref{vMF-dist-SM} vMF} & 0.5 & 1.1 & 1.3 & 2.6 & 0.3 & 1.4 & 0.8 & 0.6 & 0.1\\ 
      & 1 & 0.0 & 0.1 & 0.3 & 1.6 & 0.0 & 0.1 & 0.0 & 1.1\\ 
      & 1.5 & 0.0 & 0.0 & 0.0 & 0.0 & 0.0 & 0.0 & 0.0 & 0.0\\ 
      & 2 & 0.0 & 0.0 & 0.0 & 0.0 & 0.0 & 0.0 & 0.0 & 0.0\\
      \hline
      \multirow{4}{6em}{\ref{C-dist-SM} Ca} & 0.5 & 0.5 & 0.2 & 0.0 & 0.0 & 0.6 & 0.0 & 0.1 & 0.0\\ 
      & 1 & 0.0 & 0.0 & 0.0 & 0.0 & 0.0 & 0.0 & 0.0 & 0.0\\ 
      & 1.5 & 0.0 & 0.0 & 0.0 & 0.0 & 0.0 & 0.0 & 0.0 & 0.0\\ 
      & 2 & 0.0 & 0.0 & 0.0 & 0.0 & 0.0 & 0.0 & 0.0 & 0.0 \\
      \hline
      \multirow{4}{6em}{\ref{W-dist-SM} Wa} & 0.5 & 0.6 & 0.4 & 0.0 & 0.3 & 0.5 & 0.5 & 0.4 & 0.1\\ 
      & 1 & 1.6 & 2.1 & 1.2 & 0.8 & 0.5 & 0.4 & 0.0 & 0.1\\ 
      & 1.5 & 0.2 & 1.2 & 0.8 & 0.2 & 0.0 & 0.1 & 0.3 & 0.5\\ 
      & 2 & 0.0 & 0.3 & 0.6 & 0.0 & 0.0 & 0.1 & 0.8 & 1.1\\
      \hline
      \multirow{4}{6em}{\ref{SC-dist-SM} SC} & 0.5 & 3.5 & 1.7 & 1.9 & 0.7 & 1.1 & 0.9 & 0.4 & 0.1 \\ 
      & 1 & 0.5 & 0.6 & 1.3 & 1.0 & 0.0 & 1.4 & 1.4 & 0.0 \\ 
      & 1.5 & 0.0 & 0.7 & 2.5 & 2.5 & 0.1 & 0.1 & 1.1 & 0.6 \\ 
      & 2 & 0.0 & 0.2 & 1.1 & 0.0 & 0.0 & 0.4 & 0.9 & 1.5 \\
      \hline
      \multirow{4}{6em}{\ref{MvMF-dist-SM} MvMF} & 6 & 0.0 & 0.1 & 0.2 & 0.2 & 0.0 & 0.0 & 0.7 & 0.2\\ 
      & 9 & 0.5 & 0.6 & 0.6 & 0.5 & 0.0 & 0.0 & 0.0 & 0.9\\ 
      & 12 & 0.0 & 0.0 & 0.0 & 0.1 & 0.0 & 0.0 & 0.1 & 0.1\\ 
      & 15 & 0.0 & 0.0 & 0.0 & 0.0 & 0.0 & 0.0 & 0.0 & 0.0\\
      \hline
      \multirow{4}{6em}{\ref{MC-dist-SM} MCa} & 6 & 0.0 & 1.1 & 0.3 & 0.4 & 1.3 & 0.0 & 0.0 & 1.7\\ 
      & 9 & 1.1 & 1.3 & 0.0 & 0.2 & 0.0 & 0.0 & 0.5 & 0.2\\ 
      & 12 & 0.7 & 0.3 & 0.1 & 0.0 & 0.4 & 0.0 & 0.1 & 0.0\\ 
      & 15 & 0.0 & 0.1 & 0.0 & 0.0 & 0.0 & 0.0 & 0.0 & 0.0\\
      \bottomrule
\end{tabular}
}
\fi
    \caption{\small Power difference ($\%$) between empirical oracle $\kappa^{\ast}_{\mathcal{H}_1}$ and $\rho^{\ast}_{\mathcal{H}_1}$ parameters and oracle $\tilde{\kappa}_{\mathcal{H}_1}$ and $\tilde{\rho}_{\mathcal{H}_1}$ parameters for $T_{n,1}$ and $T_{n,2}$, respectively, for alternatives \ref{vMF-dist-SM}--\ref{MC-dist-SM} with $\kappa_\mathrm{dev}$ acting in \eqref{eq:kdev1}--\eqref{eq:kdev2}.}
    \label{tbl:opt-gap}
\end{table}

\begin{table}[H]
\iffigstabs
    \centering
    \small
    \scalebox{0.85}{
    \begin{tabular}{ m{4em} L{0.75cm} L{1.5cm} L{1.5cm} L{1.5cm} L{1.5cm} | L{1.65cm} L{1.65cm} L{1.65cm} m{1.65cm} } 
      \toprule
      \multirow{2}{4em}{Alternative} & \multirow{2}{0.6cm}{$\kappa_{\mathrm{dev}}$} & \multicolumn{4}{c}{$\tilde{\kappa}_{\mathcal{H}_1}$ ($\kappa^{\ast}_{\mathcal{H}_1}$)} & \multicolumn{4}{c}{$\tilde{\rho}_{\mathcal{H}_1}$ ($\rho^{\ast}_{\mathcal{H}_1}$)} \\ 
      \cline{3-10}
       &  & $q=1$ & $q=2$ & $q=3$ & $q=5$ & $q=1$ & $q=2$ & $q=3$ & $q=5$ \\
      \midrule
      \multirow{4}{6em}{\ref{vMF-dist-SM} vMF} & 0.5 & 0.1 (0.4) & 0.01 (0.1) & 0.01 (0.2) & 0.01 (0.1) & 0.02 (0.10) & 0.02 (0.06) & 0.02 (0.08) & 0.02 (0.04)\\ 
      & 1 & 0.2 (0.01) & 0.2 (0.8) & 0.2 (0.5) & 0.1 (0.5) & 0.06 (0.02) & 0.04 (0.18) & 0.02 (0.02) & 0.02 (0.12)\\ 
      & 1.5 & 0.5 (0.01) & 0.5 (0.01) & 0.2 (0.01) & 0.2 (1.1) & 0.12 (0.02) & 0.08 (0.02) & 0.06 (0.02) & 0.04 (0.08)\\ 
      & 2 & 0.7 (0.01) & 0.7 (0.01) & 0.5 (0.01) & 0.5 (0.01) & 0.18 (0.02) & 0.14 (0.02) & 0.08 (0.02) & 0.06 (0.02)\\
      \hline
      \multirow{4}{6em}{\ref{C-dist-SM} Ca} & 0.5 & 0.2 (0.01) & 0.5 (1) & 0.5 (0.3) & 0.5 (3.1) & 0.08 (0.02) & 0.08 (0.06) & 0.06 (0.08) & 0.08 (0.02)\\ 
      & 1 & 0.6 (2.8) & 0.8 (0.01) & 0.9 (0.01) & 1.2 (0.01) & 0.14 (0.58) & 0.16 (0.02) & 0.14 (0.02) & 0.14 (0.02)\\ 
      & 1.5 & 1.0 (1.4) & 1.0 (0.01) & 1.2 (4.1) & 1.7 (0.01) & 0.20 (0.02) & 0.20 (0.02) & 0.20 (0.02) & 0.20 (0.02)\\ 
      & 2 & 1.0 (0.01) & 1.5 (0.01) & 1.7 (0.01) & 2.2 (0.01) & 0.24 (0.28) & 0.26 (0.02) & 0.24 (0.02) & 0.24 (0.02) \\
      \hline
      \multirow{4}{6em}{\ref{W-dist-SM} Wa} & 0.5 & 6.8 (8.2) & 5.6 (4.9) & 5.4 (4.9) & 5.4 (8.2) & 0.70 (0.72) & 0.60 (0.50) & 0.54 (0.68) & 0.46 (0.28)\\ 
      & 1 & 6.8 (7.6) & 5.8 (8.0) & 5.4 (7.4) & 5.4 (7.2) & 0.70 (0.78) & 0.60 (0.62) & 0.54 (0.56) & 0.46 (0.50)\\ 
      & 1.5 & 7.0 (7.6) & 5.8 (6.2) & 5.6 (4.8) & 5.4 (7.6) & 0.70 (0.68) & 0.60 (0.58) & 0.54 (0.52) & 0.46 (0.56)\\ 
      & 2 & 7.2 (7.8) & 6.0 (4.7) & 5.6 (4.2) & 5.6 (5.6) & 0.72 (0.58) & 0.60 (0.54) & 0.54 (0.58) & 0.46 (0.42)\\
      \hline
      \multirow{4}{6em}{\ref{SC-dist-SM} SC} & 0.5 & 3.4 (4.5) & 2.6 (3.8) & 2.3 (2.8) & 1.7 (2.8) & 0.52 (0.58) & 0.38 (0.48) & 0.30 (0.40) & 0.20 (0.14) \\ 
      & 1 & 3.8 (4.2) & 2.8 (3.4) & 2.3 (2.9) & 1.7 (2.0) & 0.54 (0.54) & 0.40 (0.52) & 0.30 (0.40) & 0.20 (0.20) \\ 
      & 1.5 & 4.4 (8.2) & 3.1 (3.3) & 2.4 (3.1) & 1.7 (2.1) & 0.58 (0.44) & 0.42 (0.50) & 0.32 (0.34) & 0.18 (0.20) \\ 
      & 2 & 5.0 (1.4) & 3.3 (2.7) & 2.5 (3.6) & 1.7 (1.7) & 0.62 (0.30) & 0.44 (0.56) & 0.32 (0.40) & 0.20 (0.24) \\
      \hline
      \multirow{4}{6em}{\ref{MvMF-dist-SM} MvMF} & 6 & 30 (20) & 20 (19) & 16 (17) & 14 (16) & 0.90 (0.90) & 0.78 (0.80) & 0.70 (0.74) & 0.64 (0.58)\\ 
      & 9 & 30 (40) & 20 (14) & 17 (16) & 14 (13) & 0.84 (0.84) & 0.80 (0.80) & 0.72 (0.72) & 0.66 (0.64)\\ 
      & 12 & 30 (17) & 20 (9) & 17 (18) & 15 (12) & 0.86 (0.90) & 0.80 (0.74) & 0.76 (0.66) & 0.62 (0.64)\\ 
      & 15 & 30 (8) & 20 (6) & 18 (7) & 15 (10) & 0.86 (0.76) & 0.80 (0.68) & 0.74 (0.62) & 0.68 (0.62)\\
      \hline
      \multirow{4}{6em}{\ref{MC-dist-SM} MCa} & 6 & 30 (30) & 20 (17) & 18 (19) & 16 (14) & 0.84 (0.88) & 0.80 (0.80) & 0.72 (0.72) & 0.68 (0.64)\\ 
      & 9 & 30 (35) & 25 (20) & 20 (20) & 18 (14) & 0.88 (0.88) & 0.80 (0.80) & 0.76 (0.72) & 0.68 (0.66)\\ 
      & 12 & 35 (40) & 25 (20) & 20 (15) & 19 (13) & 0.86 (0.88) & 0.80 (0.80) & 0.76 (0.74) & 0.68 (0.66)\\ 
      & 15 & 35 (30) & 25 (30) & 20 (19) & 20 (18) & 0.88 (0.88) & 0.80 (0.80) & 0.76 (0.74) & 0.72 (0.62)\\
      \bottomrule
\end{tabular}
}
\fi
    \caption{\small Oracle $\tilde{\kappa}_{\mathcal{H}_1}$ and $\tilde{\rho}_{\mathcal{H}_1}$ (and Monte Carlo estimated oracle $\kappa^{\ast}_{\mathcal{H}_1}$ and $\rho^{\ast}_{\mathcal{H}_1}$) parameters for $T_{n,1}$ and $T_{n,2}$, respectively, for alternatives \ref{vMF-dist-SM}--\ref{MC-dist-SM} with $\kappa_\mathrm{dev}$.}
    \label{tbl:opt-params}
\end{table}


\subsection{Asymptotic null distribution and gamma match accuracies}
\label{app:null_accuracy}

Tables \ref{tbl:null_asymp_dist_01} and \ref{tbl:null_asymp_dist_10} contain extended simulation results for the investigation of null tail probabilities computation for $\alpha=0.01$ and $\alpha=0.10$, using the asymptotic null distribution computed by Imhof's method, the gamma-match proposed in Section \ref{sec:exact_approx}, and the Monte Carlo ($M=10^5$) approximation. The simulation setting is the same as described in Section \ref{sec:asymp_dist}.

\begin{table}[htpb!]
\iffigstabs
\centering
\small
\scalebox{1}{
\begin{tabular}{ m{0.35cm} m{0.35cm} m{0.9cm} R{1.2cm} R{1.2cm} R{1.2cm} R{1.3cm} R{1.3cm} R{1.3cm} } 
\toprule
$q$ & $n$ & Type & $\kappa=0.1$ & $\kappa=1$ & $\kappa=5$ & $\rho=0.25$ & $\rho=0.5$ & $\rho=0.75$ \\
  \midrule
\multirow{9}{0.6cm}{$1$} & \multirow{3}{0.6cm}{$10$} & Asymp. & $0.0067$ & $0.0076$ & $0.0087$ & $0.0073$ & $0.0082$ & $\mathbf{0.0097}$ \\ 
& & Gamma & $\mathbf{0.0093}$ & $0.0110$ & $0.0124$ & $0.0116$ & $0.0131$ & $0.0149$ \\ 
& & MC & $\mathbf{0.0104}$ & $\mathbf{0.0105}$ & $\mathbf{0.0107}$ & $\mathbf{0.0105}$ & $\mathbf{0.0107}$ & $\mathbf{0.0106}$ \\ 
\cline{2-9}
& \multirow{3}{0.6cm}{$50$} & Asymp. & $\mathbf{0.0099}$ & $0.0091$ & $\mathbf{0.0098}$ & $\mathbf{0.0096}$ & $\mathbf{0.0092}$ & $\mathbf{0.0100}$ \\ 
& & Gamma & $\mathbf{0.0101}$ & $0.0114$ & $0.0113$ & $0.0119$ & $0.0126$ & $0.0135$ \\ 
& & MC & $\mathbf{0.0102}$ & $\mathbf{0.0101}$ & $\mathbf{0.0099}$ & $\mathbf{0.0101}$ & $\mathbf{0.0104}$ & $\mathbf{0.0106}$ \\ 
\cline{2-9}
& \multirow{3}{0.6cm}{$200$} & Asymp. & $\mathbf{0.0102}$ & $\mathbf{0.0096}$ & $\mathbf{0.0097}$ & $\mathbf{0.0101}$ & $\mathbf{0.0099}$ & $\mathbf{0.0099}$ \\ 
& & Gamma & $\mathbf{0.0103}$ & $0.0119$ & $0.0118$ & $0.0120$ & $0.0122$ & $0.0127$ \\ 
& & MC & $\mathbf{0.0101}$ & $\mathbf{0.0100}$ & $\mathbf{0.0097}$ & $\mathbf{0.0100}$ & $\mathbf{0.0097}$ & $\mathbf{0.0097}$ \\ 
\hline
\multirow{9}{0.6cm}{$2$} & \multirow{3}{0.6cm}{$10$} & Asymp. & $0.0067$ & $0.0078$ & $\mathbf{0.0101}$ & $0.0076$ & $\mathbf{0.0100}$ & $0.0154$ \\ 
& & Gamma & $0.0091$ & $0.0120$ & $0.0154$ & $0.0130$ & $0.0158$ & $0.0234$ \\ 
& & MC & $\mathbf{0.0101}$ & $\mathbf{0.0102}$ & $\mathbf{0.0100}$ & $\mathbf{0.0103}$ & $\mathbf{0.0103}$ & $\mathbf{0.0097}$ \\ 
\cline{2-9}
& \multirow{3}{0.6cm}{$50$} & Asymp. & $0.0089$ & $\mathbf{0.0094}$ & $\mathbf{0.0096}$ & $\mathbf{0.0094}$ & $\mathbf{0.0099}$ & $0.0111$ \\ 
& & Gamma & $\mathbf{0.0100}$ & $0.0126$ & $0.0134$ & $0.0140$ & $0.0146$ & $0.0152$ \\ 
& & MC & $\mathbf{0.0104}$ & $\mathbf{0.0105}$ & $\mathbf{0.0108}$ & $\mathbf{0.0106}$ & $0.0110$ & $\mathbf{0.0106}$ \\ 
\cline{2-9}
& \multirow{3}{0.6cm}{$200$} & Asymp. & $\mathbf{0.0098}$ & $\mathbf{0.0097}$ & $\mathbf{0.0099}$ & $\mathbf{0.0099}$ & $\mathbf{0.0103}$ & $\mathbf{0.0102}$ \\ 
& & Gamma & $\mathbf{0.0101}$ & $0.0126$ & $0.0125$ & $0.0132$ & $0.0145$ & $0.0138$ \\ 
& & MC & $\mathbf{0.0107}$ & ${0.0109}$ & $\mathbf{0.0106}$ & ${0.0111}$ & ${0.0111}$ & $\mathbf{0.0106}$ \\ 
\hline
\multirow{9}{0.6cm}{$3$} & \multirow{3}{0.6cm}{$10$} & Asymp. & $0.0063$ & $0.0077$ & $0.0119$ & $0.0079$ & $0.0114$ & $0.0274$ \\ 
& & Gamma & $0.0088$ & $0.0125$ & $0.0180$ & $0.0143$ & $0.0199$ & $0.0346$ \\ 
& & MC & $\mathbf{0.0103}$ & $\mathbf{0.0104}$ & $\mathbf{0.0104}$ & $\mathbf{0.0102}$ & $\mathbf{0.0106}$ & $\mathbf{0.0098}$ \\ 
\cline{2-9}
& \multirow{3}{0.6cm}{$50$} & Asymp. & $0.0090$ & $\mathbf{0.0092}$ & $\mathbf{0.0101}$ & $\mathbf{0.0094}$ & $\mathbf{0.0103}$ & $0.0144$ \\ 
& & Gamma & $\mathbf{0.0099}$ & $0.0133$ & $0.0149$ & $0.0143$ & $0.0158$ & $0.0181$ \\ 
& & MC & $\mathbf{0.0103}$ & $\mathbf{0.0102}$ & $\mathbf{0.0099}$ & $\mathbf{0.0099}$ & $\mathbf{0.0101}$ & $\mathbf{0.0106}$ \\ 
\cline{2-9}
& \multirow{3}{0.6cm}{$200$} & Asymp. & $\mathbf{0.0098}$ & $\mathbf{0.0097}$ & $\mathbf{0.0101}$ & $\mathbf{0.0099}$ & $\mathbf{0.0100}$ & $0.0116$ \\ 
& & Gamma & $\mathbf{0.0106}$ & $0.0133$ & $0.0143$ & $0.0150$ & $0.0153$ & $0.0139$ \\ 
& & MC & $\mathbf{0.0098}$ & $\mathbf{0.0099}$ & $\mathbf{0.0107}$ & $\mathbf{0.0101}$ & $\mathbf{0.0102}$ & $\mathbf{0.0097}$ \\ 
\hline
\multirow{9}{0.6cm}{$5$} & \multirow{3}{0.6cm}{$10$} & Asymp. & $0.0063$ & $0.0075$ & $0.0133$ & $0.0087$ & $0.0173$ & $0.0238$ \\ 
& & Gamma & $\mathbf{0.0093}$ & $0.0128$ & $0.0215$ & $0.0168$ & $0.0274$ & $0.0250$ \\ 
& & MC & $\mathbf{0.0105}$ & $\mathbf{0.0104}$ & $\mathbf{0.0104}$ & $\mathbf{0.0106}$ & $\mathbf{0.0104}$ & $\mathbf{0.0095}$ \\ 
\cline{2-9}
& \multirow{3}{0.6cm}{$50$} & Asymp. & $\mathbf{0.0094}$ & $\mathbf{0.0095}$ & $\mathbf{0.0105}$ & $\mathbf{0.0095}$ & $0.0120$ & $0.0304$ \\ 
& & Gamma & $\mathbf{0.0099}$ & $0.0137$ & $0.0157$ & $0.0158$ & $0.0177$ & $0.0315$ \\ 
& & MC & $\mathbf{0.0100}$ & $\mathbf{0.0102}$ & $\mathbf{0.0102}$ & $\mathbf{0.0104}$ & $\mathbf{0.0097}$ & $\mathbf{0.0107}$ \\ 
\cline{2-9}
& \multirow{3}{0.6cm}{$200$} & Asymp. & $\mathbf{0.0097}$ & $\mathbf{0.0094}$ & $\mathbf{0.0105}$ & $\mathbf{0.0102}$ & $\mathbf{0.0105}$ & $0.0156$ \\ 
& & Gamma & $\mathbf{0.0101}$ & $0.0132$ & $0.0147$ & $0.0155$ & $0.0153$ & $0.0172$ \\ 
& & MC & $\mathbf{0.0106}$ & $\mathbf{0.0106}$ & $\mathbf{0.0101}$ & $\mathbf{0.0106}$ & $\mathbf{0.0106}$ & $\mathbf{0.0104}$ \\ 
  \bottomrule
      \end{tabular}
}
\fi
    \caption{\small Empirical rejection proportion for significance level $\alpha = 0.01$ of $T_{n,1}(\kappa)$ and $T_{n,2}(\rho)$ computed with $M = 10^5$ Monte Carlo samples using each approximation method: asymptotic distribution computed by Imhof's method, our gamma-match approximation, and Monte Carlo ($M=10^5$) approximation. Boldface denotes that the empirical rejection proportion lies within the $95\%$ confidence interval $(0.0092, 0.0108)$.\label{tbl:null_asymp_dist_01}}
\end{table}

\begin{table}[htpb!]
\iffigstabs
\centering
\small
\scalebox{1}{
\begin{tabular}{ m{0.35cm} m{0.35cm} m{0.9cm} R{1.2cm} R{1.2cm} R{1.2cm} R{1.3cm} R{1.3cm} R{1.3cm} } 
\toprule
$q$ & $n$ & Type & $\kappa=0.1$ & $\kappa=1$ & $\kappa=5$ & $\rho=0.25$ & $\rho=0.5$ & $\rho=0.75$ \\
\midrule
\multirow{9}{0.6cm}{$1$} & \multirow{3}{0.6cm}{$10$} & Asymp. & $0.0960$ & $0.0959$ & $0.0903$ & $0.0941$ & $0.0925$ & $0.0893$ \\ 
& & Gamma & $0.1041$ & $\mathbf{0.1000}$ & $0.0957$ & $0.0983$ & $0.0968$ & $0.0979$ \\ 
& & MC & $\mathbf{0.1007}$ & $\mathbf{0.1013}$ & $\mathbf{0.1013}$ & ${0.1017}$ & $0.1016$ & ${0.1021}$ \\ 
\cline{2-9}
& \multirow{3}{0.6cm}{$50$} & Asymp. & $\mathbf{0.0995}$ & $\mathbf{0.0993}$ & $0.0981$ & $\mathbf{0.0991}$ & $\mathbf{0.0985}$ & $0.0979$ \\ 
& & Gamma & $\mathbf{0.0998}$ & $\mathbf{0.0988}$ & $0.0969$ & $0.0955$ & $0.0970$ & $\mathbf{0.0989}$ \\ 
& & MC & $\mathbf{0.1013}$ & $\mathbf{0.1007}$ & $\mathbf{0.1006}$ & $\mathbf{0.1010}$ & $\mathbf{0.1010}$ & $\mathbf{0.1010}$ \\ 
\cline{2-9}
& \multirow{3}{0.6cm}{$200$} & Asymp. & $\mathbf{0.1005}$ & $\mathbf{0.1006}$ & $\mathbf{0.0993}$ & $\mathbf{0.1013}$ & $\mathbf{0.1006}$ & $\mathbf{0.0991}$ \\ 
& & Gamma & $\mathbf{0.0999}$ & $0.0978$ & $\mathbf{0.0984}$ & $0.0972$ & $0.0968$ & $\mathbf{0.0990}$ \\ 
& & MC & $\mathbf{0.0989}$ & $\mathbf{0.1003}$ & $\mathbf{0.1015}$ & $\mathbf{0.1001}$ & $\mathbf{0.1012}$ & $\mathbf{0.1015}$ \\ 
\hline
\multirow{9}{0.6cm}{$2$} & \multirow{3}{0.6cm}{$10$} & Asymp. & $0.0961$ & $0.0957$ & $0.0891$ & $0.0931$ & $0.0907$ & $0.0909$ \\ 
& & Gamma & $0.1030$ & $\mathbf{0.1003}$ & $0.0969$ & $0.0980$ & $\mathbf{0.0987}$ & $0.1022$ \\ 
& & MC & $\mathbf{0.1006}$ & $\mathbf{0.1013}$ & ${0.1025}$ & ${0.1016}$ & ${0.1022}$ & $\mathbf{0.0991}$ \\ 
\cline{2-9}
& \multirow{3}{0.6cm}{$50$} & Asymp. & $\mathbf{0.0990}$ & $\mathbf{0.0990}$ & $\mathbf{0.1004}$ & $\mathbf{0.0984}$ & $0.0978$ & $\mathbf{0.1000}$ \\ 
& & Gamma & $\mathbf{0.0991}$ & $0.0972$ & $\mathbf{0.0988}$ & $0.0982$ & $\mathbf{0.0990}$ & $0.1019$ \\ 
& & MC & $\mathbf{0.1015}$ & $\mathbf{0.1015}$ & ${0.1022}$ & ${0.1016}$ & ${0.1031}$ & ${0.1027}$ \\ 
\cline{2-9}
& \multirow{3}{0.6cm}{$200$} & Asymp. & $\mathbf{0.1005}$ & $\mathbf{0.0985}$ & $\mathbf{0.0995}$ & $\mathbf{0.0995}$ & $0.0982$ & $\mathbf{0.0993}$ \\ 
& & Gamma & $\mathbf{0.0987}$ & $\mathbf{0.1005}$ & $\mathbf{0.0995}$ & $0.0967$ & $\mathbf{0.1001}$ & $\mathbf{0.1006}$ \\ 
& & MC & $\mathbf{0.0999}$ & $\mathbf{0.1010}$ & ${0.1027}$ & $\mathbf{0.1005}$ & ${0.1019}$ & ${0.1023}$ \\ 
\hline
\multirow{9}{0.6cm}{$3$} & \multirow{3}{0.6cm}{$10$} & Asymp. & $0.0943$ & $0.0937$ & $0.0895$ & $0.0914$ & $0.0911$ & $0.0973$ \\ 
& & Gamma & $0.1020$ & $\mathbf{0.1004}$ & $0.0973$ & $\mathbf{0.0989}$ & $\mathbf{0.1000}$ & $0.1041$ \\ 
& & MC & $\mathbf{0.1005}$ & $\mathbf{0.1006}$ & $\mathbf{0.1003}$ & $\mathbf{0.1014}$ & $\mathbf{0.1001}$ & $\mathbf{0.1000}$ \\ 
\cline{2-9}
& \multirow{3}{0.6cm}{$50$} & Asymp. & $0.0983$ & $\mathbf{0.0996}$ & $\mathbf{0.0984}$ & $\mathbf{0.0970}$ & $\mathbf{0.0984}$ & $\mathbf{0.1001}$ \\ 
& & Gamma & $\mathbf{0.0992}$ & $\mathbf{0.0996}$ & $\mathbf{0.1004}$ & $\mathbf{0.1002}$ & $\mathbf{0.1007}$ & $0.1018$ \\ 
& & MC & ${0.1029}$ & ${0.1025}$ & ${0.1039}$ & ${0.1032}$ & ${0.1035}$ & ${0.1017}$ \\ 
\cline{2-9}
& \multirow{3}{0.6cm}{$200$} & Asymp. & $\mathbf{0.0997}$ & $0.0983$ & $\mathbf{0.1006}$ & $\mathbf{0.0994}$ & $\mathbf{0.0994}$ & $0.1018$ \\ 
& & Gamma & $\mathbf{0.1000}$ & $0.0965$ & $0.0974$ & $0.0983$ & $\mathbf{0.0997}$ & $0.1027$ \\ 
& & MC & $\mathbf{0.0996}$ & $\mathbf{0.0995}$ & $\mathbf{0.1002}$ & $\mathbf{0.1002}$ & $\mathbf{0.1004}$ & $\mathbf{0.1002}$ \\ 
\hline
\multirow{9}{0.6cm}{5} & \multirow{3}{0.6cm}{$10$} & Asymp. & $0.0930$ & $0.0917$ & $0.0921$ & $0.0911$ & $0.0943$ & $0.0489$ \\ 
& & Gamma & $0.1028$ & $\mathbf{0.1014}$ & $0.1037$ & $0.1016$ & $0.1039$ & $0.0509$ \\ 
& & MC & $\mathbf{0.1002}$ & $\mathbf{0.1009}$ & $\mathbf{0.1011}$ & $\mathbf{0.1012}$ & $\mathbf{0.1008}$ & $\mathbf{0.0994}$ \\ 
\cline{2-9}
& \multirow{3}{0.6cm}{$50$} & Asymp. & $0.0976$ & $0.0977$ & $0.0982$ & $0.0970$ & $\mathbf{0.0998}$ & $\mathbf{0.1004}$ \\ 
& & Gamma & $0.1027$ & $\mathbf{0.1004}$ & $0.1043$ & $\mathbf{0.0988}$ & $0.1036$ & $0.1023$ \\ 
& & MC & $\mathbf{0.0985}$ & $\mathbf{0.0988}$ & $\mathbf{0.1004}$ & $\mathbf{0.0989}$ & $\mathbf{0.1007}$ & $\mathbf{0.0999}$ \\ 
\cline{2-9}
& \multirow{3}{0.6cm}{$200$} & Asymp. & $\mathbf{0.0995}$ & $\mathbf{0.1003}$ & $0.0980$ & $\mathbf{0.0995}$ & $\mathbf{0.1002}$ & $0.1042$ \\ 
& & Gamma & $\mathbf{0.0997}$ & $\mathbf{0.1001}$ & $0.1018$ & $\mathbf{0.1008}$ & $0.1019$ & $0.1047$ \\ 
& & MC & ${0.1016}$ & ${0.1020}$ & $\mathbf{0.0998}$ & $\mathbf{0.1002}$ & $\mathbf{0.0983}$ & $\mathbf{0.1004}$ \\ 
\bottomrule
      \end{tabular}
}
\fi
    \caption{\small Empirical rejection proportion for significance level $\alpha = 0.10$ of $T_{n,1}(\kappa)$ and $T_{n,2}(\rho)$ computed with $M = 10^5$ Monte Carlo samples using each approximation method: asymptotic distribution computed by Imhof's method, our gamma-match approximation, and Monte Carlo ($M=10^5$) approximation. Boldface denotes that the empirical rejection proportion lies within the $95\%$ confidence interval $(0.0984, 0.1015)$.\label{tbl:null_asymp_dist_10}}
\end{table}

\subsection{\texorpdfstring{Further experiments on the estimation of $\tilde{\lambda}_{\Hcal_1}$}{Further experiments on the estimation of tilde{lambda}H1}}
\label{app:empirical_split}

As introduced in Section \ref{sec:estimated-stat}, a first attempt to develop a test based on $T_{n, \ell}$ is to use the two-subsample split approach. The following definition formalizes this procedure.

\begin{definition}[Single-partition test of uniformity based on $T_{n, \ell}$] Given an iid sample $\mathcal{S}=\{\mathbf{X}_1,\ldots,\mathbf{X}_n\}$, the $p\cdot100\%$-test based on $T_{n, \ell}$ of $\Hcal_0$ at significance level $\alpha$ proceeds as follows:

    \begin{enumerate}
    \item Split $\mathcal{S}$ into two (disjoint) subsamples $\mathcal{S}_1$ and $\mathcal{S}_2$ of size $\vert\mathcal{S}_1\vert = (1-p)\,n$ and $\vert\mathcal{S}_2\vert = p\,n$, respectively.
    \item Compute $\widehat{\lambda}(\mathcal{S}_1) = \arg \max_{\lambda\in\Lambda} T_{\vert\mathcal{S}_1\vert, \ell}(\lambda)\big/\sqrt{\mathbb{V}\mathrm{ar}_{\mathcal{H}_0}[T_{ \vert\mathcal{S}_1\vert, \ell}(\lambda)]}$, the estimate of $\tilde{\lambda}_{\mathcal{H}_1}$ using $\mathcal{S}_1$.
    \item Use the subsample $\mathcal{S}_2$ to perform the test based on $T_{\vert\mathcal{S}_2\vert,\ell}(\widehat{\lambda}(\mathcal{S}_1), \mathcal{S}_2)$ at significance level $\alpha$.
\end{enumerate} 
\end{definition}

A power investigation of these tests is presented in Figure \ref{fig:empirical-split} where the $80\%$, $75\%$, and $50\%$-tests are performed (using gamma-match $\alpha$-critical values) on $10^3$ Monte Carlo replicates of size $n=100$ for distributions \ref{vMF-dist-SM}--\ref{MC-dist-SM} depending on $\kappa_{\mathrm{dev}}$. As the curves suggest, the loss of information invested in estimating $\tilde{\lambda}_{\mathcal{H}_1}$ reduces the power of the test with respect to the optimal one, especially in multimodal distributions.

\begin{figure}[htpb!]
\iffigstabs
    \centering
    \scalebox{0.9}{
        \subfloat[][]{
        	\includegraphics[width=0.5\textwidth]{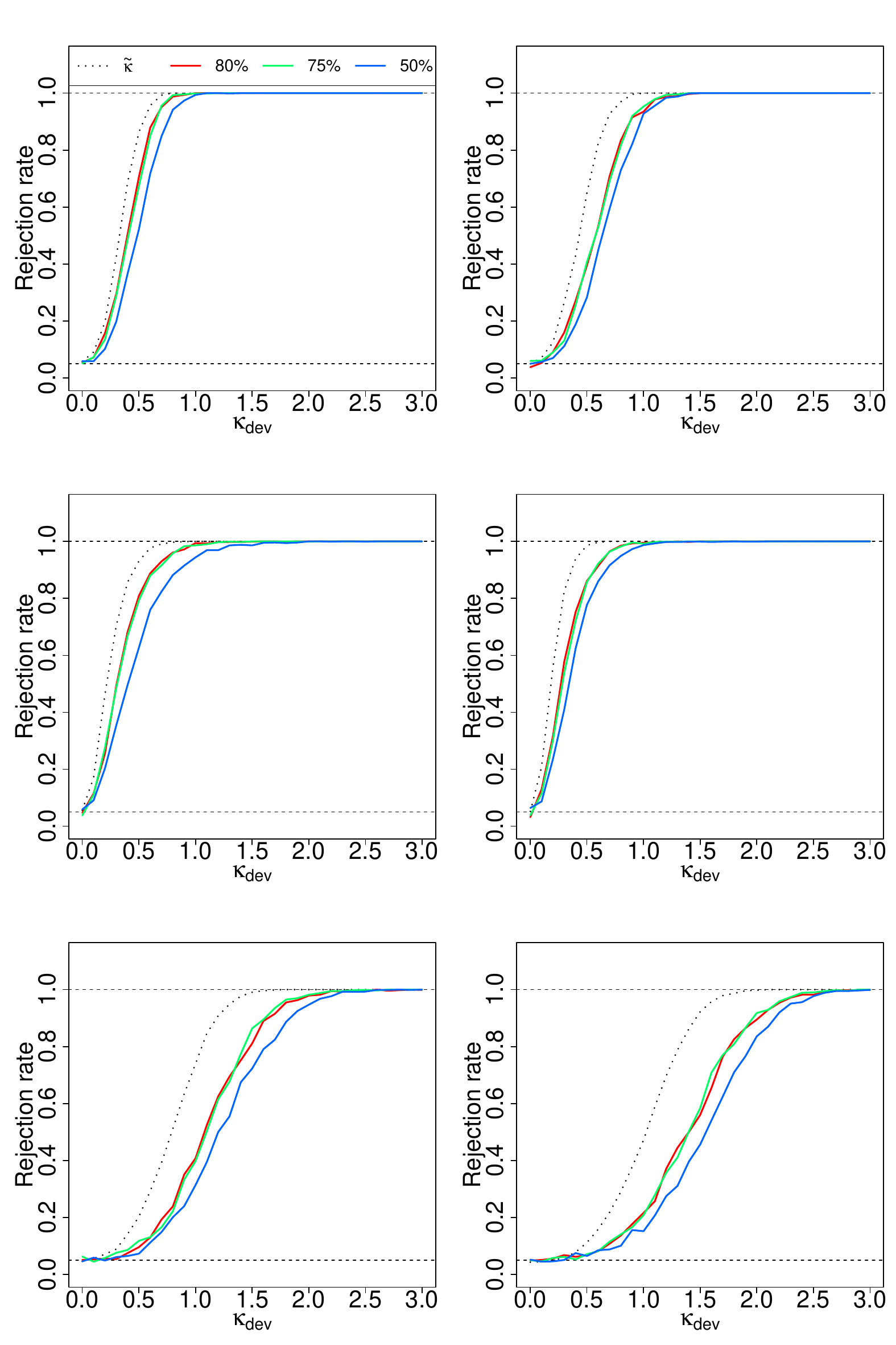}
                \vrule
        	\includegraphics[width=0.5\textwidth]{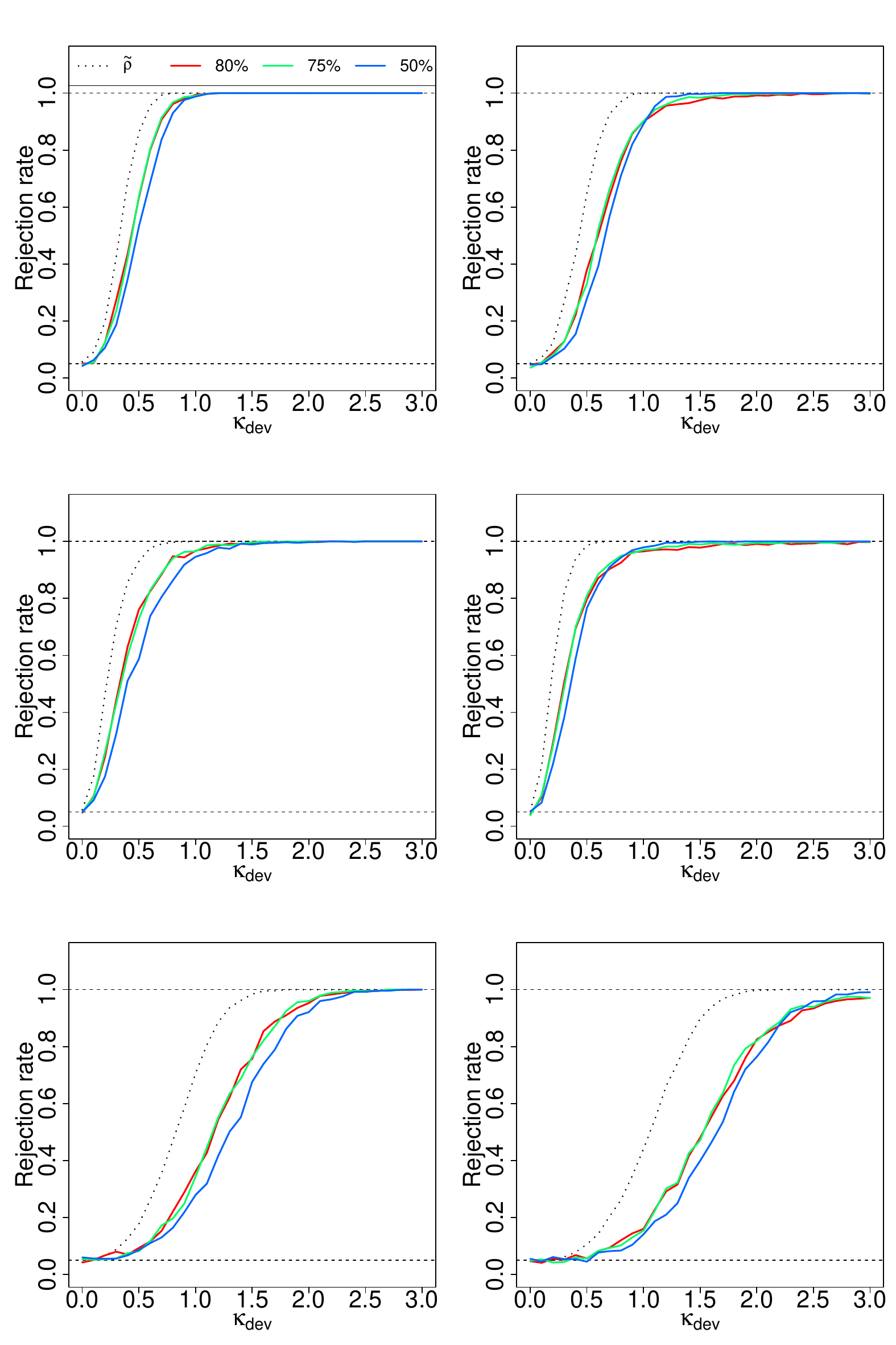}
        }
    }
    
    \scalebox{0.9}{
        \subfloat[][]{
    	\includegraphics[width=0.5\textwidth]{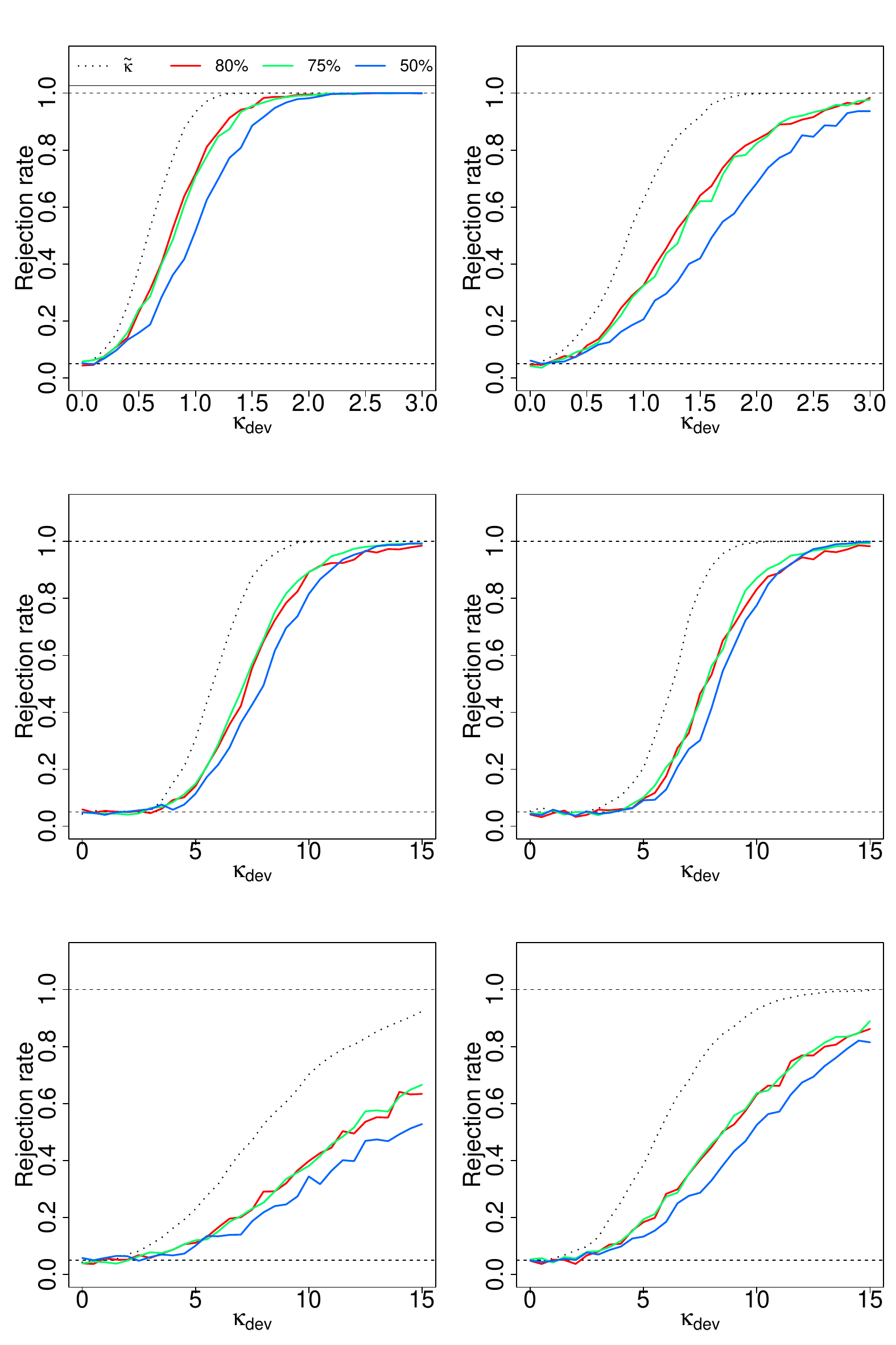}
            \vrule
    	\includegraphics[width=0.5\textwidth]{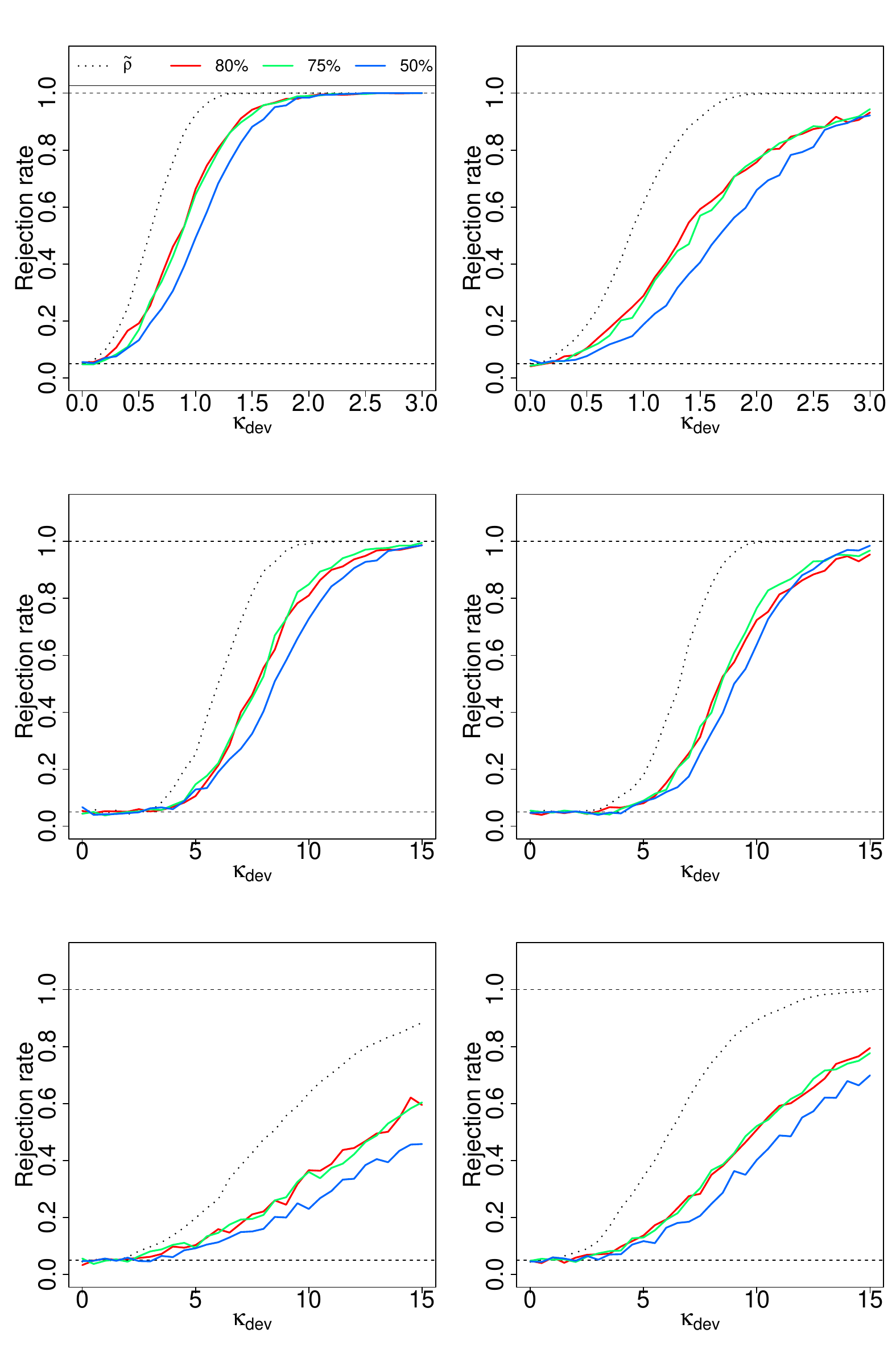}
        }
    }
\fi
	\caption{\small Empirical rejection frequency of the $p\%$-test based on $T_{n, 1}$ (left panel) and $T_{n, 2}$ (right panel) at significance level $\alpha=0.05$ for alternative distributions \ref{vMF-dist-SM}--\ref{MC-dist-SM} with concentration $\kappa_\mathrm{dev}$. Proportion $p$ is indicated in the legend. $M=10^3$ samples of size $n=100$ were drawn from the alternative distribution. Dotted curves indicate the power of the oracle test based on $T_{n, \ell}(\tilde{\lambda}_{\mathcal{H}_1})$.} \label{fig:empirical-split}
\end{figure}

%


\fi

\end{document}